\documentclass[preprint]{aastex}
\begin{document}

\title{Evolution and environment of early-type galaxies}
\author{Mariangela Bernardi$^{1}$, Robert C. Nichol$^{2}$, 
        Ravi K. Sheth$^1$, C. J. Miller$^3$ and J. Brinkmann$^4$}

\begin{abstract}
  We study the photometric and spectral properties of 39320 
  early-type galaxies within the Sloan Digital Sky Survey (SDSS), 
  as a function of both local environment and redshift. 
  The distance to the nearest cluster of galaxies (scaled by the 
  virial radius of the cluster) and the distance to the $10^{th}$ 
  nearest luminous neighbor ($M_r<-21.5$) were used to define two 
  extremes in environment.  The properties of early-type galaxies 
  are weakly but significantly different in these two extremes.  
  In particular, the Fundamental Plane of early--type galaxies in the 
  lowest density environment is systematically brighter in surface 
  brightness (by $\sim 0.08$~mag/arcsec$^2$ in $r$) compared to the 
  high density environment. A similar brightening is seen in the 
  SDSS $g$ and $i$ photometric passbands. Although the Fundamental 
  Plane is slightly thicker in the bluer passbands, we do not find 
  any significant correlation between the thickness and the environment. 

  Chemical abundance indicators are studied using composite spectra, 
  which we provide in tabular form.  Tables of line strengths measured 
  from these spectra, and parameters derived from these line strengths 
  are also provided.  From these we find that, at fixed luminosity, 
  early-type galaxies in low density environments are slightly bluer, 
  with stronger O{\small II} emission and stronger H$\delta$ and 
  H$\gamma$ Balmer absorption lines, indicative of star--formation in 
  the not very distant past.  
  These galaxies also tend to have systematically weaker D4000 indices.  
  The Lick indices and $\alpha$--element abundance indicators correlate 
  weakly but significantly with environment.  For example, at fixed 
  velocity dispersion, Mg is weaker in early--type galaxies in low 
  density environments by 30\% of the rms scatter across the full sample, 
  whereas most Fe indicators show no significant environmental dependence. 

  The galaxies in our sample span a redshift range which corresponds 
  to lookback times of $\sim$1~Gyr.  We see clear evidence for 
  evolution of line-index strengths over this time.  
  Since the low redshift population is almost certainly a passively aged 
  version of the more distant population, age is likely the main driver 
  for any observed evolution.
  We use the observed redshift evolution as a model independent clock 
  to identify indicators which are more sensitive to age than to other 
  effects such as metallicity.  
  In principle, for a passively evolving population, comparison of the 
  trends with redshift and environment constrain how strongly the 
  luminosity-weighted ages and metallicities depend on environment.  
  We develop a method for doing this which does not depend upon the 
  details of stellar population synthesis models.  Our analysis suggests 
  that the galaxies which populate the densest regions in our sample are 
  older by $\sim 1$~Gyrs than objects of the same luminosity in the 
  least dense regions, and that metallicity differences are negligible.  

  We also use single burst stellar population synthesis models, which 
  allow for non-solar $\alpha$-element abundance ratios, to interpret 
  our data.  The combination H$\beta$, Mg$b$ and 
  $\langle$Fe$\rangle$ suggests that age, metallicity and 
  $\alpha$-enhancement all increase with velocity dispersion.  
  The objects at lower redshifts are older, but have the same 
  metallicities and $\alpha$-enhancements as their counterparts of the 
  same $\sigma$ at higher redshifts, as expected if the low redshift 
  sample is a passively-aged version of the sample at higher redshifts.  
  In addition, objects in dense environments are less than $1$~Gyr older and 
  $\alpha$-enhanced by $\sim 0.02$ relative to their counterparts of the same 
  velocity dispersion in less dense regions, but the metallicities 
  show no dependence on environment.  This suggests that, in dense 
  regions, the stars in early-type galaxies formed at slightly earlier 
  times, and on a slightly shorter timescale, than in less dense regions.  
  Using H$\gamma_{\rm F}$ instead of H$\beta$ leads to slightly younger 
  ages, but the same qualitative differences between environments.  
  In particular, we find no evidence that objects in low density regions 
  are more metal rich.  

\end{abstract}  
\keywords{galaxies: elliptical --- galaxies: evolution --- 
          galaxies: fundamental parameters --- galaxies: photometry --- 
          galaxies: stellar content}

\footnotetext[1] {Department of Physics and Astronomy, 
                  University of Pennsylvania, Philadelphia, PA 19104}
\footnotetext[2] {Institute of Cosmology and Gravitation (ICG), 
                  Mercantile House, Hampshire Terrace, 
                  University of Portsmouth, Portsmouth, PO1 2EG, UK}
\footnotetext[3] {Cerro-Tololo Inter-American Observatory, NOAO, Casilla 603, 
                  La Serena, Chile}
\footnotetext[4] {Apache Point Observatory, 2001 Apache Point Road, 
                  P.O. Box 59, Sunspot, NM 88349-0059}

\section{Introduction}
There are two competing scenarios for the formation and evolution of giant
early-type galaxies:  Either they formed from an early monolithic collapse, 
and have evolved passively since, or they formed from the stochastic
mergers of smaller systems.  Both scenarios predict that the observed 
properties of galaxies should correlate with their environments.  

In the first scenario, correlations with environment may 
arise because of a host of plausible physical effects associated with 
dense environments:  these include ram pressure stripping of gas due to 
the hot, intra-cluster medium (Gunn \& Gott 1972), 
galaxy harassment (Moore et al. 1999), tidal interactions 
(Bekki, Couch, \& Shioya 2001) and strangulation (Balogh et al. 2001). 
In the second scenario also such effects may be important, but there 
is another natural reason to expect correlations with environment. 
In hierarchical clustering models, the oldest stars in the present day 
universe likely formed in the most massive systems present at high 
redshift (White \& Rees 1978; White \& Frenk 1991), 
those massive high-redshift systems merged with one another to form 
the most massive systems at the present time (Lacey \& Cole 1993), and, 
at any given time, the most massive systems populate the densest regions 
(Mo \& White 1996).  Hence, the oldest stars, and the most massive 
galaxies, should populate the densest regions; since mass and age 
influence early-type galaxy observables, a correlation with environment 
is expected (Kauffmann 1996; Kauffmann \& Charlot 1998).  
Since the two scenarios may make different predictions for trends 
with environment, it is interesting to quantify such trends.  
Our goal in what follows is not so much to distinguish between these 
models, as to develop techinques which allow one to quantify and 
interpret environmental trends, since both models predict that trends 
should exist.  

Many early-type galaxy observables correlate with one another.  
Amongst the best-studied are 
the color-magnitude relation (e.g. Sandage \& Viswnathan 1978a,b; 
Bower, Lucey \& Ellis 1992a,b), 
the luminosity-velocity dispersion relation (Poveda 1961; 
Faber \& Jackson 1977), 
the Fundamental Plane (Dressler et al. 1987; Djorgovski \& Davis 1987), 
and correlations between chemical abundance indicators such as Mg and 
the velocity dispersion (e.g. J{\o}rgensen 1997; Bernardi et al. 1998; 
Colless et al. 1999; Kuntschner et al. 2001).   
The search for correlations with environment has generally taken 
the form of measuring some of these correlations, and then trying 
to quantify if the relation is different in dense cluster-like 
regions than in less dense regions.  
For instance, the Mg-$\sigma$ relation shows only a weak dependence 
on environment (Bernardi et al. 1998), and recent work with the SDSS 
indicates that the color-magnitude relation also shows little 
dependence on environment (Bernardi et al. 2003d; Hogg et al. 2004;
Balogh et al. 2004; Wake et al. 2005).  

Interpreting these weak differences is more complicated.
It is not clear if the small differences reflect genuinely small 
differences in the ages and metallicities of galaxies in low and high 
density environments, 
or if large changes in age are compensated-for by changes in metallicity 
(e.g. Worthey et al. 1994; Kuntschner et al. 2001), leaving the 
observables essentially unchanged. In addition, because early-types 
evolve relatively rapidly only when they are younger than a few Gyrs, 
small differences in formation times lead to only small differences 
in observables at later times.

In this study, we attempt to separate out the effects of age from other 
effects in two ways.  In the first method, we study the spectroscopic 
and photometric properties of massive early-type galaxies over a wide 
range in environment and over a relatively small range in redshift.  
The small redshift range helps ensure that the galaxy population at the 
low redshift end of the sample is essentially a passively aged version 
of the population at higher redshifts.  
Comparison of the observed evolution with the observed dependence 
on environment provides a relatively model independent estimate of 
the typical age differences between environments.  
Our second method is more model dependent.  We compare a variety of 
absorption line-strengths in the spectra of early-type galaxies with 
the latest generation of stellar population synthesis models.  
These models account for the fact that early-type galaxies are 
$\alpha$-enhanced relative to solar (e.g. Tripico \& Bell 1995; 
Trager et al. 2000; Thomas, Maraston \& Bender 2003 (TMB03); 
Thomas, Maraston, Korn 2004 (TMK04); Tantalo \& Chiosi 2004).

The requirements of relatively small redshift coverage, but 
relatively large numbers of galaxies with well calibrated 
photometry and spectroscopy (so that small evolutionary 
and environmental trends can be detected) make the Sloan Digital Sky 
Survey (SDSS; York et al. 2000) the ideal database for this study.  
Section~\ref{sample} describes various aspects of the dataset:  
how the early-type galaxy sample was selected (similar to 
Bernardi et al. 2003a), how composite spectra suitable for line-index 
measurements were assembled (similar to Bernardi et al. 2003d), and 
how estimates of the local environment for each galaxy were made.  
Section~\ref{fp} presents evidence from the Fundamental Plane that 
cluster galaxies are slightly different from their counterparts in 
low density regions.  
Correlations between environment and various chemical abundance 
indicators are studied in Section~\ref{chemicals}.  
Section~\ref{evenv} compares the dependence on environment with that 
on redshift, and discusses how these observed trends can begin to 
distinguish age effects from those associated with changes in 
metallicity.  It then provides a more quantitive argument 
which does not depend on the use of stellar population synthesis models.  
Such models are used to interpret our data in Section~\ref{sspmodels}.   
A final section summarizes our findings, and discusses some implications.  
Appendices~\ref{selev} and~\ref{fluxcalib} discuss how we correct for 
selection effects and flux-calibration issues, respectively.  
Tables of the composite spectra we use, the line-strengths measured 
from them, and the ages, metallicities, and $\alpha$-element 
abundance ratios derived from the line-strengths are available, 
in their entirety, in the electronic version of the journal.  
Where necessary, we assume a background cosmological model which is flat, 
with matter accounting for thirty percent of the critical density, and a 
Hubble constant at the present time of $H_0=70$~km~s$^{-1}$~Mpc$^{-1}$.  

\section{Sample selection}\label{sample}
All the objects we analyze were selected from the Sloan Digital Sky 
Survey (SDSS) database.
See York et al. (2000) for a technical summary of the SDSS project; 
Stoughton et al. (2002) for a description of the Early Data Release; 
Abazajian et al. (2003) et al. for a description of DR1, the First Data 
Release; Gunn et al. (1998) for details about the camera; 
Fukugita et al. (1996), Hogg et al. (2001) and Smith et al. (2002) 
for details of the photometric system and calibration; 
Lupton et al. (2001) for a discussion of the photometric data reduction 
pipeline; 
Pier et al. (2002) for the astrometric calibrations;
Blanton et al. (2003) for details of the tiling algorithm; 
Strauss et al. (2002) and Eisenstein et al. (2001)
for details of the target selection. 

We selected all objects targeted as galaxies and having Petrosian
(1976) apparent magnitude $14.5 \le r_{\rm Pet}\le 17.75$.  To extract 
a sample of early-type galaxies we then chose the subset with the
spectroscopic parameter {\tt eclass < 0}, which classifies the
spectral type based on a Principal Component Analysis, and the
photometric parameter {\tt fracDev$_r$ > 0.8}, which is a
seeing-corrected indicator of morphology. {\tt fracDev$_r$} is
obtained by taking the best fit exponential and de Vaucouleurs fits to
the surface brightness profile, finding the linear combination of the
two that best-fits the image, and storing the fraction contributed by
the de Vaucouleurs fit.  We removed galaxies with problems in the
spectra (using the {\tt zStatus} and {\tt zWarning} flags).  
From this subsample, we finally chose those objects for which the 
spectroscopic pipeline had measured velocity dispersions (meaning that 
the signal-to-noise ratio in pixels between the restframe wavelengths
4200\AA\ and 5800\AA\ is S/N $>10$).  This gave a sample of 39320
objects, with photometric parameters output by version ${\tt V5.4}$ of
the SDSS photometric pipeline and ${\tt V.23}$ reductions of the
spectroscopic pipeline.  For reasons given in Bernardi et al. (2003a), 
the luminosities and sizes we use in what follows are {\em not} derived 
from Petrosian quantities, 
but from fits of deVaucoleur profiles to the images.

\subsection{Local environment}
There is some debate in the recent literature over the optimal method
for defining the local environment of galaxies (Eisenstein 2003;
Kauffmann et al. 2003; Balogh et al. 2004). The options include using
catalogs of clusters and groups of galaxies, adaptive measurements of
local galaxy density, like distance to the $n^{th}$ nearest neighbor,
and physical measurements of density based on expectation from
simulations. Each of these methods has its own set of pros and cons.
In this study, we attempt to minimize the problems associated with any
one of these measurements of environment by using a combination.

We have chosen to represent the environment of a galaxy in two ways.
One is to estimate the comoving distance to the nearest cluster, and
the other defines a local density proportional to the inverse of the
volume which encloses the nearest ten galaxies at $M_r < -21.5$
(Petrosian magnitudes). As our sample is magnitude limited, and
the abundance of luminous galaxies drops exponentially at the bright
end, the sample is much sparser at high redshifts.  As a result,
distances to the $n$th nearest neighbour will all be larger at high
redshifts, unless we also specify that all $n$ neighbours were
sufficiently luminous that they would have been seen at all redshifts
in our catalog.  Our brightness cut was chosen to satisfy the
competing constraints of having enough objects from which to estimate
distances, and of ensuring that those objects would satisfy the SDSS
magnitude limits over as wide a range as possible.  Since such objects
cannot be seen beyond $z=0.15$, we limit our catalog to $z\le 0.14$.

For each galaxy in our sample, we estimate comoving 
(three-dimensional) distances to all the objects which are more
luminous than $1.75\times 10^{11}h_{100}^{-2}\, L_\odot$ in the C4 
cluster catalog (Miller et al. 2004).  
This catalog is more than 90\% complete out to $z=0.14$.  
Estimates of the mean redshift $z_{\rm cl}$, virial radius $r_{\rm cl}$, 
and velocity dispersion $\sigma_{\rm cl}$, are available for each C4 
cluster.  Given the redshift, the virial radius defines an angular scale, 
$\theta_{\rm cl}$.  We label as cluster galaxies all those which lie 
within $\theta_{\rm cl}$ and $3\sigma_{\rm cl}$ of a C4 cluster. 

The typical absolute magnitude of the early-type galaxies in our
catalog is $M_r \simeq -21$.  The absolute magnitude of the Sun is
$4.62$ in this band, so the luminosity of a typical early-type is
$\sim 1.77\times 10^{10}h_{70}^{-2}L_\odot$: the C4 clusters are $\sim
10$ times more luminous than a typical early-type galaxy.  This means
that smaller groups are {\em not} included in the C4 catalog;
early-type galaxies in such groups will be assigned to denser
environments only if such groups typically cluster around C4 clusters.
Since such group members inhabit environments which are intermediate
between rich clusters and low density environments, if we wish to
define a sample in low density environments, we would like to include
as few group galaxies as possible.

\begin{figure}
 \centering
 \plotone{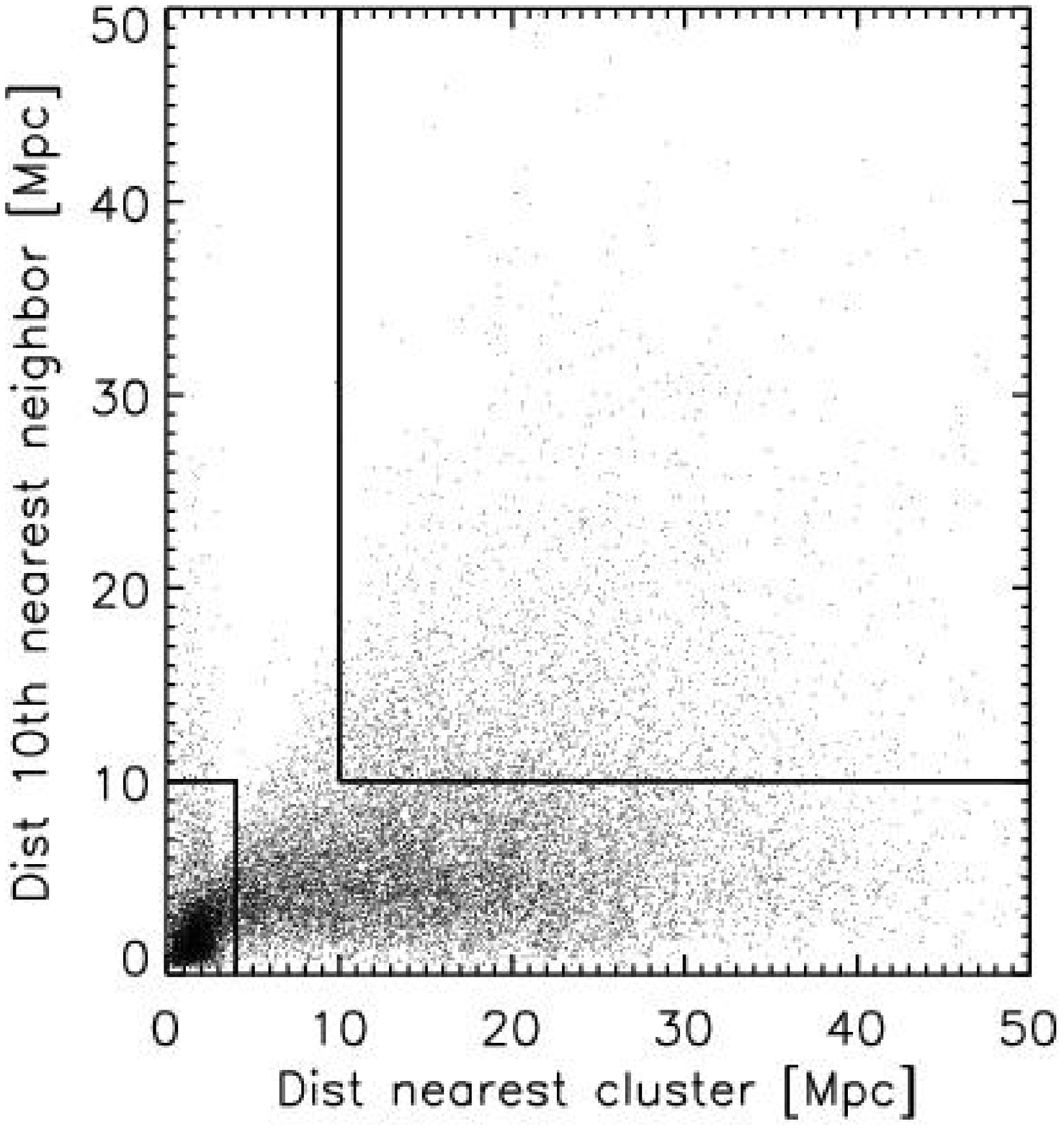}
 \caption[]{Classification into high density (lower left) and low density 
           environments (upper right) is based on the comoving distance 
           to the tenth nearest luminous ($L>3L_*$) neighbour, and on the 
           comoving distance to the nearest massive cluster in the C4 
           catalog (Miller et al. 2004; each cluster is more luminous 
           than $1.75\times 10^{11}h^{-2}L_\odot$ $\sim 10L_*$).  }
 \label{select}
\end{figure}

With this in mind, we define a sample of galaxies in less dense 
environments when the distance to the nearest cluster {\it and} the
distance to the tenth nearest neighbour galaxy is larger than 10~Mpc. 
Our cut on tenth neighbour distance is supposed to eliminate most group 
galaxies from our low density environment sample.  
Thus, only objects in the densest (bottom left) and least dense (top right) 
environments of Figure~\ref{select} are used in the analysis which 
follows: in all there were 3112 and 5711 early-types in the two 
environments at $z\le 0.14$.  

There is a supercluster in the SDSS sample at $z\sim 0.08$.  
In what follows, we see some peculiarities in the redshift bin which 
contains this structure, so it may be that our estimates of 
environmental effects are altered by the supercluster.

\subsection{Line indices}
\begin{figure}
 \centering
 \plottwo{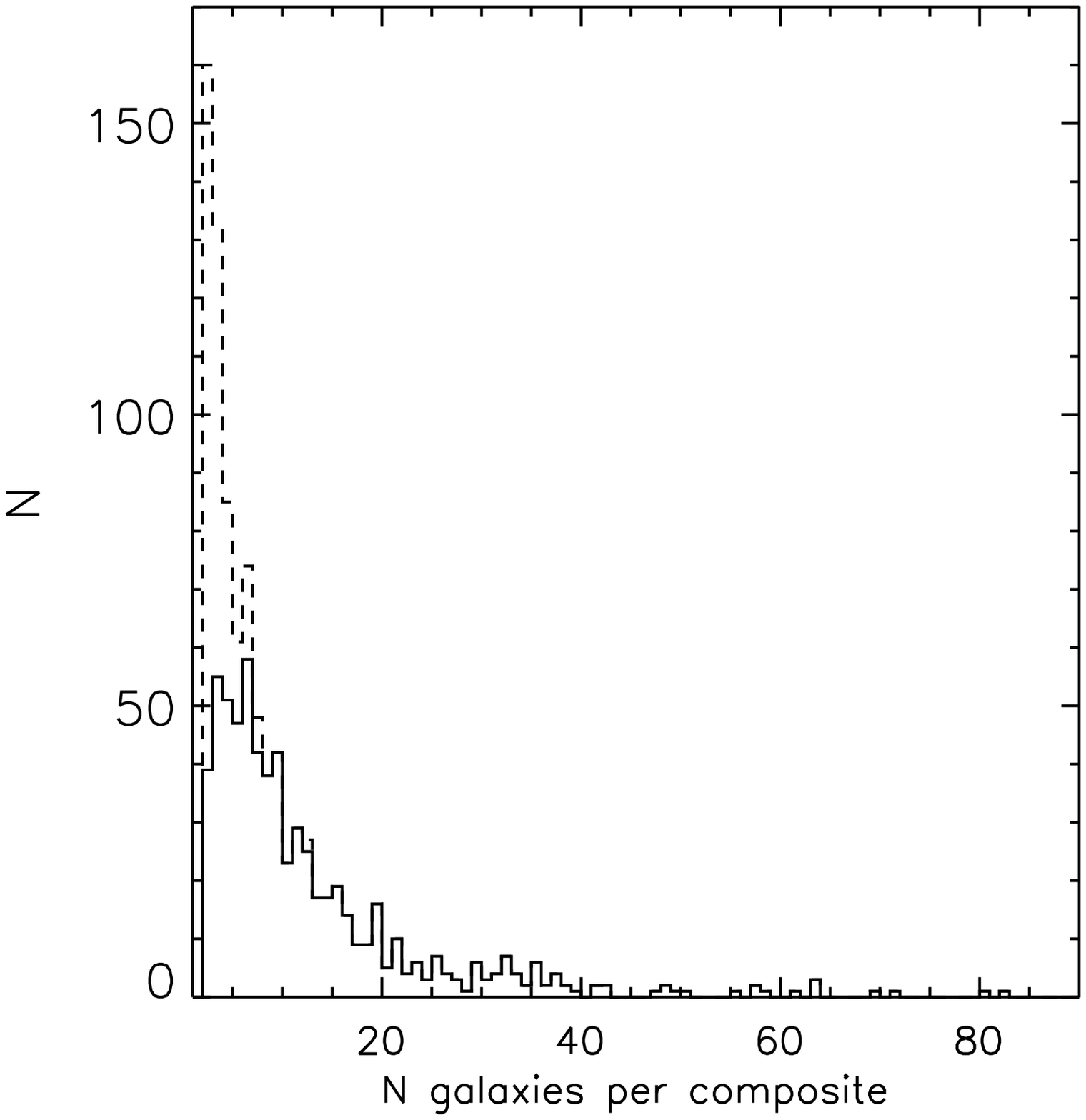}{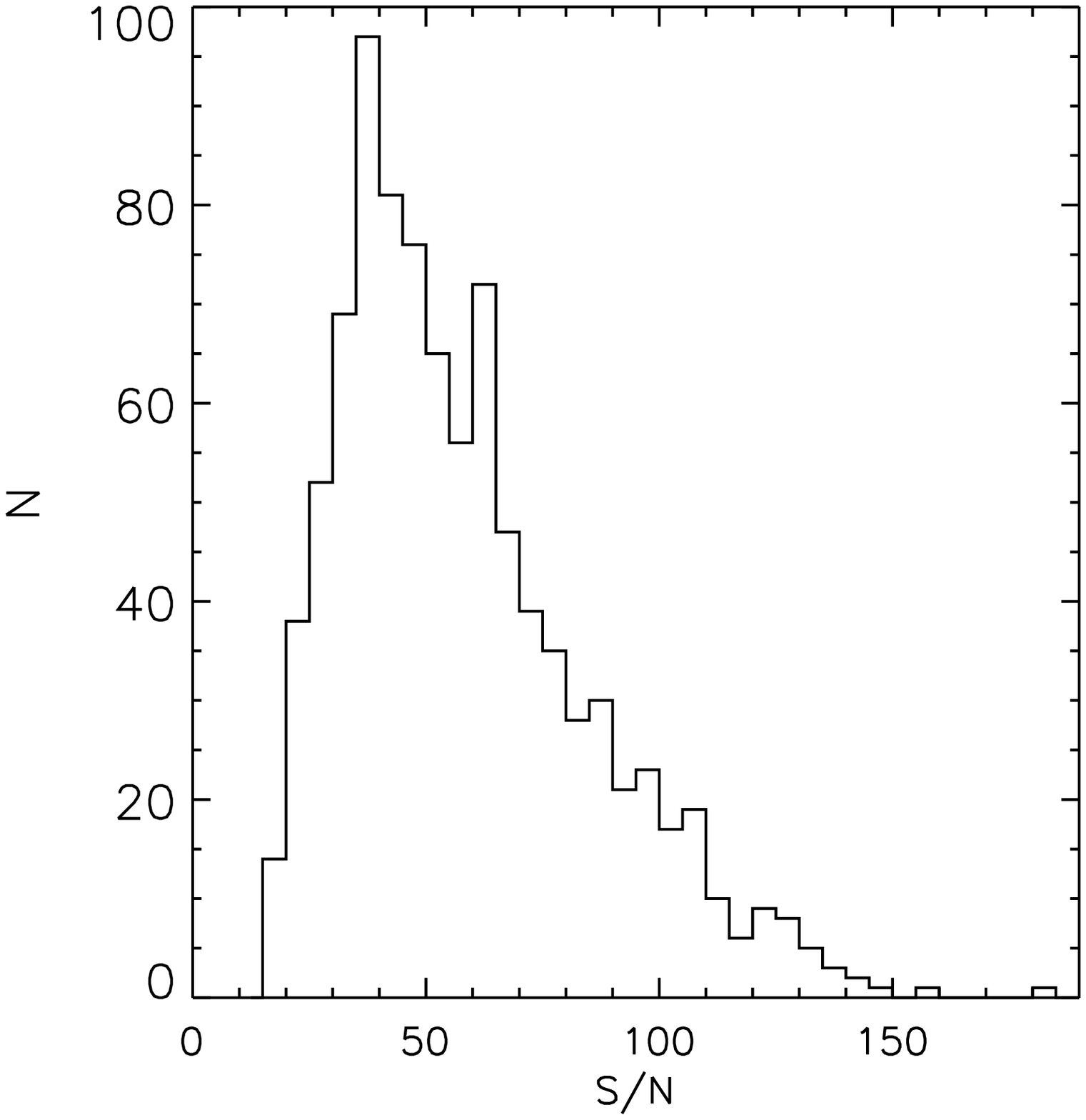}
 \caption{Distribution of the number of galaxies in a composite 
          spectrum (left), and the signal-to-noise ratios of the 
          composites (right).  
          Dashed histogram in left panel shows 
          all composites, and solid histogram shows those with 
          S/N$\ge 40$.
          }
 \label{s/n-hist}
\end{figure}

Later in this paper, we will study how various chemical abundance 
indicators of the early-type galaxy population depend on redshift and 
environment.  The typical signal-to-noise ratio of an individual SDSS 
spectrum in our sample, $\sim$18, is considerably smaller than the 
value ($\sim$50) required to make reliable estimates of line-strengths.  
Therefore, for each environment, we construct high S/N composite spectra, 
suitable for line-index measurements, by co-adding the spectra of similar 
objects. We use narrow bins in luminosity, size, velocity dispersion, and 
redshift, chosen so that the resulting composite spectra had 
signal-to-noise ratios of $\sim$100.  
See Bernardi et al. (2003d) for details of the co-addition procedure. 
Briefly, spectra are normalized to have the same flux between 
3900 and 7000\AA, and then co-added, weighting by the observed error 
estimate in each pixel.  
Table~\ref{composites} describes the bins we have chosen.  
The 925 composite spectra themselves are available from the electronic 
edition of the journal.  
Figure~\ref{s/n-hist} shows the distribution of the number of objects 
per composite, as well as the distribution of S/N ratios.  
The results which follow have been obtained by using all the 
composites, although we have checked that the main conclusions 
do not change if only the subset with S/N$>40$ is used.  The 
left panel of the Figure shows the difference between the full 
set of composites, and the higher signal-to-noise subset.  

We measure the strengths of the original Lick indices 
(Worthey et al. 1994; Trager et al. 1998), as well as a number of 
other spectral features in each composite.  This is because, 
with the exception of H$\beta$, none of the Lick indices are 
particularly sensitive to recent star-formation.  Although we do not 
expect to find objects with large star-formation rates in our sample, 
some of the objects in it may have been forming stars at relatively 
small look-back times.  
Therefore, we also study some Balmer lines in absorption, 
H$\delta_{\rm A}$, H$\delta_{\rm F}$, H$\gamma_{\rm A}$ and 
H$\gamma_{\rm F}$, defined following Worthey \& Ottaviani (1997), 
which are expected to indicate star-formation activity in the less 
distant past.  
Our estimates of the Balmer line strengths should be treated as lower 
limits because they may be filled-in by emission for which we have not 
corrected.  However, we do correct H$\beta$ by adding 0.05 O{\small II} 
to the measured value.  The standard correction uses O{\small III} (e.g.
Trager et al. 1998), 
but this is very noisy in our sample:  O{\small II} shows a cleaner 
correlation.  

In addition, many of these lines are close to the edge of the SDSS 
spectrograph, where flux calibration problems may bias our 
measurements.  We discuss this more fully shortly.  

Although our sample selection excludes strong emission lines, some weak 
emission is permitted.  We find weak emission in O{\small II} which 
indicates very recent star formation, and/or AGN activity.  
We also study the strength of the break at 4000\AA, and combinations 
of indices which are expected to be indicators of the metallicity and 
the relative abundances of $\alpha$-elements.

Where available (e.g. J{\o}rgensen 1997), small aperture corrections 
have been applied to the measured values.  These corrections are 
typically of the form $(8r/r_e)^\alpha$ with $|\alpha|\sim 0.05$ 
(where $r_e$ is the half-light radius).  
Where no prescription was available in the literature, we used 
$|\alpha| = 0.05$. This correction only matters for the 
trends with redshift which are presented in Section~\ref{evenv}; 
but they are unimportant for the trends with environment, because we 
always study environmental effects at fixed redshift, and we see no 
significant dependence of galaxy size on environment.  
We have also corrected all line indices for the effects of velocity 
dispersion calibrated using the Bruzual \& Charlot (2003) models 
and assuming Gaussian velocity distributions.  In principle, the 
correction is different for non-Gaussian velocity distributions 
(e.g. Kuntschner 2004); we have chosen the Gaussian because our 
analysis is based on composite spectra, so choosing the appropriate 
non-Gaussian model to make the correction is not straightforward.  

Tables~\ref{linestrengths}--\ref{lick2} give our measurements of the 
index strengths in these composite spectra.  In Section~\ref{sspmodels}
we use stellar population synthesis models to interpret these measurements.  
These assume measurements at Lick rather than SDSS resolution, so, in 
that section only, we smooth the spectra to Lick resolution before 
correcting for the effects of velocity dispersion.  The line-strengths 
at Lick resolution, for the indices used in that section, are given 
in Table~\ref{lickres}.

\section{The Fundamental Plane}\label{fp}

\begin{figure}
 \centering
 \plottwo{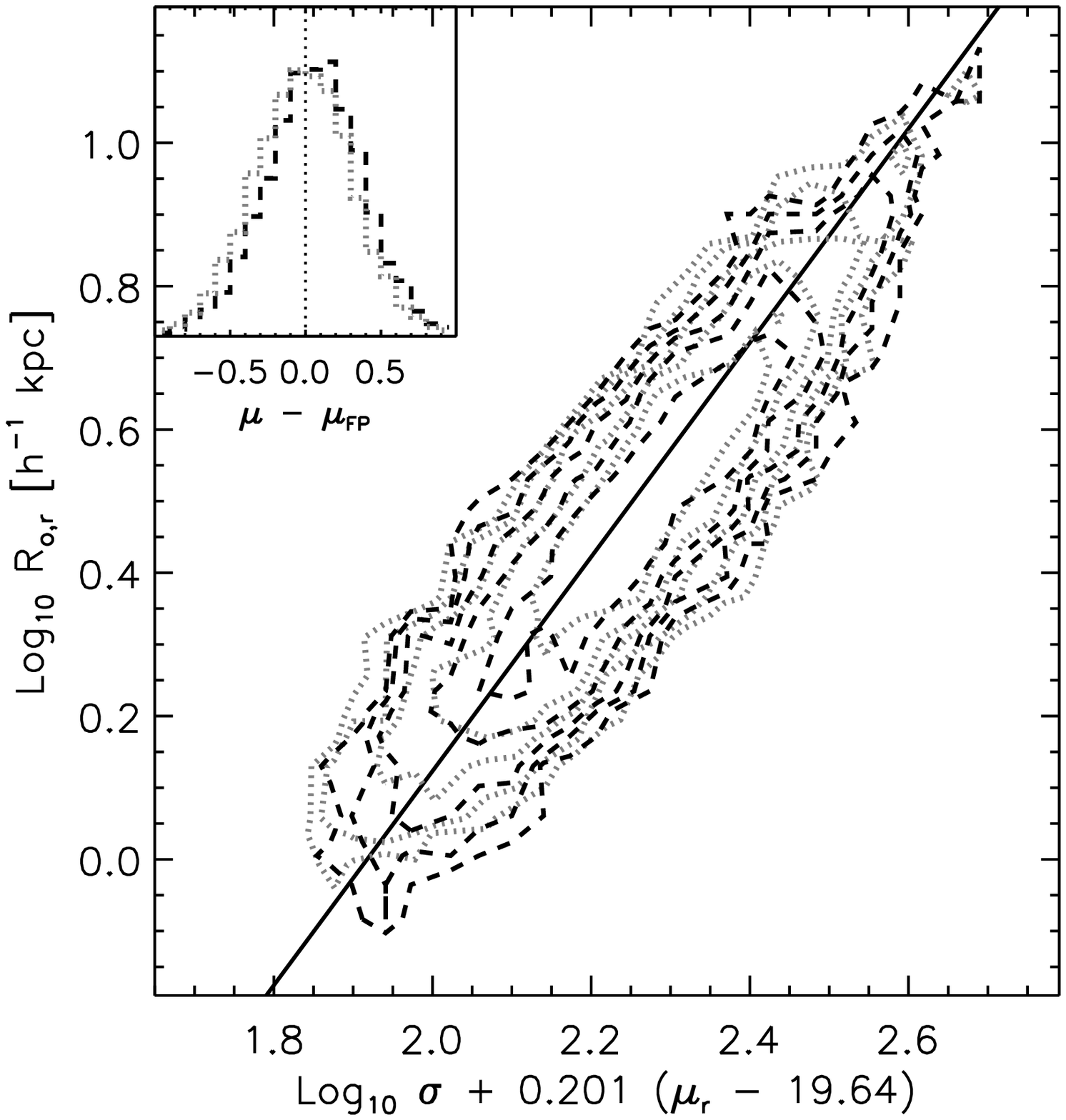}{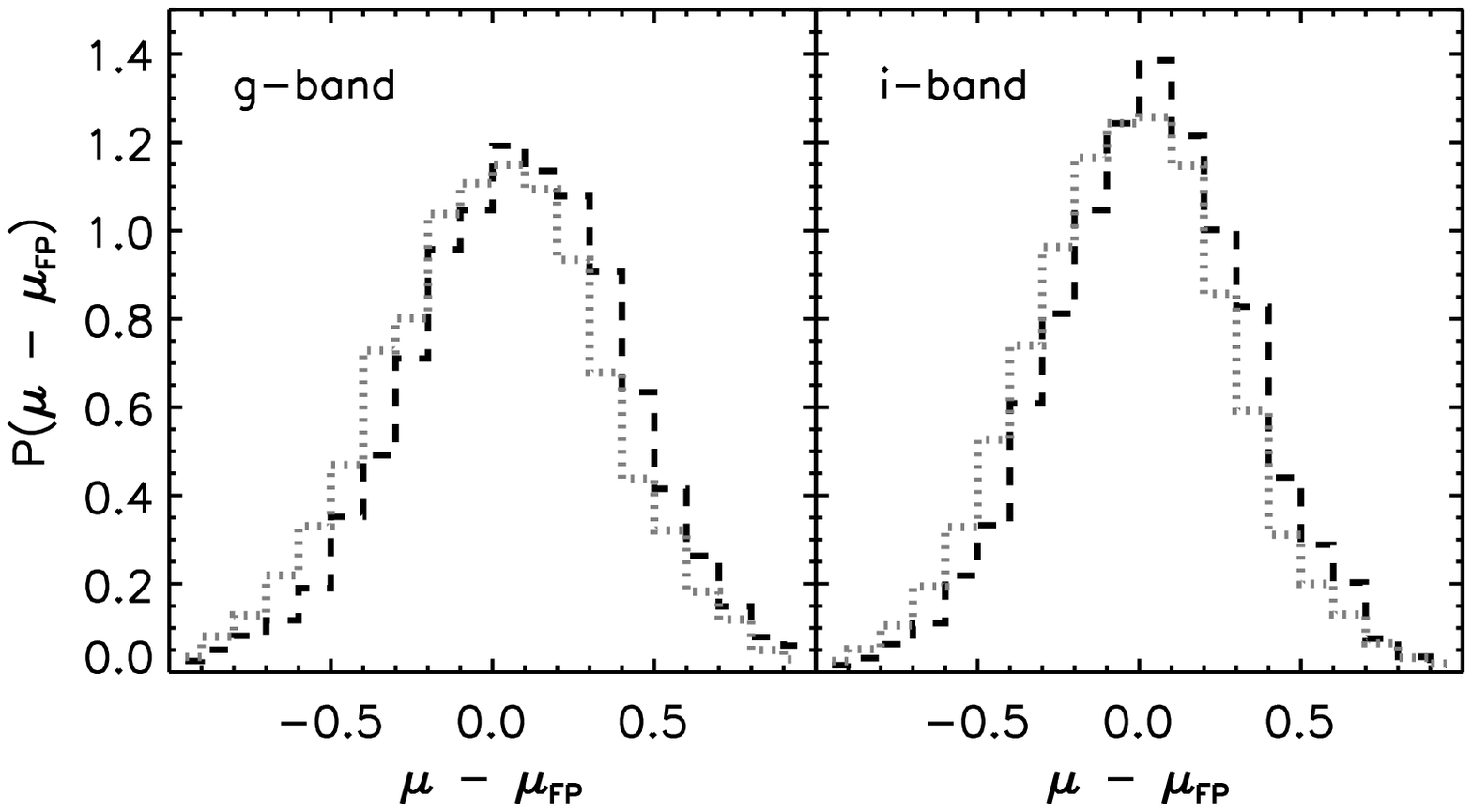}
 \caption{Top:  The Fundamental Plane in the $r$ band.  
          Solid line shows the fit from Bernardi et al. (2003c).  
          Dashed and dotted lines represent the subset of galaxies 
          which populate dense and less dense regions.  Inset shows 
          the distribution of residuals in surface brightness: galaxies 
          in dense regions tend to be $\sim 0.08$~mag fainter than 
          those in the least dense regions.  
          Bottom panels show the distribution of residuals in 
          surface brightness with respect to the FP in the $g$ and $i$ 
          bands; the offset between low density and high density samples is 
          similar to that in the $r$ band.  Notice that the width of the 
          distribution of residuals is approximately independent of 
          environment. }
 \label{fpgi}
\end{figure}

There are a number of small but significant differences between our 
early-type galaxy samples in low and high-density environments.  
A traditional way to search for trends with environment is to study 
the distribution of galaxies in the space of log(size), 
log(velocity dispersion) and log(surface brightness).  
In this space, early-type galaxies populate a Fundamental Plane 
(e.g., Dressler et al. 1987; Djorgovski \& Davis 1987).  
The solid line in the top panel of Figure~\ref{fpgi} shows the best 
{\it orthogonal-fit} (determined following the methods described in 
Bernardi et al. 2003c) Fundamental Plane in the $r$ band using all 
early-type galaxies in our sample, whatever their environment. 
(Although, the old photometric reductions output by the SDSS pipeline 
{\tt photo} were incorrect, using the new corrected photometry has not 
changed significantly the coefficients of the Fundamental Plane from 
those reported in Bernardi et al. 2003c). The dashed contours show 
the distribution of the subset of cluster galaxies in dense regions
around the plane, while the dotted contours represent galaxies in less 
dense regions. The dependence on environment is weak.  To show the 
dependence more clearly, the inset shows the distribution of residuals 
from the plane:  evidently, galaxies in dense regions tend to have 
slightly ($0.075\pm 0.008$~mag) fainter surface brightnesses than 
their counterparts in less dense regions.  The bottom panel shows a 
similar analysis of Fundamental Plane residuals in the $g$ and $i$ 
bands, for which the shifts are $0.081\pm 0.008$~mags and 
$0.069\pm 0.007$~mags, respectively.  
Note that the typical scatter around the plane is not significantly 
smaller in the cluster sample than in the low density sample.  
(Table~\ref{fptable} gives the mean offsets in each band for the 
two environments, as well as the rms residual.)  


Our findings are consistent with a model in which the stars in any given 
early-type galaxy formed in a single burst, and, for the galaxies which 
are now in dense regions, this burst happened at higher redshifts.  
But a model in which the chemical abundances (e.g. metallicities) 
depend on environment would also work.  
In Section~\ref{sspmodels} we use single burst stellar population synthesis 
models to interpret these trends in terms of age and/or metallicity 
differences between the two populations.  
A more model independent analysis is the subject of Section~\ref{evenv}. 

\section{Additional trends with environment}\label{chemicals}
This section studies how various structural parameters and chemical 
abundance indicators depend on environment.  
In the figures that follow, we will often group together observables 
that are expected to trace similar physics.  To facilitate comparison 
of different observables with one another, we standardize as follows:  
for each observable $Y$, we compute the mean $\langle Y\rangle$ and 
the rms simply by summing over the entire sample of composite spectra 
weighting each by the number of galaxies in the composite.  
We then show $(Y - \langle Y\rangle)/{\rm rms}(Y)$, rather than $Y$ 
itself.  To simplify interpretation of selection effects, we show 
results from subsamples in different redshift bins.
The values of $\langle Y\rangle$  and rms$(Y)$ are provided in 
Table~\ref{evenvtab}.

The mean and rms values we quote are obtained by number weighting 
the composites when averaging, rather than averaging over the galaxies 
themselves. We have chosen to do this because line strengths computed 
from individual rather than composite spectra are very unreliable, 
so for most choices of $Y$ we must use composites.  
Therefore, although the mean values we quote should be accurate, 
the quoted rms values may underestimate the true scatter, because 
they do not include the contribution from the scatter among objects 
which make up each composite.  However, because our composites are 
constructed from sufficiently small bins in the parameters which 
correlate most strongly with index strengths, it is likely that the 
scatter within a bin is negligible compared to the scatter between 
bins.  

As a check, we performed all the analysis which follows using 
individual rather than composite spectra, subtracting the measurement 
errors in quadrature when computing rms values.  This showed that our 
quoted rms values actually do not underestimate the true scatter 
substantially.  Using the individual spectra rather than the 
composites has no effect on any of our qualitative conclusions, 
but because quantitative measurements based on composites are more 
robust, we have chosen to present all results using the composites only.  

Note that a variety of emission and absorption lines, 
as well as the Lick indices, are known to correlate with luminosity 
and/or velocity dispersion.  We have measured a number of such 
correlations (for a selection, see Bernardi et al. 2003d as 
well as Appendix~\ref{fluxcalib}):  
for most, the primary correlation is with velocity dispersion, 
the correlation with luminosity (if present) being primarily a 
consequence of the index-$\sigma$ and luminosity-$\sigma$ correlations.  
In this respect, many line-strengths behave similarly to color 
(Bernardi et al. 2003d, 2005).  Therefore, if we find that 
these line-strengths vary with environment, it is important to check 
if the environmental trend is entirely a consequence of a correlation 
between luminosity and/or velocity dispersion and environment, or if 
the environment did indeed play an additional role.

\subsection{Structural parameters}
\begin{figure}
 \centering
 \plottwo{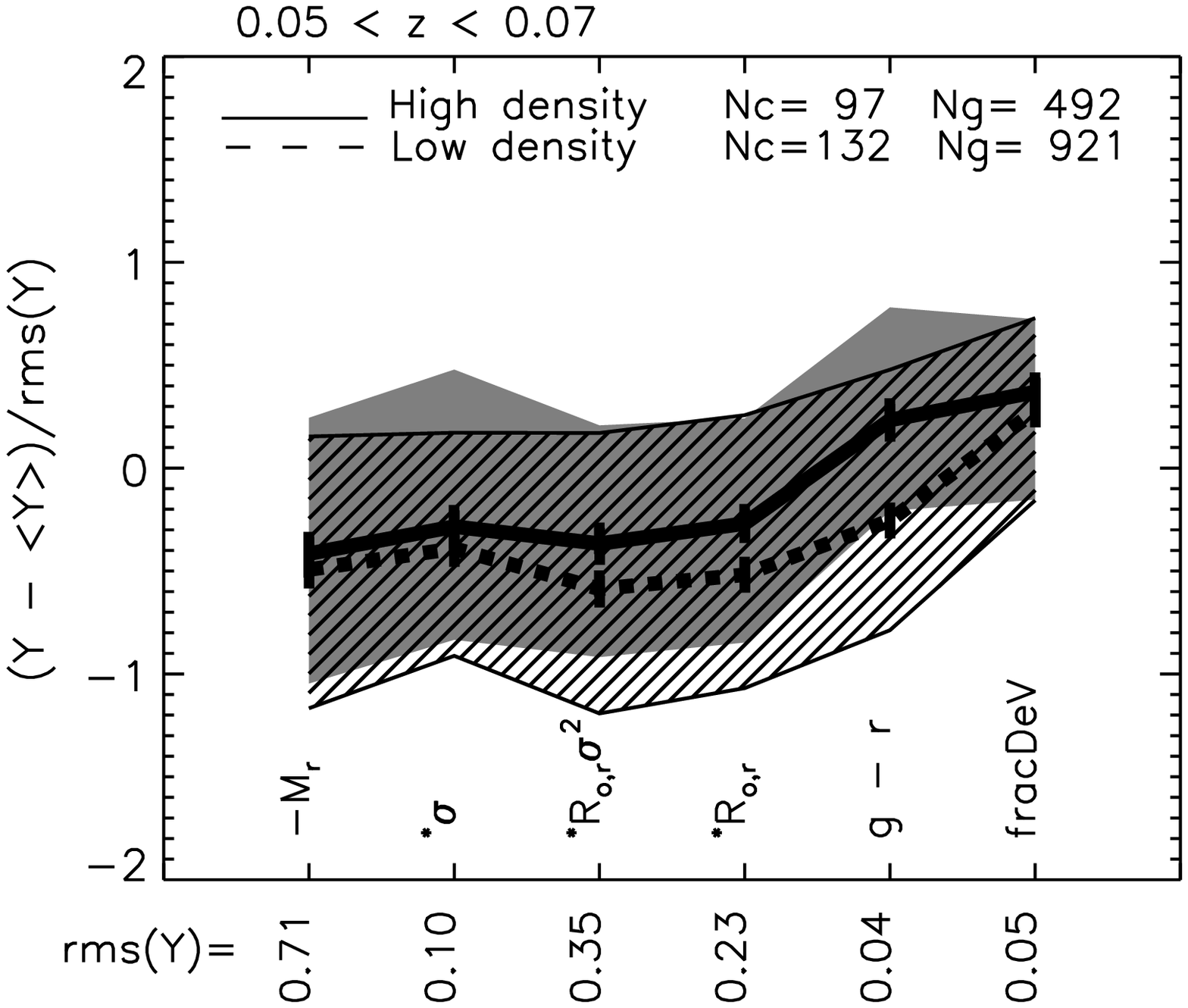}{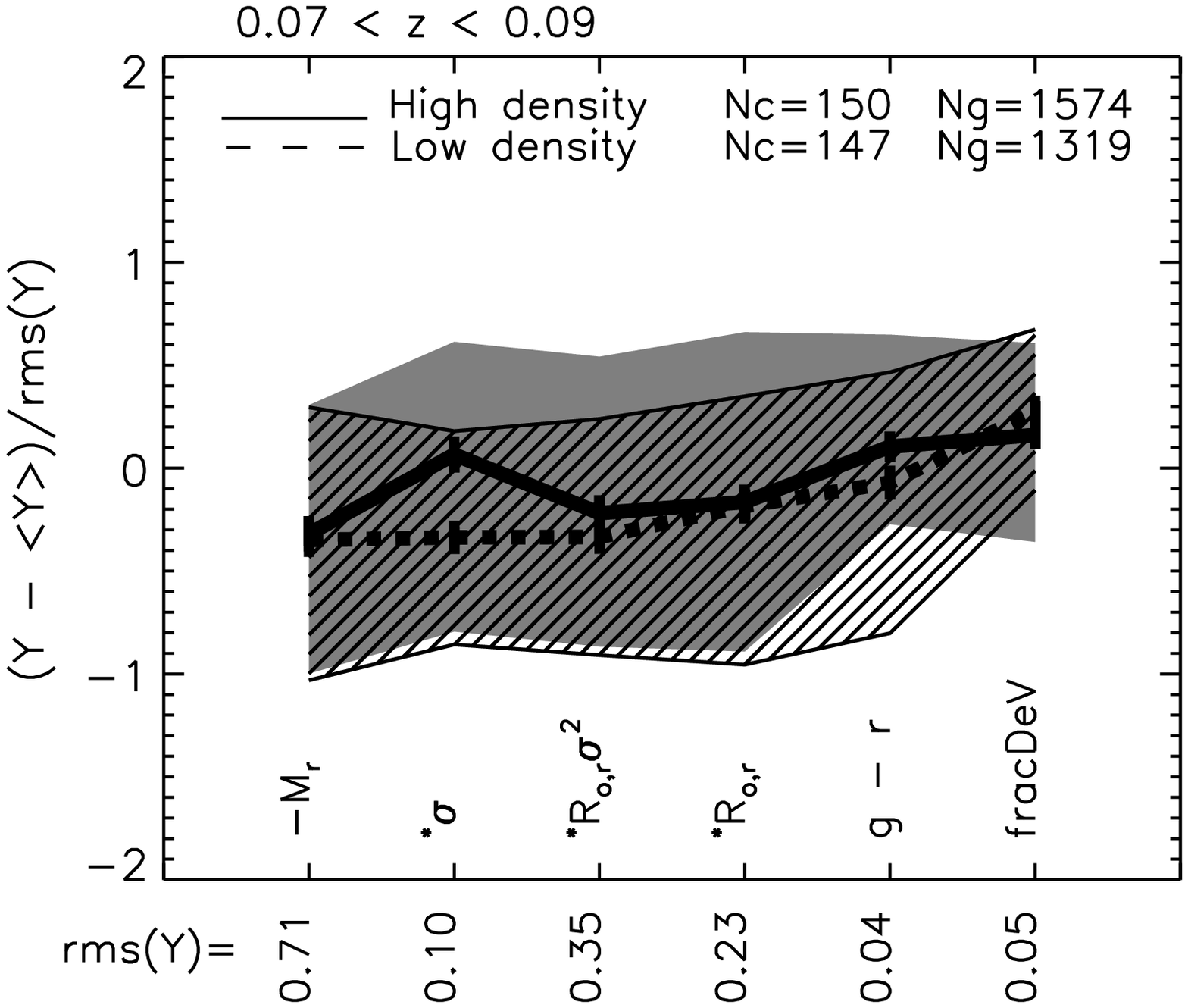}
 \plottwo{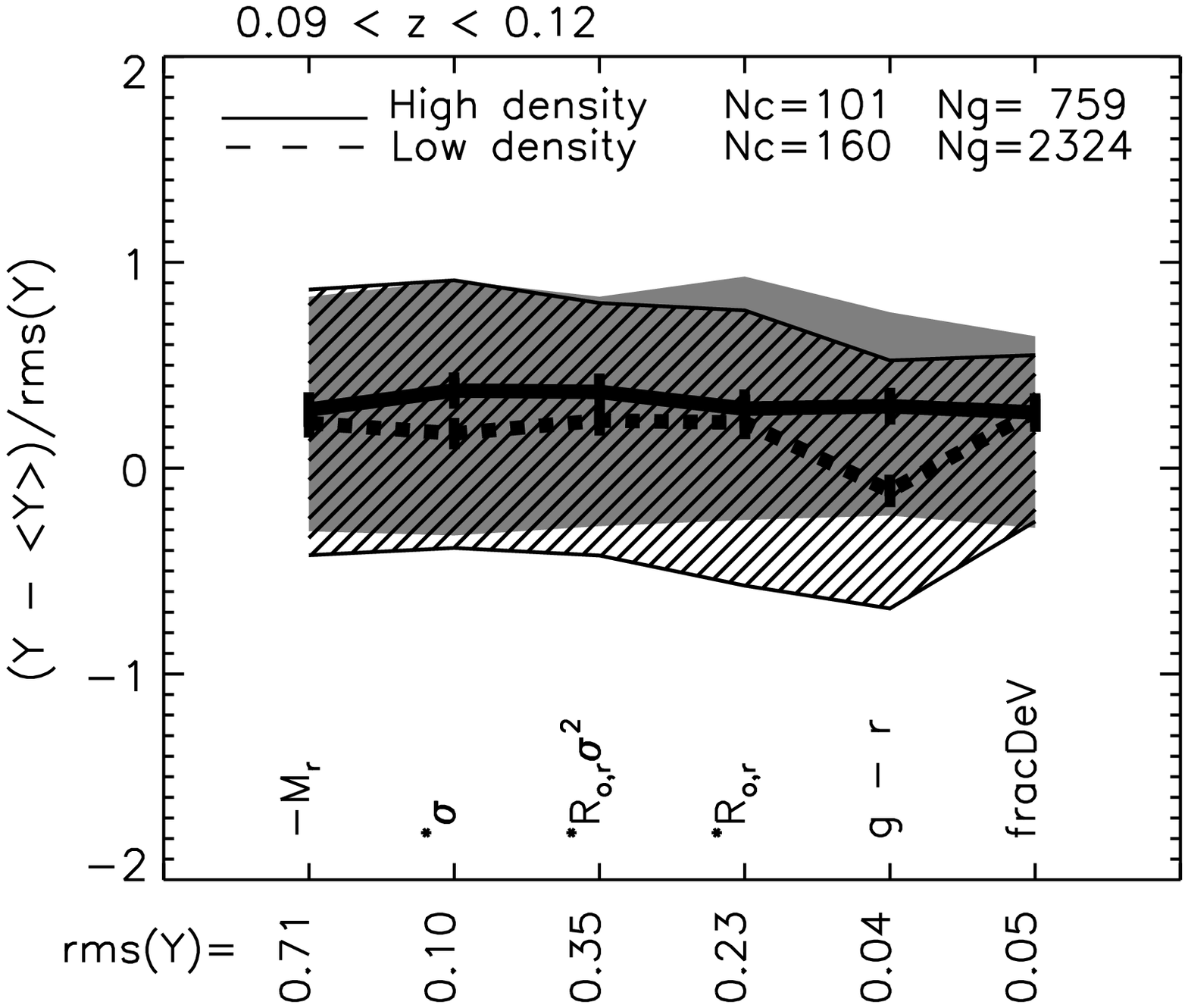}{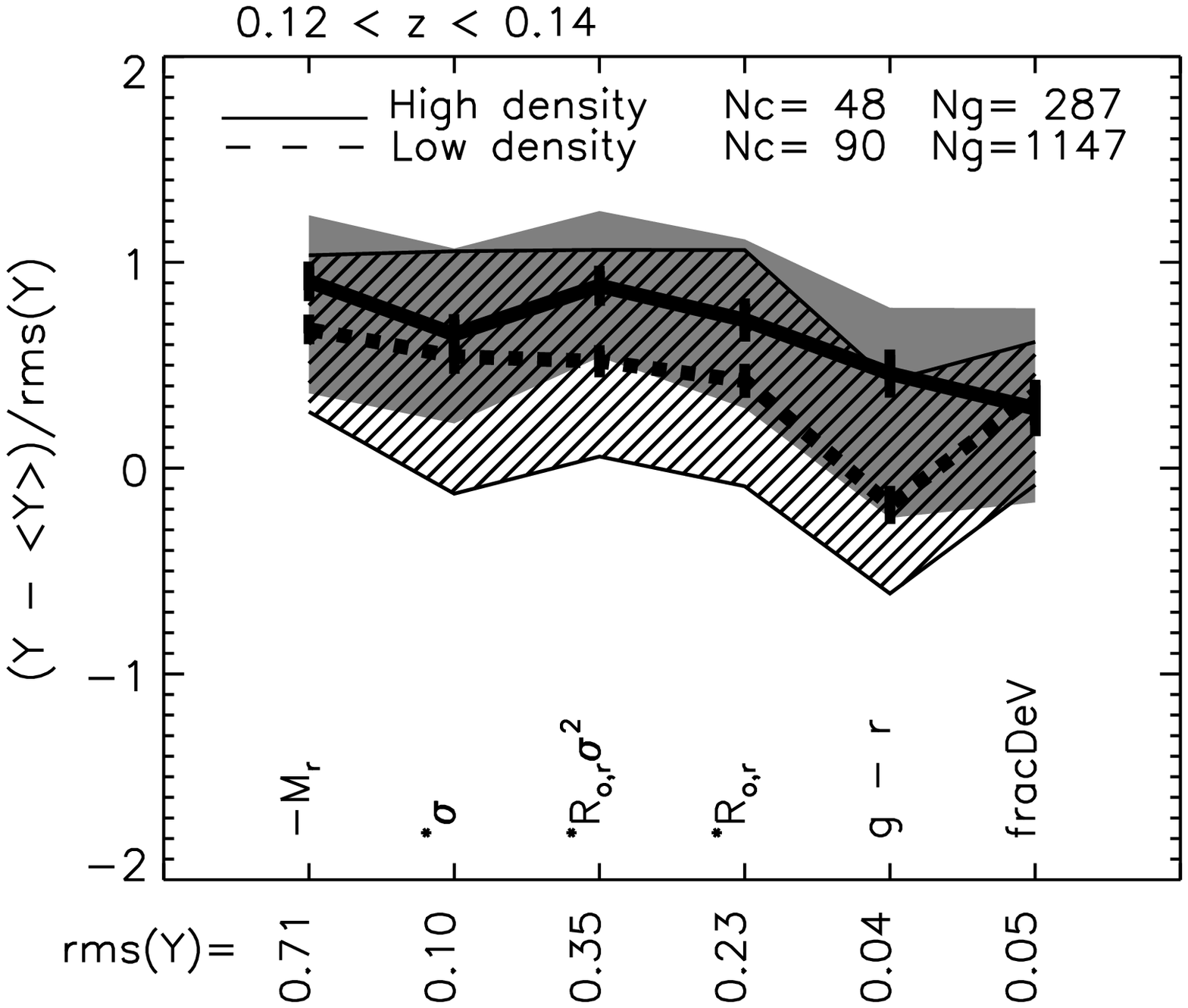}
 \caption{Median values of luminosity, velocity dispersion, mass, size, 
          color, and light-profile 
          shape {\tt fracDev} (i.e. morphological type)
          in dense and underdense regions.  
          To facilitate comparison with one another, all observables 
          have been rescaled by subtracting the mean and then dividing 
          by the rms spread across the entire sample (i.e. across all 
          redshifts and luminosities).  The x-axis lists the observable 
          and the value of the rms.  An asterisk denotes that the quoted 
          rms is for $\log_{10}$ of the parameter.             
          Shaded regions show the 25th and 75th percentile values.  
          Different panels show subsamples at $0.05<z<0.07$, 
          $0.07\le z <0.09$, 
          $0.09\le z<0.12$ and $0.12\le z<0.14$.  Text in top right of 
          each panel indicates the total number of composites in each 
          redshift bin, and the total number of galaxies which made-up 
          those composites.  Top right panel contains a supercluster, 
          so environmental effects may not be accurate.}
 \label{paramenv}
\end{figure}

Since we will be interested in whether or not environment plays a role 
over and above determining the structural parameters of galaxies, 
Figure~\ref{paramenv} shows the typical values of luminosity, 
velocity dispersion, mass, size, color and light-profile shape 
{\tt fracDev} in dense and less dense regions.  
All observables have been rescaled by subtracting the mean in the entire 
sample, and then dividing by the rms.  The x-axis lists the observable 
and the value of the rms.  
Solid and dashed lines show the median values in dense and less dense 
regions, respectively.  Error bars show the error on this median value, 
and shaded regions show the 25th and 75th percentile values.
(An asterisk signifies that the quoted rms is for $\log_{10}$ of the index.)
To simplify interpretation of selection effects, we show results from 
subsamples at $0.05<z<0.07$, $0.07\le z <0.09$, $0.09\le z<0.12$ and 
$0.12\le z<0.14$.  The text in the top right of each panel indicates 
the total number of composites in each redshift bin, and the total 
number of galaxies which made-up those composites. 

Note that the supercluster at $z\sim 0.08$ is obvious:  the ratio of 
the number of galaxies in high density regions to that in regions 
of lower density is considerably higher in the $0.07<z<0.09$ bin.  
In what follows, we see some peculiarities in this redshift bin.

The main point of this figure is not to compare the different panels with 
one another, but to compare the two environments in each panel with one 
another.  This is because the primary difference between the different 
panels is caused by the magnitude limit:  the higher redshift samples 
contain objects which are more luminous, have higher velocity dispersions, 
larger masses and sizes.  
However, the figure shows that, in any given redshift bin, trends 
with environment are weak:  in all cases where there is a small 
difference, the objects in the low density environments tend to be 
slightly less luminous, 
to have smaller velocity dispersions, masses and sizes, and to be 
slightly bluer, although these differences are usually less than twenty 
percent of the rms variation across the entire sample (i.e., two tickmarks), 
except in the highest redshift bin.  

\begin{figure}
 \centering
 \plottwo{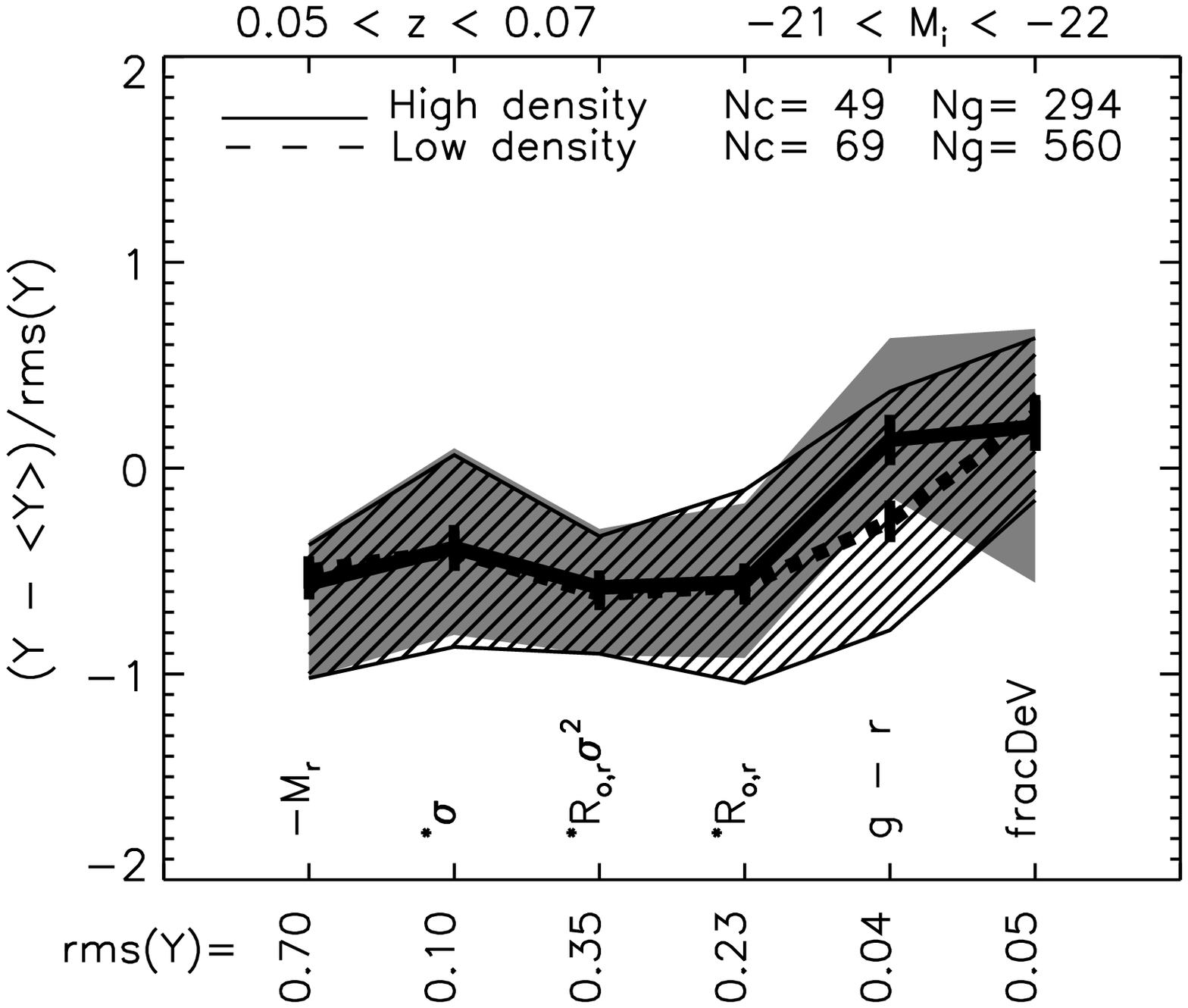}{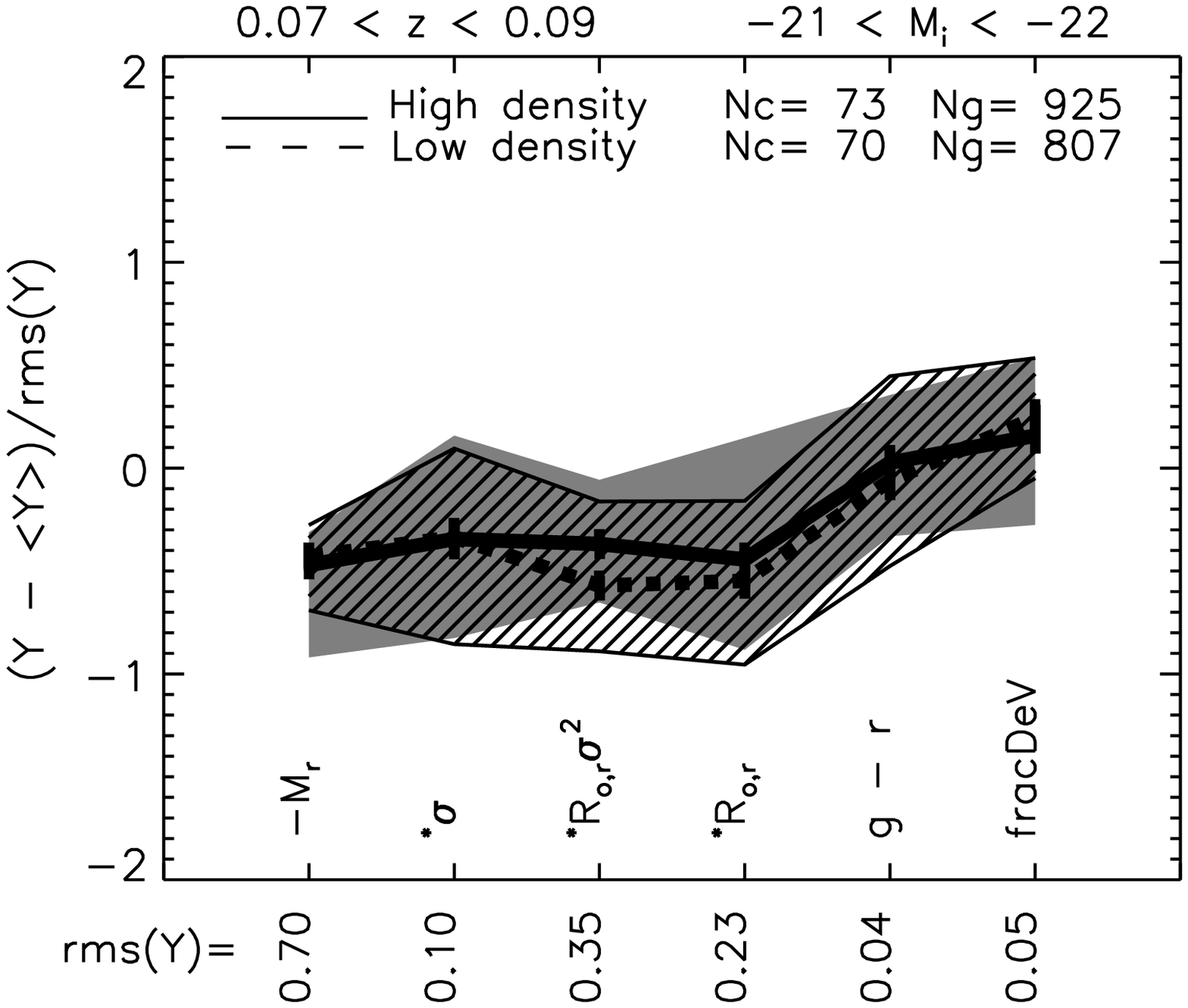}
 \plottwo{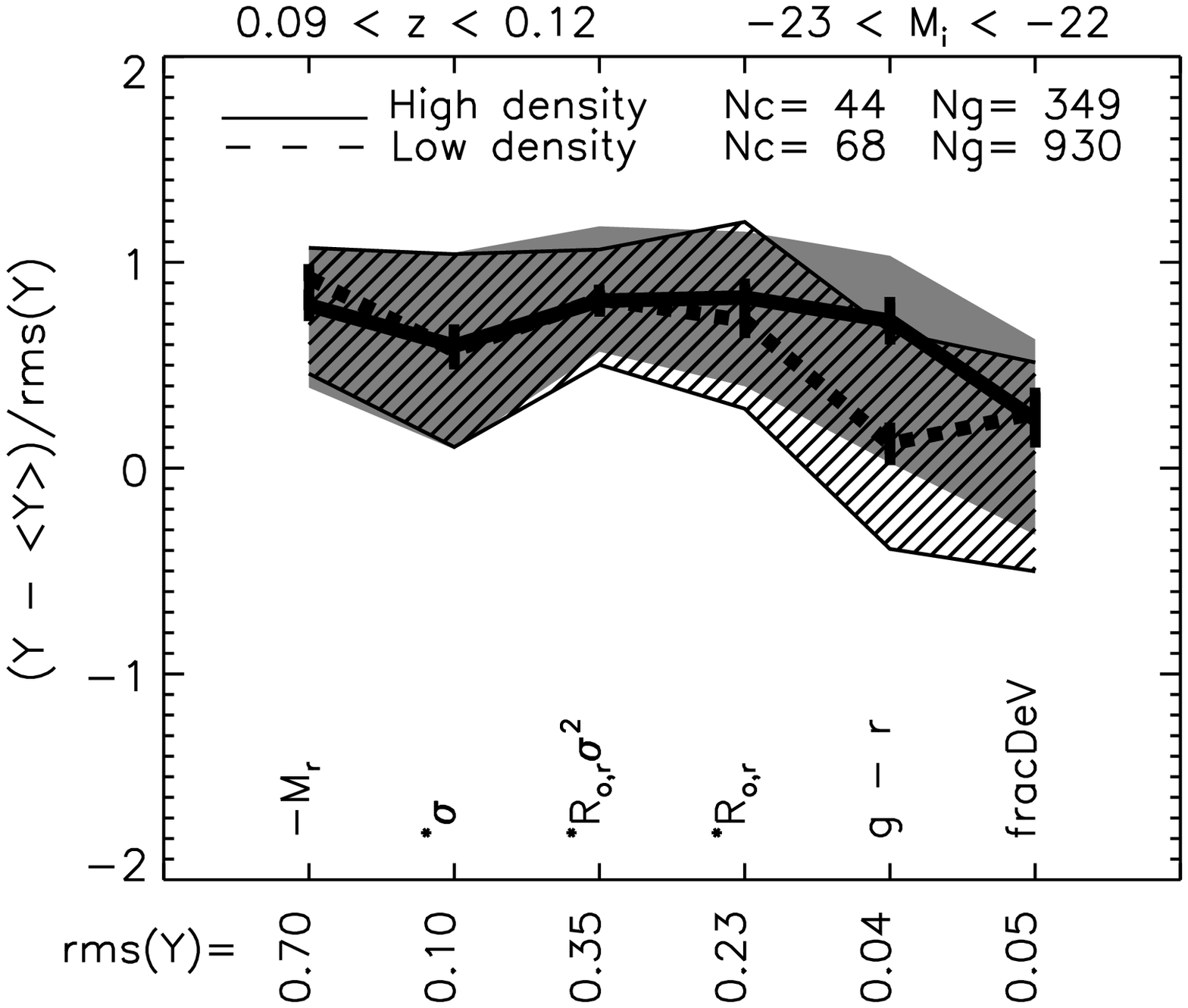}{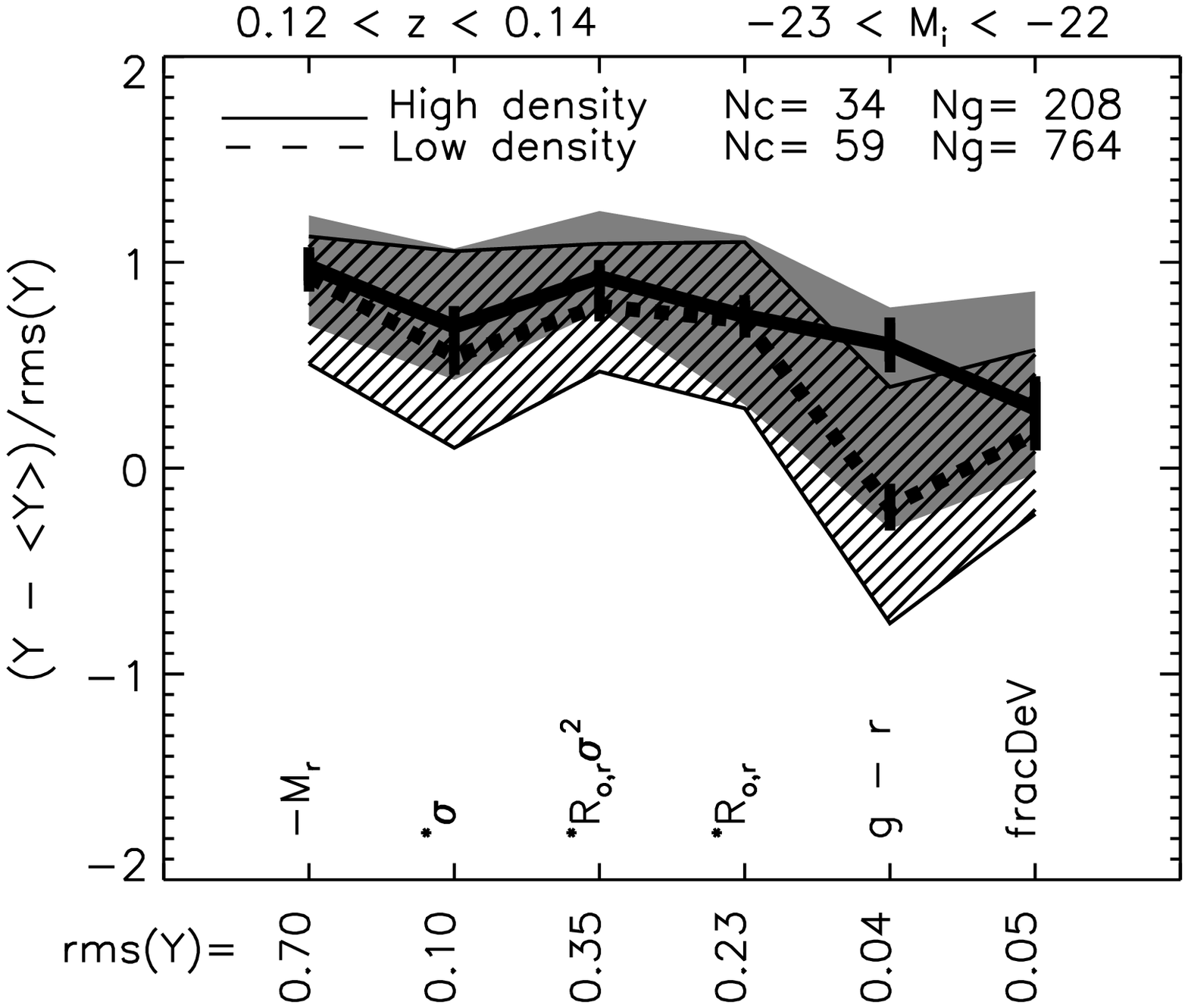}
 \caption{As for the previous figure, but now only results from objects 
          with a narrow range of luminosities (indicated in the top right 
          of each panel) are shown.  The magnitude limit of the sample 
          means that the low redshift bins contain objects with smaller 
          luminosities than the higher redshift bins.  
          (An asterisk signifies that the quoted rms is for $\log_{10}$ of 
          the index.)
          In these narrow bins of redshift and luminosity, the mean 
          velocity dispersions are similar in dense and less dense 
          environments (this is not quite true for the bottom 
          right panel, in which $\sigma$ seems to scatter to smaller 
          values in lower density regions). 
          The top-right panel is different from the others; 
          it happens to contain a supercluster, and this may have 
          compromised our estimates of environmental effects.}
 \label{paramsfb}
\end{figure}

To remove the effects of correlations with luminosity or velocity 
dispersion, we have further divided each redshift bin into narrow bins 
in luminosity.  
Figure~\ref{paramsfb} shows that the small environmental trends evident 
in Figures~\ref{paramenv} are seen consistently in all the panels.  
The top two panels of Figure~\ref{paramsfb} show the values of 
the structural parameters of galaxies which are slightly more 
luminous than $L_*$:  in the lowest redshift bin (i.e., the panel on 
the left), the luminosities, velocity dispersions, masses, sizes and 
profile shapes ({\tt fracDev}) of the high and low density environment 
samples are virtually identical, whereas a small but significant 
difference is detected in the color. 
Except for the color, the panel on the right shows similar trends, 
but recall that this redshift bin contains a supercluster, and 
this may bias our results.  

The bottom two panels show results in the two higher redshift bins, 
for which the bin in luminosity is necessarily brighter.  In these 
panels, the sample in low dense regions is significantly bluer, even 
though the median luminosity and velocity dispersion are the same as 
that in the higher density sample (this is not quite true for the bottom 
right panel, in which $\sigma$ seems to scatter to smaller values in 
lower density regions).  This illustrates clearly that the environment 
plays a role in determining galaxy colors.  
A K-S test confirms what is obvious to the eye:  the only cases in 
which the distributions of the parameters in low and high-density 
regions are significantly different ($p_{\rm KS}<0.05$) are for the 
$g-r$ color in the top left, and bottom two panels.    

The trends in Figure~\ref{paramsfb} are reported in Table~\ref{evenvtab}.  
Comparison of the two values of {\tt fracDev} in each panel of this 
and the preceding figure (also see Table~\ref{evenvtab}) show that the 
mean {\tt fracDev} is the same in both dense and  less dense regions.  
(We do not expect {\tt fracDev} to distinguish between S0s and 
ellipticals---for our purposes, S0s and ellipticals are both early-type 
galaxies.)  Since {\tt fracDev} is an indicator of morphology, this shows 
that any trends with environment are probably {\it not} driven by a 
correlation between morphology and density (e.g. Dressler 1980).
If some of the trends we see are indeed associated with the 
morphology-density relation, then the differences we detect in 
the properties of galaxies in high and low density regions are 
overestimates of the true differences.  This places even tighter 
limits on the possible role played by the environment.

\subsection{O{\small{II}}, Balmer lines, D4000, and $\alpha$-elements}
This subsection studies the correlation between emission line 
strengths (O{\small II}), absorption lines which are not part of the 
original Lick system (Worthey et al. 1994; Trager et al. 1998), i.e. 
the Balmer line indices H$\delta$ and H$\gamma$ (Worthey \& Ottaviani 1997), 
the strength, D4000, of the break in the spectrum at 4000\AA\ (e.g. 
Balogh et al. 1999), and some combinations of the Mg and Fe lines, i.e. 
[MgFe]$^\prime$ (e.g. TMB03) and 
Mg/$\langle$Fe$\rangle$, with the environment.  
The O{\small II} emission line is an indicator either of very recent 
star-formation or of AGN activity, 
the Balmer (absorption) line indices are sensitive to star formation, 
the 4000\AA\ break is an indicator of age (although all these 
observables also depend on metallicity and $\alpha$-element abundance 
ratios of the stellar population), 
Mg/$\langle$Fe$\rangle$ is an indicator of the $\alpha$-element 
abundance ratios (in early-type galaxies, this ratio is enhanced 
relative to the solar value), and [MgFe]$^\prime$ is an index which 
is an indicator of metallicity that is not expected to be very 
sensitive to the $\alpha$-element abundance ratios.

\begin{figure}
 \centering
 \plottwo{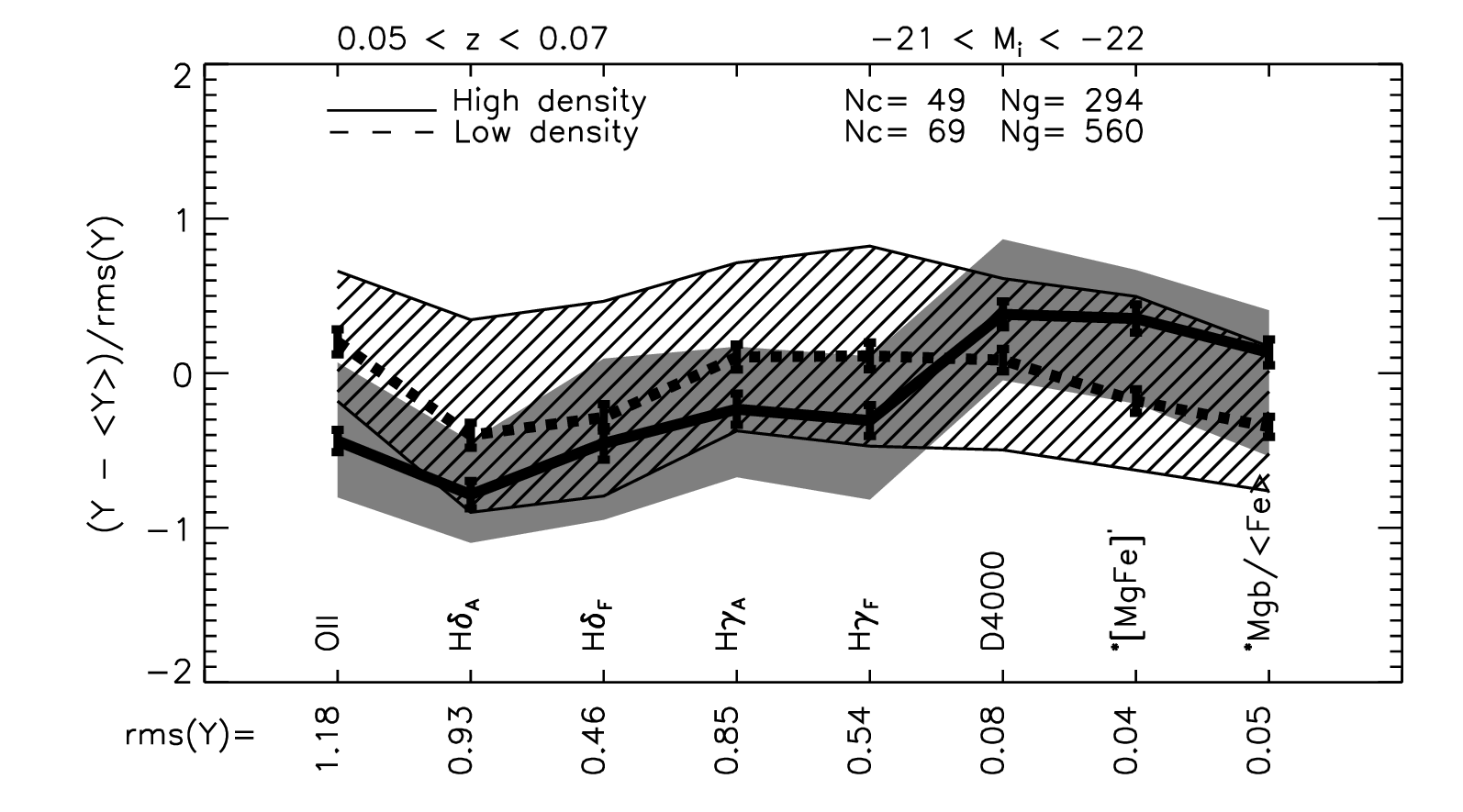}{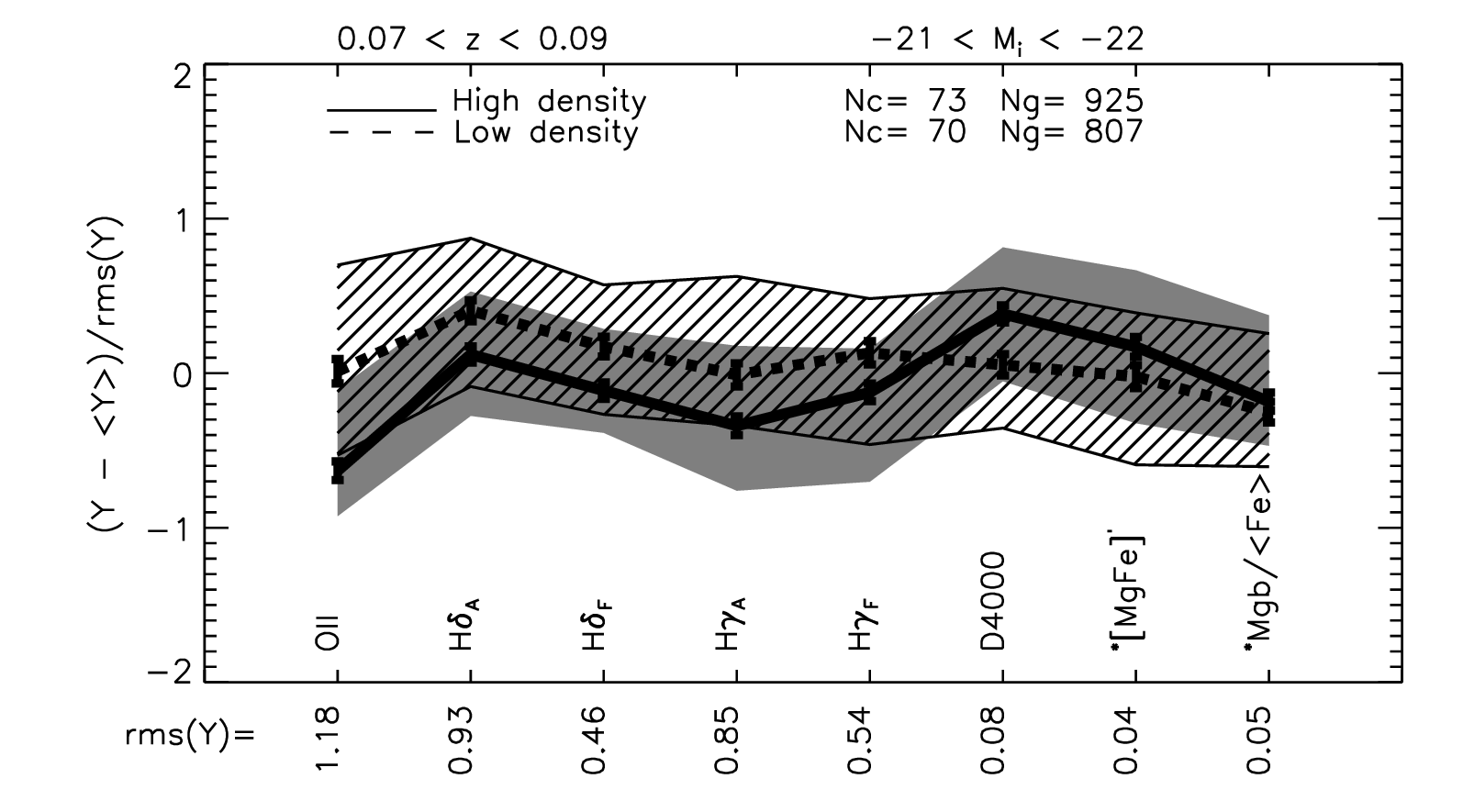}
 \plottwo{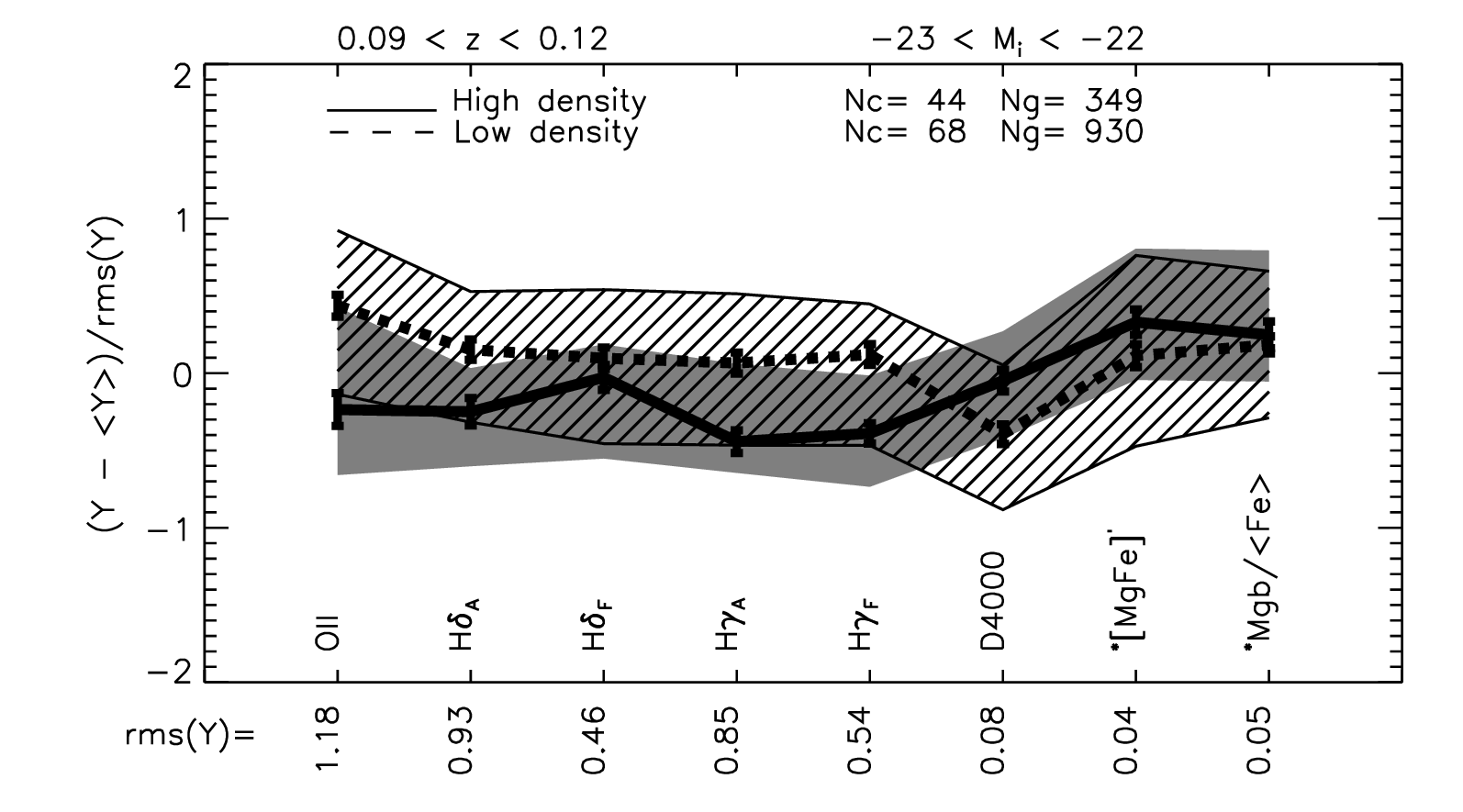}{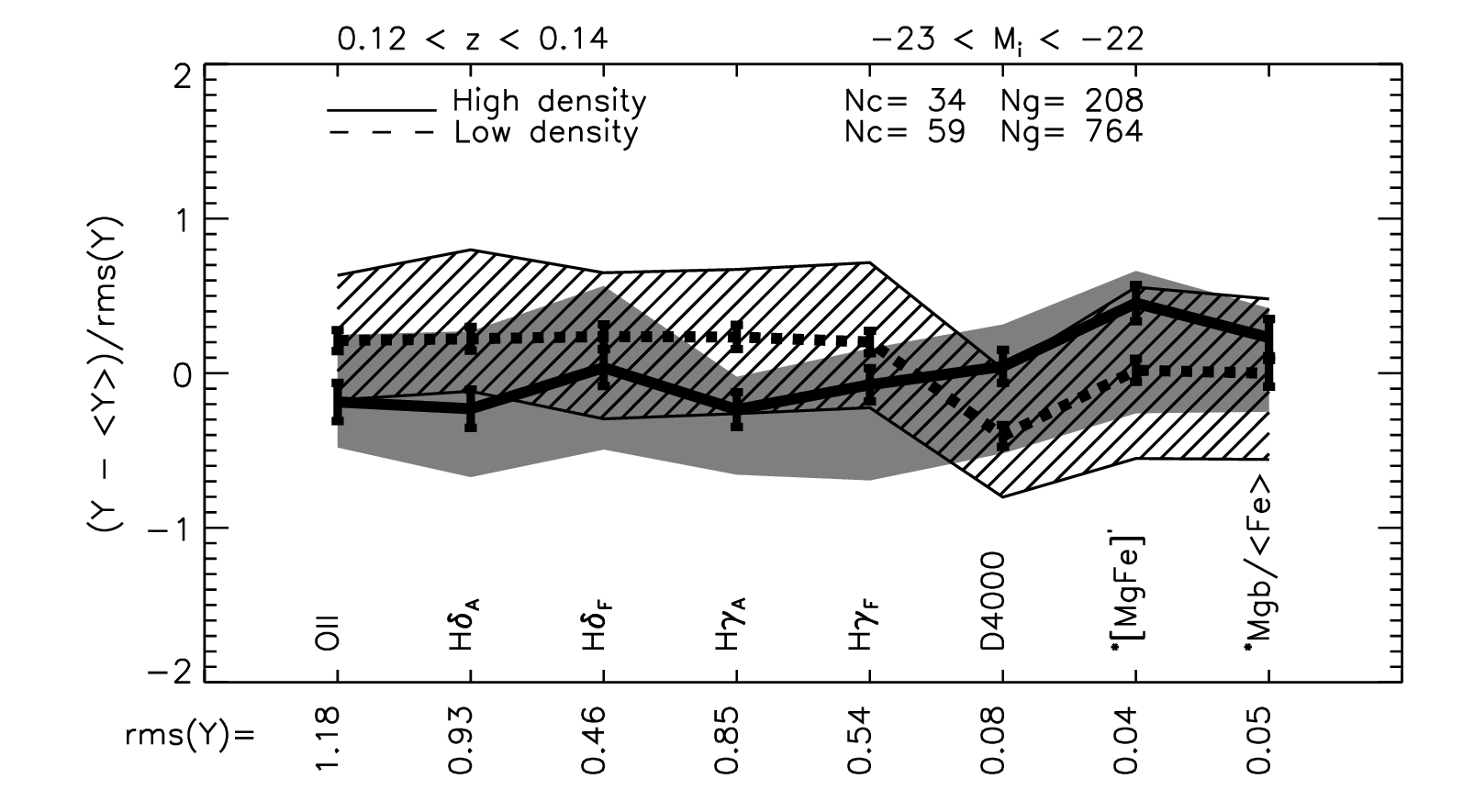}
 \caption{Emission line strengths (O{\small II}), 
          star-formation indicators (H$\delta$, H$\gamma$), 
          the strength of the break at 4000\AA (D4000), and 
          the combinations [MgFe]$^\prime$ and Mg/$\langle$Fe$\rangle$, 
          for objects with the same narrow range of redshifts and 
          luminosities as in the previous figure. 
          An asterisk signifies that the quoted rms is for $\log_{10}$ of 
          the index.}
 \label{linesfb}
\end{figure}

To simplify interpretation of the results, Figure~\ref{linesfb} presents 
measurements in the same narrow bins in redshift and luminosity as were 
used in Figure~\ref{paramsfb}.  
Recall that, for these bins, the effects of correlations between 
luminosity, velocity dispersion and environment are unimportant.  
(Plots similar to Figure~\ref{paramenv} show similar trends to those 
in Figure~\ref{linesfb}, but are harder to interpret).  
Galaxies in low density regions have stronger emission lines, stronger 
Balmer lines, and weaker 4000\AA\ breaks.  They also tend to have 
lower values of [MgFe]$^\prime$ and Mg/$\langle$Fe$\rangle$.  
(KS tests indicate that the distributions are not significantly 
different if the small error bars shown for each parameter overlap.)  
In all observables, the high and low density samples differ 
by about three tickmarks, indicating that the difference is about 
30\% of the rms spread across the entire sample.  
Table~\ref{evenvtab} gives more precise values for these differences.  

Recall that the mean luminosities and velocity dispersions are the
same in dense and less dense environments.  Therefore, trends with
environment in Figure~\ref{linesfb} are {\em not} the result of
index-$\sigma$ and $\sigma$-environment correlations.  
We have already argued that these trends are probably not due to the 
morphology-density relation either.

\subsection{Lick indices}
Lick indices (defined as in Trager et al. 1998) for objects with the 
same narrow range of redshifts and luminosities as in the previous 
figure are shown in Figure~\ref{lickf}.
The indices are arranged in order of increasing wavelength, so the 
figure can be thought of as illustrating how (little) the spectrum 
depends on environment.  Notice that the Mg line strengths tend to be 
slightly weaker in the low dense regions (about 30\% of the rms across 
the full sample, making Mg$_2$ and Mg$b$ weaker by about 
$0.01\pm 0.004$~mags and $0.2\pm 0.06$~\AA, respectively).  
On the other hand, most of the Fe indicators show no significant 
dependence on environment.  Once again, KS tests indicate that the 
error bars provide a reasonably accurate guide to the significance 
of the difference between environments:  overlapping error bars 
indicate no significant difference.

Table~\ref{evenvtab} quantifies the trends we have detected; 
recall that they are {\em not} the result of index-$\sigma$ and 
$\sigma$-environment correlations, and that they are unlikely to 
have arisen from a morphology-density relation.  
Presumably, age, metallicity and $\alpha$-element abundance differences 
play some role in the dependence on environment.  By studying which 
indices behave similarly in these plots, one can begin to identify 
which elements respond similarly to variations in age, metallicity, 
and $\alpha$-abundance.  We discuss this in the next section.

\section{Evolution and environment}\label{evenv}
The previous section showed that, at any given redshift, many 
line strengths depend weakly, but significantly, on environment.  
On the other hand, many of the line strengths have evolved 
significantly between $z=0.05$ and $z=0.15$ (Bernardi et al. 2003d).  
Section~\ref{sspmodels} uses stellar population synthesis models 
to interpret our measurements.  This section describes a relatively 
model independent interpretation of what our measurements mean.  

\subsection{Measurements of evolution}
The luminosity function of this sample changes slightly with redshift 
(Bernardi et al. 2003b).  The observed evolution can be accounted-for 
if one assumes that the number densities are not changing, but the 
higher redshift population is more luminous than that at lower 
redshifts:  the change is $1.15z$, $0.85z$ and $0.75z$ in the 
$g$-, $r$- and $i$-bands.  
Since a pure luminosity evolution model is consistent with the data, 
it is plausible that the population at lower redshifts is a passively 
evolved sample of the population at higher redshifts.  If so, then the 
observed evolution with redshift can be used as a clock.  
The argument is as follows.

The dashed lines in Figures~\ref{paramfbev}--\ref{lickbev} show how 
the structural parameters and line-strengths vary as the redshift 
changes.  These dashed lines (the same in all four panels of each 
figure) represent the evolution of galaxies with 
$2.35\le\log_{10}\sigma\le 2.4$ over the redshift range 
$0.06\le z\le 0.17$.  
The actual values are reported in Table~\ref{evenvtab}.  
(The redshift limits were set by requiring that these objects be seen 
over the entire redshift range, so as to minimize selection effects.  
Appendix~\ref{selev} discusses why selection effects are important, 
why we chose this bin in $\sigma$, and why our estimates of evolution 
are almost certainly upper-limits.)

By using the same dashed curve in all four panels (note that the solid 
curves in the top two panels in each figure are for a lower luminosity 
bin), we are implicitly assuming that evolution is not differential 
(i.e., all galaxies evolve similarly).  
Since age is likely the main driver for the dependence on redshift, 
indices with similar values traced by the dashed line may have a 
similar fraction of the total scatter across the whole sample arising 
from changes in age.  
For reference, the range $0.06\le z\le 0.17$ corresponds to a time 
interval of $\sim 1.3$~Gyr. 

Before we consider these figures in detail, recall that the 
Fundamental Plane indicates that galaxies in cluster-like environments 
are fainter than their counterparts in less dense regions 
(c.f. Table~\ref{fptable}).
Pure luminosity evolution between $0.06\le z\le 0.17$ gives approximately 
the same shift in magnitudes as those listed in Table~\ref{fptable} 
(e.g. in the $r$-band, the difference between environments is 
$\sim 0.08$~mags, and the evolution is $0.85\Delta z\sim 0.09$~mags).  
If the difference between the two environments is entirely due to age 
effects, then the objects in cluster-like environments must be older 
by 1.3~Gyrs.  If the other structural parameters and line-strengths 
show environmental differences which are similar to those seen by 
comparing populations at redshifts which are separated by 1.3~Gyrs, 
then this would constrain the roles played by age and metallicity in 
determining environmental differences.  

Figure~\ref{paramfbev} indicates that color is different from the 
other parameters such as size and velocity dispersion:  1.3~Gyrs 
of evolution appears to account for the entire spread in color 
values, but accounts for a negligible fraction of the spread in 
velocity dispersions, sizes and masses.  Evidently, color responds 
very differently to age than do the other parameters.  We discuss 
this in more detail shortly.  

The dashed lines in Figure~\ref{linesfbev} indicate that the Balmer 
lines are weaker at low redshift, whereas D4000, [MgFe]$^\prime$ and 
Mg/$\langle {\rm Fe}\rangle$ are slightly stronger.  
The sign of the evolution is consistent with that of an aging 
single-burst population.  
Taken at face value, Figure~\ref{linesfbev} indicates that an age 
difference of 1.3~Gyrs can produce a change in the Balmer line strengths, 
and in D4000, which is almost as large as the rms spread across the 
whole sample.  However, this apparently dramatic evolution should be 
treated cautiously.  At low redshifts, the Balmer lines, and D4000, 
are close to the edge of the SDSS spectrograph where there are known 
to be flux-calibration problems at the 3\% level.  
Evidence that this may have compromised our estimates of evolution 
is presented in Appendix~\ref{fluxcalib}:  the apparent evolution 
of H$\delta_{\rm A}+$H$\gamma_{\rm A}$ is about three times larger than 
measurements from the literature (Kelson et al. 2001) suggest.  
Hence, the dotted lines in Figure~\ref{linesfbev} show the result of 
dividing all estimates of evolution for the Balmer lines and D4000 
by a factor of three.  When this is done, 1.3~Gyrs of evolution 
produces a similar fraction of the rms spread ($\sim 30\%$) for 
O{\small II}, [MgFe]$^\prime$ and Mg/$\langle {\rm Fe}\rangle$.  
Note that this is a smaller fraction of the rms spread than it 
was for color.  

The dashed lines in Figure~\ref{lickfev} show that evolution accounts 
for different fractions of the rms spread across the sample for the 
different Lick indices.  (We have again corrected 
for potential flux calibration problems by reducing the measured 
values, for indices with rest wavelengths shorter than 4350\AA, by 
a factor of three:  the dotted curves show these corrected values).  
For instance, an age difference of 1.3~Gyrs produces a change in Mg 
which is 50 percent of the spread in Mg line-strengths.  If the spread 
in ages across the sample is 1.3~Gyrs, as color indicates, then we must 
conclude that other effects (such as metallicity) are responsible for 
the remaining scatter.  
Similarly, evolution changes the Fe index strengths by a negligible 
fraction of the spread in Fe index values (typically about 0.06~dex).  
Evidently, this spread must be due to effects other than (or in 
addition to) age.  In this respect, our results indicate that Fe and 
Mg respond differently to age and metallicity, and that C$_2$4668 
and NaD are more similar to Mg than to Fe.  

Notice that H$\beta$ appears to evolve slightly less than the other 
Balmer lines shown in the previous figure.  Although it is more 
sensitive to fill-in by emission than the other Balmer lines, we have 
attempted to correct for this (recall we add 0.05 O{\small II} to the 
measured value).  Since emission is more likely in the higher redshift 
population, it may be that the deeping of the H$\beta$ absorption 
feature with redshift is compensated by fill-in due to emission, 
and our correction has not completely accounted for this.  

\subsection{Dependence on environment}
The solid lines in Figures~\ref{paramfbev}--\ref{lickbev} show 
variations with environment.  Each panel shows environmental 
differences for galaxies in small bins in luminosity and redshift 
(as indicated).  Before we discuss individual parameters and indices 
further, notice how similar the solid curves are in each set of panels:  
environmental effects are approximately the same in all our redshift 
and luminosity bins (with the possible exception of the $z\sim 0.08$ 
bin which includes the supercluster).  
Table~\ref{evenvtab}, which quantifies trends with redshift, also 
quantifies these enviromental trends.  


Some of the differences between indices arise because the different 
observables correlate differently with luminosity (e.g. mass correlates 
more strongly with luminosity than size or velocity dispersion, and they 
all correlate more strongly with luminosity than does color).  
If an observable correlates strongly with luminosity, the rms spread 
reported along the bottom of each panel may be dominated by the rms 
spread in luminosities, rather than by the scatter at fixed 
luminosity (we discuss this again in the next section).  
The solid curves in the different panels show results for galaxies in 
a narrow bin in luminosity.  In this case, even if age effects account 
for the full scatter in index strength at fixed luminosity, they will 
not account for the full (i.e., the one computed using the full range 
of luminosities) rms spread.  

Previous work indicates that the slope of the color magnitude relation 
does not evolve out to redshifts of order unity, and this has been used 
to argue that residuals from the color magnitude relation are indicators 
of age (e.g. Kodama et al. 1998; Blakeslee et al. 2003).  
Since the various panels in Figure~\ref{paramfbev} are for a small range 
in luminosity, in essence, the value of the dashed line for color shows 
the mean color residual from the color magnitude relation.  Hence, it is 
an age indicator.  Since it shows the same variation as the solid line 
in the Figure (the difference between cluster and lower density 
environments), the mean age difference between environments is 
approximately the same as the mean age difference between the two 
indicated redshifts:  on average, objects in dense regions are 
less than $\sim$1.3~Gyr older.  

Like color, Mg shows approximately the same trends with environment as 
with redshift (Figure~\ref{lickfev}), although the magnitude of both 
trends for Mg appears to be approximately half that for color.  
However, Mg correlates strongly with luminosity, and a substantial 
part of this apparent difference is because we are only considering 
Mg for a small range in luminosity.  If we account for this 
difference, then Mg and color are remarkably similar.  If the age 
difference between cluster and low density environments is $\sim$1.3~Gyr, 
as suggested by color, then the similarity between the solid and dashed 
lines for Mg leaves little room for, e.g., metallicity effects.  
The next section provides a more quantitative model of these trends.

\subsection{Interpretation:  Evolution as a clock}\label{model}
The typical age, metallicity and $\alpha$-abundance may change 
with environment.
If trends with redshift primarily reflect age effects, then a 
comparison of the dashed and solid lines allows one to constrain 
the relative roles of age and metallicity and/or $\alpha$-abundance 
on the different indices.  A quantitative model is developed below.  

%
%

Suppose that the spread in index strength $I$ is 
\begin{equation}
 \sigma^2_{II}(z) = \Bigl[f_T^I(z)\, \sigma_T(z)\Bigr]^2 
                   + \Bigl[f_X^I(z)\,\sigma_X(z)\Bigr]^2
 \label{sigmaItx}
\end{equation}
where $T=\log_{10}$(age) and $X$ represents other effects 
(e.g., it could be $\log_{10}$(metallicity), 
$f_T^I$ describes how sensitive index $I$ is to age, 
$f_X^I$ describes how sensitive index $I$ is to other effects 
(e.g. metallicity), 
and $\sigma_T$ and $\sigma_X$ denote the spread in ages and the spread 
in everything else.  Note that it is the combination $f_T\sigma_T$, 
rather than the two terms individually, which determines the 
contribution of age effects to the spread in index values.  
Also note that $\sigma_T$ and $\sigma_X$ are the same for all indices.  
(The expression above really follows from assuming that index strengths 
are determined by some, possibly degenerate, combination of age and 
other effects.  For instance, if $X$ is mainly sensitive to metallicity, 
then the expression above is a consequence of the age-metallicity 
degeneracy.  Models suggest that, in this case, $f_X/f_T=3/2$.)  

If the only difference between the population at two redshifts 
is age, then the change in index strength is 
\begin{equation}
 \Delta I_{evol} = f_T^I\, \Delta T_{evol}, 
 \label{DeltaI}
\end{equation}
where $\Delta T$ denotes the change in $T$ between the two epochs.  
(This actually assumes that $f_T^I$ is the same for the two epochs, 
and that the other effects such as metallicity are also.  
This is unlikely to be correct if the age difference between the two 
epochs is large, so this really assumes that the two epochs are 
close in units of the timescale over which the relations between 
index strength and age and metallicity change.)  
This provides an estimate for $f_I^T$ in terms of observables.  

Consider what this implies for color.  
It is often argued that residuals from the color magnitude relation 
are indicators of age.  In such a model, the full spread in colors 
comes from the scatter in color at fixed magnitude, plus a term 
which accounts for additional effects:  
\begin{equation}
 \sigma_{CC}^2 = \sigma_{C|M}^2 + (f_X^C\sigma_X)^2
\end{equation}
Therefore, 
\begin{eqnarray}
 f_T^C\sigma_T &=& \sigma_{C|M} \equiv \sigma_{CC}\sqrt{1 - \xi_{CM}^2}
               \quad {\rm and}\nonumber\\
 f_X^C\sigma_X &=& \sqrt{\sigma_{CC}^2-\sigma_{C|M}^2} 
               \equiv \sigma_{CC}\xi_{CM}. 
\end{eqnarray}
(The previous section included a discussion of the relative roles of 
the scatter at fixed luminosity, and the slope of the index-luminosity 
relation.  In the present context of the scatter in color, these are 
$\sigma_{C|M}$ and $\xi_{CM}\sigma_{CC}$:  the scatter in colors at 
fixed magnitudes dominates if $|\xi_{CM}|\ll 1$.)

Since $\Delta C = f_T^C\, \Delta T$, $\sigma_T$ equals that $\Delta T$ 
for which $\Delta C = \sigma_{C|M}$.  To illustrate, 
in the SDSS dataset studied in the main text, $\Delta C_{evol} = 0.031$ 
when $\Delta T_{evol}$ corresponds to a time interval of 1.3~Gyrs.  
Therefore the time interval required to produce a color change of 
$\sigma_{C|M}$ is 
\begin{equation}
 \sigma_T = (\Delta T_{evol}/\Delta C_{evol})\,\sigma_{C|M}.
 \label{sigmaT}
\end{equation}
Since $\sigma_{C|M}$ is also measurable (the data indicate 
$\sigma_{C|M}=0.037$), one has calibrated the relation between 
age and color.  


Now consider other indices.  We observe a change 
 $\Delta I_{evol} = f_T^I\,\Delta T_{evol}$, so, over the range 
$\sigma_T$, the change in index strength would have been 
\begin{eqnarray}
 f_T^I\,\sigma_T &=& f_T^I\,\Delta T_{evol}\,(\sigma_T/\Delta T_{evol}) 
                  = \Delta I_{evol}\,(\sigma_T/\Delta T_{evol})\nonumber\\ 
                 &=& \Delta I_{evol}\,(\sigma_{C|M}/\Delta C_{evol}).  
 \label{ftst}
\end{eqnarray}
All the terms on the right hand side are observables, so 
$f_T^I\,\sigma_T$ can be estimated from the data.  
Since $f_T^I\,\sigma_T$ is the first term on the right hand side of 
equation~(\ref{sigmaItx}), and the left hand side of 
equation~(\ref{sigmaItx}) is also measured, 
\begin{equation}
 f_X^I\,\sigma_X = 
 \sqrt{\sigma_{II}^2-\Delta I_{evol}^2\,(\sigma_{C|M}/\Delta C_{evol})^2}
\end{equation}
can also be estimated from the data.  

The relative importance of age to other effects on the distribution of 
index strengths is given by the ratio 
\begin{eqnarray}
 {f_X^I\sigma_X\over f_T^I\sigma_T}  
  &=& \sqrt{\left({\sigma_{II}\over\Delta I_{evol}}
                  {\Delta T_{evol}\over\sigma_T}\right)^2 - 1}\nonumber\\
  &=& \sqrt{\left(1.74 {0.33\sigma_{II}\over \Delta I_{evol}}
            {\Delta C_{evol}/\sigma_{C|M}\over 0.031/0.037}\right)^2 - 1},
 \label{fxsxftst}
\end{eqnarray}
where the first line follows from equations~(\ref{sigmaItx}) 
and~(\ref{DeltaI}), and the second line uses the fact that 
 $\sigma_{C|M}/\Delta C_{evol} = \sigma_T/\Delta T_{evol}$ 
(equation~\ref{sigmaT}), and writes everything in units of 
typical observed values (e.g., Mg, [MgFe'], D4000) all have 
$\Delta I_{evol}/\sigma_{II} \approx 0.4$).  
Since $\Delta T_{evol}/\sigma_T$ is the same for all indices, 
larger values of $\Delta I_{evol}/\sigma_{II}$ indicate that 
a larger fraction of the spread in index strengths comes from 
age-related effects.  Thus, age determines a larger fraction of 
the observed spread in colors than it does for Mg, and ages matter 
even less for the spread in Fe values.  

In addition to comparing the relative roles of age and other 
effects for a given index, we can also compare indices with one 
another.  For instance, although $\sigma_X$ is not known, we know that 
it is the same for all indices $I$.  Therefore, if we measure 
 $\sigma_{II}^2-\Delta I_{evol}^2\,(\sigma_{C|M}/\Delta C_{evol})^2$ 
for two indices and compute the ratio, then the result equals 
the ratio of $(f_X^I)^2$ for the two indices.  
This allows a calibration of the relative sensitivities of the 
two indices to effects included in $X$.  Thus, 
\begin{equation}
 {f_X^{{\rm Mg}b}\over f_X^C} 
    = {\sqrt{0.06^2 - (0.020\,\sigma_{C|M}/0.031)^2\over 
            \sigma_{CC}^2 - \sigma_{C|M}^2}}
    = 3.44,
\end{equation}
whereas 
\begin{equation}
 {f_X^{{\rm Mg}b}\over f_X^{\rm Fe}} 
   = \sqrt{0.06^2 - (0.020/0.031)^2\sigma_{C|M}^2\over 
      0.05^2 - (0.005/0.031)^2\,\sigma_{C|M}^2} = 1.11.
\end{equation}
This indicates that color is less sensitive to effects such as 
metallicity than are Mg or Fe (or, e.g., metallicity and 
$\alpha$-enhancement effects on color cancel each other, 
whereas they do not cancel as strongly for Mg and Fe).  

We can proceed further if we are willing to make more model-dependent 
assumptions.  In models where residuals from the color-magnitude 
relation are age indicators, it is usually assumed that metallicity 
$\propto L^\beta$.  Therefore, $\sigma_Z=0.4\beta\,\sigma_{MM}$, where 
$\sigma_{MM}$ denotes the rms spread in magnitudes.  
If $X$ is mainly sensitive to metallicity, then 
\begin{equation}
 f_X^I 
    = {2.5\over\beta}{\sigma_{CC}\over\sigma_{MM}}\,
       \sqrt{\left({\sigma_{II}\over\Delta I_{evol}}
       {\Delta C_{evol}\over\sigma_{CC}}\right)^2 - 1 + \xi_{C|M}^2}.
\end{equation}
If we set $\sigma_{CC}=0.04$, $\sigma_{MM}=0.71$, $\xi_{C|M}=0.4$, 
$\Delta C_{evol} = 0.031$ and $\Delta I_{evol}/\sigma_{II}\approx 0.33$ 
(as it is for Mg, [MgFe]$^\prime$ and D4000), then 
\begin{equation}
 {f_X^I\over f_T^I} 
    = {2.5\over\beta}{0.04\over 0.71}\,{2.16\over f_T^I} 
    = {0.30\over\beta}{\Delta T_{evol}\over\Delta C_{evol}}
    = {0.44\over\beta}.
\end{equation}
To obtain $f_X=(3/2)f_T$ as models suggest would require $\beta = 0.3$:  
this is not inconsistent with single-burst models 
(e.g. Bruzual \& Charlot 2003) of the color-magnitude relation 
(Bernardi et al. 2004).  The implied spread in metallicity across the 
population is $\sigma_Z = 0.4\beta\sigma_{MM} = 0.08$.  

The behaviour of Fe is difficult to explain in such a model.  
Figure~\ref{lickfev} indicates that the trends with 
redshift and environment are weak.  The lack of evolution is suggestive 
of Fe being sensitive to effects other than age.  
But if there is a metallicity-luminosity correlation, as suggested 
above, one would expect to see an Fe-luminosity correlation.  
There is no such correlation in our dataset (e.g., see 
Figure~\ref{MgFesigma}).  

\subsection{Comparison of evolution and environment}
The previous subsection studied the effects of fixing the metallicity 
and varying the redshift by a small amount.  Here, we fix the redshift 
and vary the environment.  In this case, 
\begin{equation}
 \Delta I_{env} = f_T^I\sigma_T\,{\Delta T_{env}\over \sigma_T} + 
                  f_X^I\sigma_X\,{\Delta X_{env}\over \sigma_X}.
\end{equation}
This illustrates that age differences can be compensated-for by 
associated changes in $X$.  

Many of the indices have $\Delta I_{env}=\Delta I_{evol}$.  
In this case
\begin{equation}
 f_T^I\,\Delta T_{evol} = f_T^I\sigma_T\,{\Delta T_{env}\over \sigma_T} + 
                  f_X^I\sigma_X\,{\Delta X_{env}\over \sigma_X},
\end{equation}
so 
\begin{equation}
 {\Delta T_{env}\over \sigma_T} = {\Delta T_{evol}\over \sigma_T} - 
     {f_X^I\sigma_X\over f_T^I\sigma_T}\,{\Delta X_{env}\over\sigma_X}.  
 \label{dTenv}
\end{equation}
If $\Delta X_{env}=0$, then 
the mean age difference between environments is the same as the mean 
age difference between two epochs, but in general, a larger or smaller 
age difference can be compensated-for by an associated change in $X$, 
and $(f_X\sigma_X/f_T\sigma_T)$ governs how large this change must be.  

For the indices where 
 $\Delta I_{env}\approx \Delta I_{evol}\approx 0.33\sigma_{II}$ 
(e.g. Mg), 
 $(\Delta T_{env}/\sigma_T) = (\Delta T_{evol}/\sigma_T) - 
     1.42\,({\Delta X_{env}/\sigma_X}) 
  = 0.84 - 1.42\,({\Delta X_{env}/\sigma_X})$; 
evidently, objects in the densest regions of our sample are older 
than those in the least dense regions, unless they have larger $X$ 
values (e.g., they are more metal rich) by $(0.84/1.42)\sigma_X$.  
Alternatively, if the objects in dense regions have smaller $X$ values 
(e.g., they are metal poor), then the age difference between environments 
can be larger than 0.84 times the rms age variation across the sample.  

The age-$X$ degeneracy can be broken if we again consider color.  
Our estimate of environmental effects uses a small range of magnitudes 
in any given redshift bin, so, in effect, we are studying residuals from 
the color magnitude relation in different environments.  If these 
residuals are only sensitive to age, they have $f_X^{C|M}=0$.  
Therefore, $\Delta T_{env} = \Delta C_{env}/f_T^{C|M} 
            = \Delta C_{env}/(\Delta C_{evol}/\Delta T_{evol})$.  
It happens that $\Delta C_{env}\approx\Delta C_{evol}$, so 
$\Delta T_{env} \approx \Delta T_{evol}$.  
This age difference must be independent of the index which was used 
to estimate it.  Now consider other indices for which 
$\Delta I_{env}\approx \Delta I_{evol}$ (e.g. Mg, Fe, D4000).  
Since equation~(\ref{dTenv}) describes these indices, and 
$f_X^I\sigma_X/f_T^I\sigma_T$ is of order unity (equation~\ref{fxsxftst}), 
it must be that $\Delta X_{env}/\sigma_X\ll 1$.  

The fact that Mg behaves similarly to color suggests that the entire 
environmental variation of Mg can be accounted for by age effects: 
metallicity effects are not very important.  
This leads to a provocative conclusion:  claims that metallicity 
effects are important for explaining the weak environmental 
dependence of Lick index strengths like Mg are incompatible 
with currently popular interpretations of the color-magnitude 
relation.

\section{Comparison with stellar population synthesis models}\label{sspmodels}
The previous section argued that our data are consistent with the 
hypotheses that the stellar populations of early-type galaxies 
in dense regions are slightly older than in less dense regions, 
and that metallicity differences between environments are 
negligible.  
This section shows that using single-burst stellar population 
synthesis models to interpret our data results in qualitatively 
similar conclusions.

\subsection{The models}
The models are characterized by three numbers: age, metallicity, and 
$\alpha$-enhancement.  In most of the plots which follow, these models 
are evaluated on a grid of  age = 2, 3, 5, 10 and 15~Gyrs, 
 metallicity [Z/H] = $-0.33$, 0, 0.35 and 0.67, 
 and [$\alpha/$Fe] = 0, 0.3 and 0.5, and we interpolate linearly 
between these grid points.

The effects of non-solar [$\alpha/$Fe] values have been considered 
by Tripico \& Bell (1985), and incorporated into models by 
Trager et al. (2000), TMB03, TMK04 and Tantalo \& Chiosi (2004).  
In what follows, we have chosen to concentrate on the models of 
TMB03-TMK04.
However, the interpretations which follow are necessarily model 
dependent, and we caution that different models sometimes defer 
substantially from one another, so the resulting interpretations 
should be treated with caution.  Indeed, Tantalo \& Chiosi (2004) 
show that their $\alpha$-enhanced models can differ substantially 
from those of other groups, and have argued that uncertainties in how 
one treats [Ti/Fe] can have important effects.  Hence, they argue 
strongly that the tendency to draw strong conclusions from such models 
is unwarranted.  

To illustrate the point that all quantitative conclusions which 
follow are model dependent, Figure~\ref{models} compares the evolution 
of H$\beta$, H$\gamma_{\rm F}$, Mg$b$, and Fe, in the models of 
TMB03-TMK04 with the models of Bruzual \& Charlot (2003).  
The model tracks are for solar [$\alpha$/Fe], because the 
Bruzual-Charlot models do not yet include non-solar [$\alpha$/Fe].  
The solid and dashed curves in each panel show the Bruzual-Charlot 
models for two values of the metallicity 
(solar and greater than solar).  The three dotted curves in each 
panel show the TMB03-TMK04 models for three different metallicities 
(solar and above, all at solar $\alpha$-enhancement).  
In all panels, the models agree at solar metallicity (the solid 
curve is close to the dotted one).  However, in some panels the 
Bruzual-Charlot models are more like the highest metallicity 
TMB03-TMK04 models (e.g. H$\beta$), whereas in others, they are 
more like the intermediate metallicity models (e.g. H$\gamma_{\rm F}$).  

To see that these differences matter, consider the two curves close to 
the center in the top right panel.  Although the two models differ 
by 10 percent at fixed age, they differ by 100 percent at fixed 
$\langle$Fe$\rangle$.  Since it is $\langle$Fe$\rangle$ which is 
observed, any quantitative conclusions about metallicity and age 
are strongly model dependent.  
The Figure also shows that the differences between the ages 
infered from the models will depend on whether one uses H$\beta$ or 
the higher-order Balmer lines H$\gamma_{\rm F}$, 
and that these differences can be substantial.  
The plot above suggests that one would infer younger ages from 
H$\gamma_{\rm F}$ than from H$\beta$, a point we will return to later.

\subsection{The method}

Figure~\ref{modelgrids} shows the distribution of 
H$\beta$ vs Mg (top), $\langle$Fe$\rangle$ vs. Mg (middle) 
and H$\beta$ vs Mg/$\langle$Fe$\rangle$ (bottom).  
Table~\ref{lickres} gives these index strengths at Lick, rather 
than SDSS, resolution.  
(Recall that $\langle$Fe$\rangle =$(Fe5270$+$Fe5335)$/2$.)  
In this, and the Figures which follow, we have divided the sample 
into the same small bins in redshift as in the previous figures, 
so as to be able to separate out the effects of evolution.  
Black, green, red and magenta points represent objects at 
successively higher redshifts ($0.05<z<0.07$, $0.07\le z <0.09$, 
$0.09\le z<0.12$ and $0.12\le z<0.14$).  In addition, symbol sizes 
have been scaled to reflect the number of galaxies which made-up 
the composites they represent.  

The grids in each panel show the TMB03 models.  
The solid and dashed grids in the top panels show age and 
metallicity at $\alpha$-enhancements which are 0 (solar) 
and 0.3.  This choice of observables separates out 
age and metallicity quite clearly.  
The observables in the middle panel separate out metallicity and 
$\alpha$-enhancement nicely:  solid and dashed grids show metallicity 
and $\alpha$-enhancement for ages of 10 and 15~Gyrs.  
And the solid and dashed grids in the bottom panels show ages and 
$\alpha$-enhancements when the metallicity is solar and 0.35 
respectively.

We determine ages, metallicities and $\alpha$-enhancements for each 
data point from these grids as follows.  We begin with a guess for 
the $\alpha$-enhancement (say, solar).  Then the H$\beta$-Mg plot 
provides estimates of the age and metallicity (by linear interpolation 
in the model grid).  We then use the $\langle$Fe$\rangle$-Mg grid 
associated with the age estimate to estimate a metallicity and 
$\alpha$-enhancement.  
If this new $\alpha$-enhancement differs substantially from our 
initial guess, we return to the H$\beta$-Mg plot, but now use the 
new $\alpha$-enhancement to re-estimate the age and metallicity from the 
model, and use the new age estimate to determine the appropriate 
$\langle$Fe$\rangle$-Mg model with which to estimate [Z/H] and 
[$\alpha$/Fe] for the data point.  
We continue iterating this procedure until consistent ages, 
metallicities and $\alpha$-enhancements have been obtained.  
Typically, this happens after about five iterations.  
These provide our first estimates of self-consistent values 
of age, [Z/H] and [$\alpha$/Fe].  

We then repeat this process, but now using the $\langle$Fe$\rangle$-Mg 
and H$\beta$-Mg/$\langle$Fe$\rangle$ plots.  
(I.e. we begin with a guess for the age, 
determine metallicity and $\alpha$-enhancement from the model grid 
for $\langle$Fe$\rangle$-Mg, and we use this metallicity to determine 
the model grid to use for the H$\beta$-Mg/$\langle$Fe$\rangle$ 
diagnostic.  In turn, this provides revised estimates for the age and 
$\alpha$-enhancement; the new age estimate is then used in the 
$\langle$Fe$\rangle$-Mg diagram.  
This is repeated until a new self-consistent triple of age, [Z/H] and 
[$\alpha$/Fe] has been found.)  
Finally, we repeat all this using the H$\beta$-Mg/$\langle$Fe$\rangle$ 
and H$\beta$-Mg plots.  Thus, we have three estimates of age, 
[Z/H] and [$\alpha$/Fe] which we can use to test for self-consistency 
(see next subsection).  

It is particularly important to estimate the typical uncertainty on 
the derived parameters which are due to the measurement errors
because the model grids are not aligned with the axes of observables 
in Figure~\ref{modelgrids}, so these uncertainties will be correlated 
(e.g. Trager et al. 2000).  We do this as follows.  
We compute the typical uncertainties in the measurements of 
H$\beta$, Mg and $\langle$Fe$\rangle$ (the errors on 
Mg/$\langle$Fe$\rangle$ are easily derived from those on Mg and 
$\langle$Fe$\rangle$).  
We then generate twenty seven mock `galaxies' by adding and subtracting 
the reported measurement errors to the mean observed values of H$\beta$, 
Mg and $\langle$Fe$\rangle$, and hence Mg/$\langle$Fe$\rangle$ 
(we use mean, mean plus error, and mean minus error for each of the 
three observables).  
We then run the algorithm described in the previous section for 
each of these model `galaxies'.  This provides three estimates of 
age, [Z/H] and [$\alpha$/Fe] for each of the twenty seven `objects'.  
The resulting distribution of points in the age, [Z/H] and [$\alpha$/Fe] 
plane forms our estimate of the full uncertainties in the derived 
parameters.  

\subsection{Results}
From the three combinations of plots described in the previous 
subsection, we have three different estimates of age, [Z/H] and 
[$\alpha$/Fe].  If the procedure just described is self-consistent, 
and the models are sufficiently realistic, one expects these three 
estimates to be similar.  The extent to which this is the case is 
shown in Figure~\ref{agemetalenhan}.  Associated with each data 
point is a triangle, the vertices of which represent the three 
estimates of the derived age, [Z/H] and [$\alpha$/Fe].  
The size of the triangle gives a rough estimate of the systematic 
uncertainty in the derived quantities associated with the model 
(i.e., assuming perfect measurements).  The figure indicates that 
this uncertainty does not contribute significantly to the scatter 
in the various panels.  Solid lines show the bisector fit to the 
relation in the top left panel, and the direct fit to the relation 
in the bottom panels.  
These suggest that older systems tend to have smaller metallicities 
(top panels) and slightly larger $\alpha$-enhancements (bottom panels).  

The ellipses in the left corner of each panel show the typical 
uncertainty on the derived parameters which come from the measurement 
errors; these dwarf the systematic uncertainties associated with our 
algorithm.  
In the top panels especially, the errors are strongly correlated, so 
that much of the apparent anti-correlation between age and metallicity 
is due to the errors.  This anti-correlation of the errors tracks the 
well-known age-metallicity degenaracy associated with such models.  
Table~\ref{derivedTZA} provides the derived ages, [Z/H] and 
[$\alpha$/Fe] values, with associated error estimates, for each 
composite spectrum.  

In all panels of Figure~\ref{agemetalenhan}, the distribution of 
derived parameters is slightly, but not substantially, broadened by 
the errors.  In particular, although the long-axis of the distribution 
shown in the top panels has been broadened by the correlated errors, 
the shape of the error ellipse shows that the distribution along the 
shorter axis cannot have been significantly broadened.  
Trager et al. (2000) argued that the scatter in this direction is 
primarily due to scatter in $\sigma$:  
objects with the same $\sigma$ lie parallel to this relation, with 
larger $\sigma$ objects lying at larger ages and metallicities.  
A similar statement applies if we substitute [$\alpha$/Fe] for $\sigma$.  
This is true of our dataset also:  at fixed $\sigma$ (or [$\alpha$/Fe]), 
older objects appear to be metal poor.  

The panels on the right allow a comparison between environments.  
Red lines show the fits from the panels on the left, 
and blue lines shows the result of fitting for the shift in zero-point 
while requiring the same slope as the red line.  
Black lines show the fit (bisector in top right panel, direct in 
bottom right) which allows the slope to vary as well.  
These fits indicate that there is a small offset between the 
age-metallicity correlation in the two environments, 
and that objects in dense regions tend to have slightly larger 
$\alpha$-enhancements than their counterparts of the same age 
in less dense regions.

Figure~\ref{modelsigma} shows all three derived quantities 
as a function of velocity dispersion in the two environments.  
(Because our three estimates of the derived quantities tend to be so 
similar, we have not shown the triangles associated with each point.  
Also, note the selection effect---the high redshift bins do not have 
objects at small $\sigma$).  In each panel, black solid lines show 
direct fits to the relations in the high density regions at 
$z\sim 0.06$; red solid lines show the offset in Gyrs, [Z/H] and 
[$\alpha$/Fe] required to fit the relations at $z\sim 0.11$.  
Dotted lines in the panels on the right show the result of fitting 
for the offsets (in Gyrs, [Z/H] and [$\alpha$/Fe]) from the solid 
curves which best fit the relations for the low density sample.  
These fits show that, at any given redshift and in any environment, 
objects with large $\sigma$ tend to be older (top panels), 
have larger metallicities (middle panels), 
and larger $\alpha$-enhancements (bottom).  
These fits are given in Table~\ref{sigmafits}, which also quantifies 
trends with redshift and environment.

Comparison of the red and black curves shows that the high and 
low density samples are both qualitatively consistent with the 
hypothesis of passively evolving populations:  at fixed $\sigma$, 
the high redshift objects are younger, but there is no evolution 
in [Z/H] and [$\alpha$/Fe].  Moreover, the age difference is 
consistent with the redshift difference in our assumed cosmology.  

Comparison of the solid and dotted lines in the panels on the right 
allows a study of the effects of environment.  These indicate that, at 
fixed $\sigma$, objects in dense regions tend to have the same ages 
and metallicities as their counterparts in less dense regions, but 
they have slightly larger $\alpha$-enhancements.  
In constrast, Thomas et al. (2005) found that objects in less dense regions 
are slightly younger and metal rich, but they find no 
$\alpha$-enhancement effect (see also Kuntschner et al. 2001).  

To study this further, dot-dashed lines show the fits reported by 
Thomas et al. (2005).  Whereas our metallicity-$\sigma$ and 
[$\alpha$/Fe]-$\sigma$ values tend to be slightly different from 
theirs, our age-$\sigma$ relations are very different, especially in 
dense regions.  
On the other hand, our derived correlations in the high-density 
regions in are slightly better agreement with those of the NOAO-FP 
survey (Nelan et al. 2005) of clusters (dashed curves).  

In the lower density regions, one might worry that our decision to 
compare low and high density environments by keeping the slope 
of the correlations with $\sigma$ fixed and simply fitting for an 
offset appears to be unreasonable for the [Z/H]-$\sigma$ relation.  
The slope of this relation appears to be steeper in the low density 
regions.  Allowing for this would bring it into better agreement 
with that of Thomas et al. (2005). But the significant differences at high 
densities remain.  

As a check, we have run our algorithm on the data analyzed by 
Thomas et al. (2005), and verified that the derived ages, metallicities 
and $\alpha$-enhancements we derive are within a few percent of those 
they obtained using a slightly different algorithm.  For example, for 
an object with 
(H$\beta$, Mg~$b$, $\langle$Fe$\rangle$) = (1.59\AA, 4.73\AA, 2.84\AA), 
our procedure returns an age of 10.7~Gyr, [Z/H]=0.26 and [$\alpha$/Fe]=0.24.  
Thus, differences in our methods used to derive ages, [Z/H] and 
[$\alpha$/Fe] from the model grids are unimportant.  
Hence, we do not have a good explanation for why our correlations with 
$\sigma$, and the environmental dependences of these correlations, 
in our data set differ from the correlations in theirs---while our 
definitions of environment differ in detail, it is not obvious that 
this should lead to the relatively large qualitative differences we see.

Figures~\ref{modelHdHg}--\ref{modelsigmaHdHg} show the result of 
repeating this analysis but using H$\gamma_{\rm F}$ instead of H$\beta$.  
(This choice was motivated by discussion in TMK04; 
we find similar results if we use H$\delta_{\rm A}$+H$\gamma_{\rm A}$ 
instead.)  In general, this choice results in ages that are younger 
by 1~Gyr, but similar [Z/H] and [$\alpha$/Fe].  
This is true even if one allows for the possibility that there may be 
flux calibration problems for these Balmer lines at low redshifts 
(cf. Appendix~\ref{fluxcalib}).  (If there is a problem, then the 
ages we derive for the lowest redshift bin are overestimates, 
and the H$\gamma_{\rm F}$ ages will be even smaller than those 
from H$\beta$.  In addition, the metallicities will increase slightly.  
Applying these corrections would make the relative differences 
between the low and high redshift samples, at least in dense regions, 
more similar to the differences seen when H$\beta$ was used.  
But the overall offsets between the ages and metallicities 
estimated from these two sets of plots remains.  
Note that this difference from the H$\beta$ results is what one 
might have guessed from the differences shown in Figure~\ref{models}.)  
In particular, these differences are larger than any difference 
between the two environments.  


\section{Discussion and conclusions}\label{discuss}
The properties of early-type galaxies show weak but significant 
correlations with environment.  The Fundamental Plane indicates that 
cluster galaxies are $\sim$0.08~mag/arcsec$^2$ fainter than their 
counterparts in low density environments, and the scatter around the 
plane shows no significant dependence on environment  
(Figure~\ref{fpgi} and Table~\ref{fptable}).  This is consistent with 
the hypothesis that cluster galaxies are only slightly older than 
their counterparts in the low density environments, or that metallicity 
effects on the fundamental plane parameters counteract those of age.  

Various indicators of chemical abundances also correlate weakly with 
environment:  galaxies in low density environments are slightly bluer, 
have experienced star formation more recently, and have weaker D4000 
and Mg (Figures~\ref{paramsfb}--\ref{lickb}), but these trends tend 
to be smaller than about 30\% of the rms spread across the entire 
sample: i.e. the full distribution of observed values is not 
subtantially broadened by environmental effects.  

Galaxy color, and many absorption line-indices, correlate primarily 
with velocity dispersion.  However, differences between cluster and 
low density environment populations are seen even when the velocity 
dispersion is the same in both environments---the environment plays 
an important role in determining galaxy properties. 

We discussed two methods of quantifying correlations with environment:  
both indicate that objects in clusters tend to be slightly older than those 
in less dense regions, but that metallicity differences are small.  
The first method (Section~\ref{evenv}) uses an argument which is 
relatively model independent, and proceeds as follows.  
The absorption line-strengths evolve with redshift.  Compared to their
values at $z\sim 0$, Balmer lines were stronger, D4000 was weaker, 
Mg was weaker, and $\langle$Fe$\rangle$ was not very different at 
$z\sim 0.2$.  
If the high redshift population is simply a passively evolved version 
of that locally (previous analysis suggests that this is a good 
approximation), then the observed evolution with redshift can be used 
as a clock.  In particular, comparison of the difference between cluster 
and low density environment populations with the dependence on redshift 
(Figures~\ref{paramfbev}--\ref{lickbev}, and Table~\ref{evenvtab}) 
allows one to constrain the different relative roles of age and 
metallicity/$\alpha$-abundance on the various observed properties of 
galaxies.  

For instance, our analysis shows, in a model independent way, 
that Fe is less sensitive to age than is Mg.  In addition, residuals 
from the Mg-luminosity relation show similar trends with evolution and 
environment as residuals from the color-magnitude relation.  
If we add the constraint that residuals from the color-magnitude 
relation are age indicators, then our data indicate that the rms 
spread in ages across the sample is slightly greater than $\sim1.5$~Gyrs, 
and the rms spread in log$_{10}$(metallicity) is 0.08.  
In addition, galaxies in cluster environments are $\sim1$~Gyrs older 
than their counterparts in less dense environments, and the 
metallicity difference between the two environments is a negligible 
fraction of the full range of metallicities in the sample.  

The second method (Section~\ref{sspmodels}) uses single burst stellar 
population synthesis models to interpret our data.  These indicate 
that the objects at the lower redshifts in our sample are indeed 
consistent with being passively evolved versions of the objects 
at higher redshifts:  we see evolution in age, but not in 
metallicity [Z/H] or the $\alpha$-element abundance ratio [$\alpha$/Fe].  
We find that age, [Z/H] and [$\alpha$/Fe] all increase with 
increasing velocity dispersion $\sigma$, in qualitative agreement 
with previous work.  In addition, objects in dense regions tend to be 
older by less than $1$ Gyr and have larger [$\alpha$/Fe] ($\sim 0.02$) than 
their counterparts of the same $\sigma$ in less dense regions, but 
there is no evidence that objects in low density regions are more 
metal rich (Figures~\ref{modelsigma} and~\ref{modelsigmaHdHg}).  
This suggests that, in dense regions, the stars in early-type galaxies 
formed at slightly earlier times, and on a slightly shorter timescale, 
than in less dense regions.  This is in qualitative agreement 
with a number of recent results (e.g. Thomas et al. 2005; 
Carretero et al. 2004).
Whereas these qualitative conclusions do not depend on which absorption 
lines are used to interpret the data, quantitative conclusions do:  
use of H$\beta$ leads to systematically older ages than does 
H$\gamma_{\rm F}$.  

We began with the statement that the monolithic and stochastic models 
make different predictions for how early-type galaxy properties should 
depend on environment.  While the analyses presented here do not 
answer the question of which model is correct, we feel that they 
provide a useful method for addressing this issue.  
In particular, it will be interesting to see if these models are 
able to reproduce the weak environmental dependence of formation time 
and timescale suggested by our data.  

\bigskip

Funding for the creation and distribution of the SDSS Archive has
been provided by the Alfred P. Sloan Foundation, the Participating
Institutions, the National Aeronautics and Space Administration, the
National Science Foundation, the U.S. Department of Energy, the
Japanese Monbukagakusho, and the Max Planck Society. The SDSS Web site
is http://www.sdss.org/. 

The SDSS is managed by the Astrophysical Research Consortium (ARC)
for the Participating Institutions. The Participating Institutions are
The University of Chicago, Fermilab, the Institute for Advanced Study,
the Japan Participation Group, The Johns Hopkins University, the Korean
Scientist Group, Los Alamos National Laboratory, the
Max-Planck-Institute for Astronomy (MPIA), the Max-Planck-Institute
for Astrophysics (MPA), New Mexico State University, University of
Pittsburgh, Princeton University, the United States Naval Observatory,
and the University of Washington.

\appendix
\section{Evolution and selection effects}\label{selev}
In the main text, we present estimates of how index-strengths in 
the population have evolved.  In all cases, these estimates were 
made by computing residuals from the index-$\sigma$ relation 
rather than from the index-magnitude relation.  
This was done for two reasons.  First, for almost all indices, 
the index-$\sigma$ relation is considerably tighter than the 
index-luminosity relation, so a small amount of evolution is more 
easily detected.  
Second, the velocity dispersion is expected to evolve much less than 
the luminosity, so it is a more direct surrogate for estimating 
evolution at fixed mass.  

However, the SDSS sample is magnitude limited.  Therefore, care must 
be taken to account for the effects of this selection before 
interpretting trends with redshift as being due to evolution.  
To illustrate, Figure~\ref{MgFesigma} shows the Mg-luminosity, 
Mg-$\sigma$, Fe-luminosity and Fe-$\sigma$ correlations measured using 
objects in different redshift bins (bins are adjacent with edges 
at $z=0.05$, 0.07, 0.09, 0.12, 0.14 and 0.17, as in the main text).  
The plot uses luminosities corrected for evolution to $z=0$.  
In all cases, the filled circles show the median values of the index 
(Mg or Fe) in small bins in luminosity or $\sigma$.  
Hashed regions show the range which includes sixty-eight percent 
of the objects.  
(Mg and Fe are representative of most of the other indices presented 
in the main text.)  

At any given redshift, Mg correlates with both luminosity and 
$\sigma$ whereas Fe correlates with $\sigma$ but not with luminosity.  
The Mg-$\sigma$ correlation is significantly tighter than Mg-luminosity.  
Both Mg-$\sigma$ and Fe-$\sigma$ appear to evolve.  
In the case of Fe-$\sigma$, the apparent evolution is differential.  
One might have thought that because Fe does not correlate with 
luminosity, measurement of the Fe-$\sigma$ relation can be made 
without worrying about the magnitude limit of the sample, so the 
differential evolution is real.  The following model shows that this 
is not the case.

Let $I_*$, $M_*$ and $V_*$ and $\sigma_{II}$, $\sigma_{MM}$ and 
$\sigma_{VV}$ denote the mean and rms values of index strength, 
absolute magnitude and log$_{10}$(velocity dispersion) in the entire 
population at a given redshift (i.e., not just the part which was 
selected for observation).  
Let ${\cal I} = (I-I_*)/\sigma_{II}$, 
${\cal M} = (M-M_*)/\sigma_{MM}$, and 
${\cal V} = (V-V_*)/\sigma_{VV}$ and define 
$\xi_{IM}\equiv \langle {\cal I}{\cal M}\rangle$, 
$\xi_{IV}\equiv \langle {\cal I}{\cal V}\rangle$, and 
$\xi_{VM}\equiv \langle {\cal V}{\cal M}\rangle$, 
where the angle brackets denote averages over the entire population.  
The mean value of $I$ at fixed $M$ is 
\begin{equation}
 \langle {\cal I}|{\cal M}\rangle = {\cal M}\,\xi_{IM},\quad
\end{equation}
and similar expressions hold for the other pairs of observables:  
e.g. 
 $\langle {\cal I}|{\cal V}\rangle = {\cal V}\,\xi_{IV}$, 
 $\langle {\cal V}|{\cal M}\rangle = {\cal M}\,\xi_{VM}$, 
 $\langle {\cal M}|{\cal I}\rangle = {\cal I}\,\xi_{IM}$, etc.  
Similarly, the mean $I$ at fixed $V$ and $M$ is 
\begin{eqnarray}
 \left\langle{\cal I}\Big|{\cal M},{\cal V}\right\rangle 
 &=& {\cal M}\,{\xi_{IM} - \xi_{IV}\xi_{MV}\over 1-\xi_{MV}^2} 
     \nonumber\\
 && +\ {\cal V}\,{\xi_{IV} - \xi_{IM}\xi_{MV}\over 1-\xi_{MV}^2} 
\end{eqnarray}
with obvious permutations giving 
$\langle{\cal M}|{\cal V},{\cal I}\rangle$ and 
$\langle{\cal V}|{\cal I},{\cal M}\rangle$.  

If the correlation between index and magnitude is entirely due to 
the index-$\sigma$ and magnitude-$\sigma$ correlations, 
then $\xi_{IM} = \xi_{IV}\xi_{VM}$.  
(Bernardi et al. 2004 show that this is true for the color-$\sigma$ 
relation.)  In this case, 
$\langle{\cal I}|{\cal M},{\cal V}\rangle = {\cal V}\,\xi_{IV}$ 
is independent of ${\cal M}$, so the index-$\sigma$ relation 
can be measured directly from the observed data without worrying 
about the magnitude limit (e.g. Bernardi et al. 2004).  Note that 
this is despite the fact that there is an index-magnitude relation.  

Now consider what happens if the index does not correlate with 
magnitude (as is the case for Fe):  $\xi_{IM}=0$.  In this case, 
$\langle{\cal I}|{\cal M},{\cal V}\rangle =
 ({\cal V} - \xi_{MV}{\cal M})\,\xi_{IV}/(1-\xi_{MV}^2)$.  
Because this expression depends on absolute magnitude, 
measuring the mean index strength as a function of velocity dispersion 
using galaxies observed in a fixed redshift bin of a magnitude limited 
sample will lead to a biased estimate of the true correlation.

To illustrate this bias, a mock galaxy catalog was generated by 
distributing objects uniformly in a comoving volume with 
$p(M,V,I)$ a multivariate Gaussian.  The parameters of this 
Gaussian were chosen to match the distribution of luminosities and 
velocity dispersions in the catalog (from Bernardi et al. 2003b):  
$\sigma_{MM}=0.84$, $\sigma_{VV}=0.11$, $\xi_{MV}=-0.77$, 
$\sigma_{II}=0.05$ and $\xi_{IV} = 0.25$.  (The values of $\sigma_{II}$ 
and $\xi_{IV}$ were chosen to match the Fe-$\sigma$ relation.)  

The top set of panels in Figure~\ref{biasev} shows the index-magnitude 
and index-$\sigma$ relations if $\xi_{IM}=0$.  Dashed lines show the 
input relations, and symbols show the relations measured using objects 
which have apparent magnitudes between 14.5 and 17.45 but have 
redshifts between 0.05 and 0.07 (open circles) or 0.12 and 0.14 
(filled circles).  Different sets of symbols at each bin are for 
different realizations of mock catalogs which have approximately 
the same number of objects in each redshift bin as the data 
($\sim 1500$ for the redshift bins shown).  The apparent steepening 
of the index-$\sigma$ relation with redshift is {\em entirely} due 
to the magnitude limit. 
This demonstrates that direct measurement of the index-$\sigma$ 
using objects in an apparent magnitude limited catalog can lead to 
a biased estimate of the true relation, even if the index itself 
does not correlate with absolute magnitude.  

The bottom panel is similar, except that it has 
$\xi_{IM}=\xi_{IV}\xi_{VM}$.  
In this case, the index-luminosity and index-$\sigma$ relations 
measured in the magnitude limited sample are unbiased estimates of 
the true relation, even though index-strength correlates with 
magnitude---in agreement with the model described above.  

Thus, for the Fe-$\sigma$ relation in Figure~\ref{MgFesigma}, 
most of the difference between the low and high-redshift bins 
is a selection effect.  
The Mg-luminosity correlation is slightly weaker than 
$\xi_{IV}\xi_{VM}$, so it too is biased by selection effects, 
although by a smaller amount.  In all cases, the selection effect 
is more dramatic at low $\sigma$.  
Since we are interested in estimating how these indices evolve, 
we estimate evolution from the highest bin in $\sigma$ for which 
we have at least 1000 objects over $0.06\le z\le 0.17$:  
i.e., $2.35\le\sigma\le 2.4$.  
Since even this may be slightly affected by selection effects, 
it gives an upper limit to the true evolution.

\section{SDSS flux-calibration and Balmer line-strengths}\label{fluxcalib}

The main text notes that flux calibration problems around 4000\AA\ 
may affect measurements of Balmer line-strengths and D4000.  Evidence 
that this is a serious concern is presented in Figure~\ref{Hsigma}.  
Dotted lines show the correlation between 
 H$\delta_{\rm A}+$H$\gamma_{\rm A}$ and velocity dispersion measured 
by Kelson et al. (2001) using early-type galaxies in four clusters at 
the redshifts indicated.  They find that the zero point of the relation 
evolves, but are not able to conclude if the slope does or not.  

The shaded regions show the correlation between 
 H$\delta_{\rm A}$+H$\gamma_{\rm A}$ and velocity dispersion in our 
sample in the same redshift bins used in the main text 
($0.05<z<0.07$, $0.07<z<0.09$, $0.09<z<0.12$, $0.12<z<0.14$ and 
 $0.14<z<0.20$).  
Notice that our SDSS data produce the same slope as the data from 
the literature, but the zero points are quite different at low redshift.  

What is most relevant to the analysis in the main text is the rate at 
which the line strengths evolve.  
Since flux-calibration problems affect observed wavelengths, they 
appear as redshift dependent effects when studying features at fixed 
restframe wavelength.  Figure~\ref{Hsigma} demostrates that the 
apparent evolution in the SDSS sample is about three times larger than 
that seen by Kelson et al. (2001).  Since we already have reason to 
believe that flux-calibration is difficult around 
$\lambda_{obs}=4000$\AA, so it is possible that the low redshift bins 
are more strongly affected than the bins at higher redshift, 
this strongly suggests that flux calibration systematics are affecting 
the SDSS measurement.  Therefore, in the main text, we show the apparent 
evolution using the raw measurements as well as the result of dividing 
the apparent evolution by a factor of three.

Figure~\ref{HbHgMgFesigma} shows our measurements of the 
correlations between H$\beta$, H$\gamma_{\rm F}$, Mg$b$, and 
$\langle{\rm Fe}\rangle$ with velocity dispersion.  
These are the indices we use in our study of single burst stellar 
population synthesis models (Section~\ref{sspmodels}).  Our 
measurements are in reasonable agreement with previous work, 
with the exception of Mg$b$.  Our Mg$b$ values are offset to smaller 
values than those of Bender et al. (1996); this offset is not 
due to the fact that Bender et al. aperture correct both Mg$b$ 
and $\sigma$ to a fixed physical scale, whereas we correct to 
$r_e/8$.  Shifting our values upwards so they agree with the 
literature makes little qualitative difference to our conclusions 
about the ages, metallicities and $\alpha$-enhancements in our 
sample.  

As another check, Figure~\ref{Mg2sigma} shows that the Mg$_2$-$\sigma$ 
correlation and its evolution are consistent with previous work 
(Bernardi et al. 1998).  Using Mg$_2$ in place of Mg$b$, and making 
the corresponding change in the SSP models, makes no difference to 
our findings.  


{}

\clearpage


\begin{figure}
 \centering
 \plotone{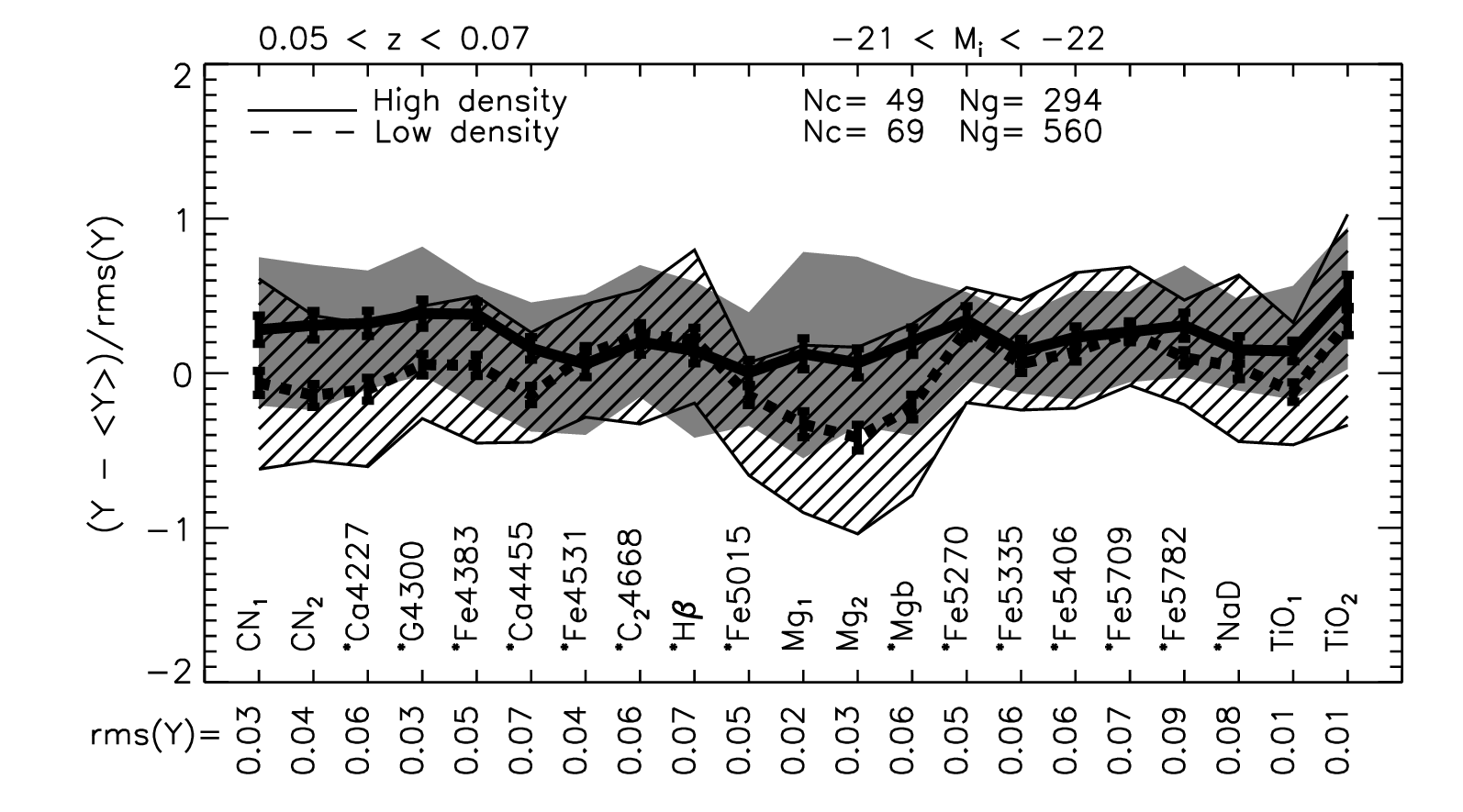}
 \plotone{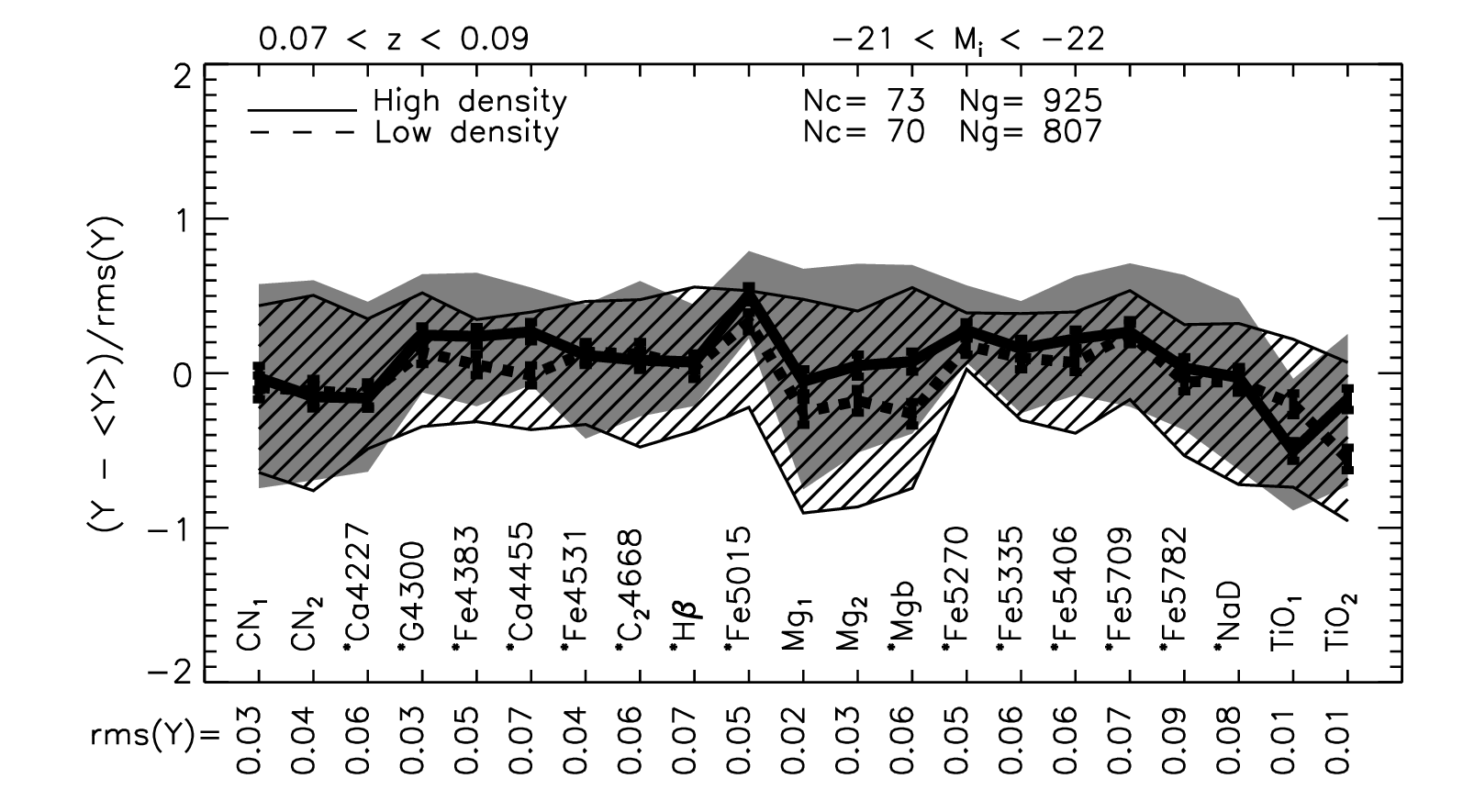}
 \caption{Lick indices for objects with the same narrow range of 
          redshifts and luminosities as in the previous figure. 
          (An asterisk signifies that the quoted rms is for $\log_{10}$ of 
          the index.)
          Recall that the mean luminosity and velocity dispersions 
          are the same in 
          dense and less dense environments, so trends with environment 
          are not caused by correlations with $\sigma$.
          Bottom panel contains a supercluster, 
          so environmental effects may not be accurate.}
 \label{lickf}
\end{figure}

\setcounter{figure}{6}

\begin{figure}
 \centering
 \plotone{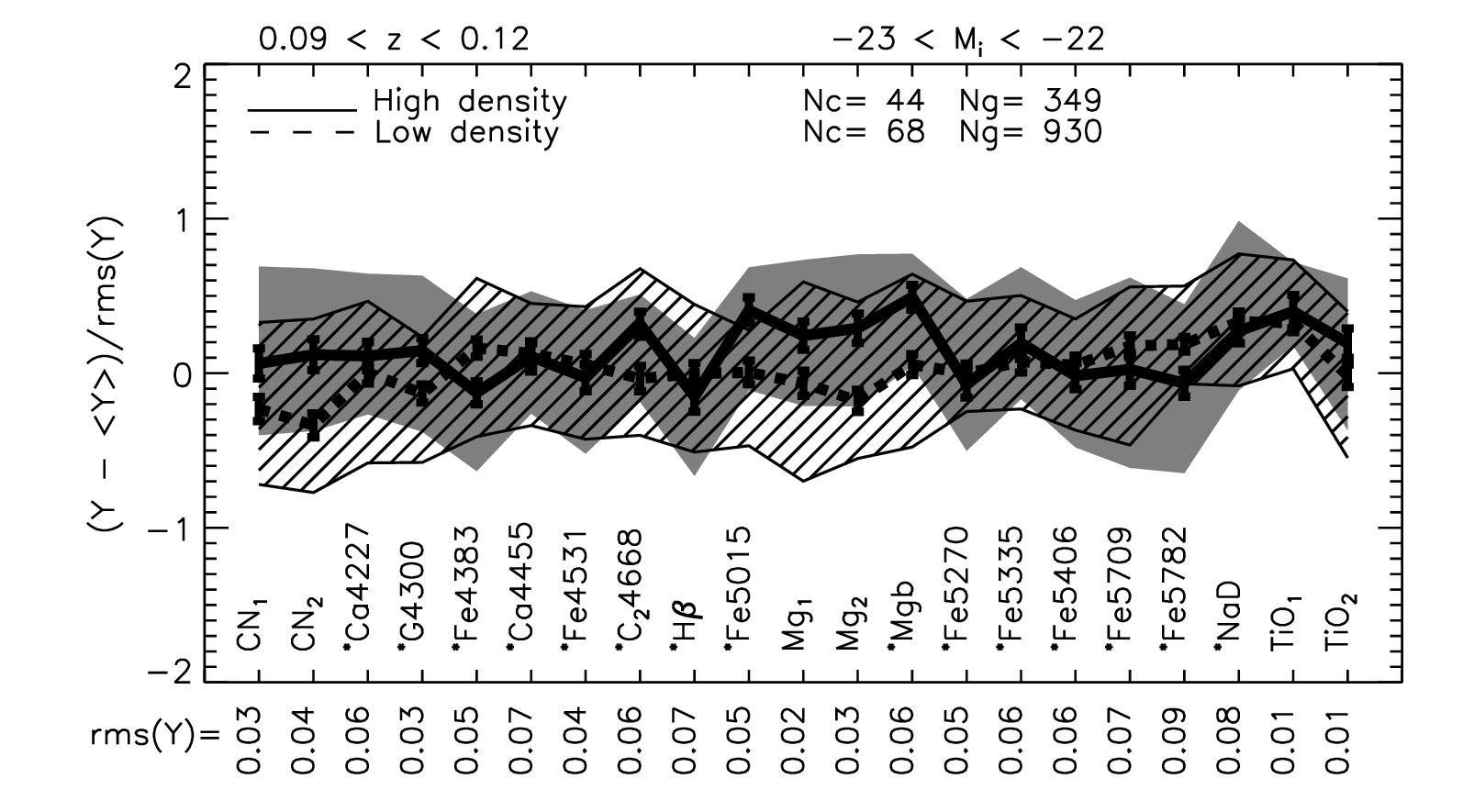}
 \plotone{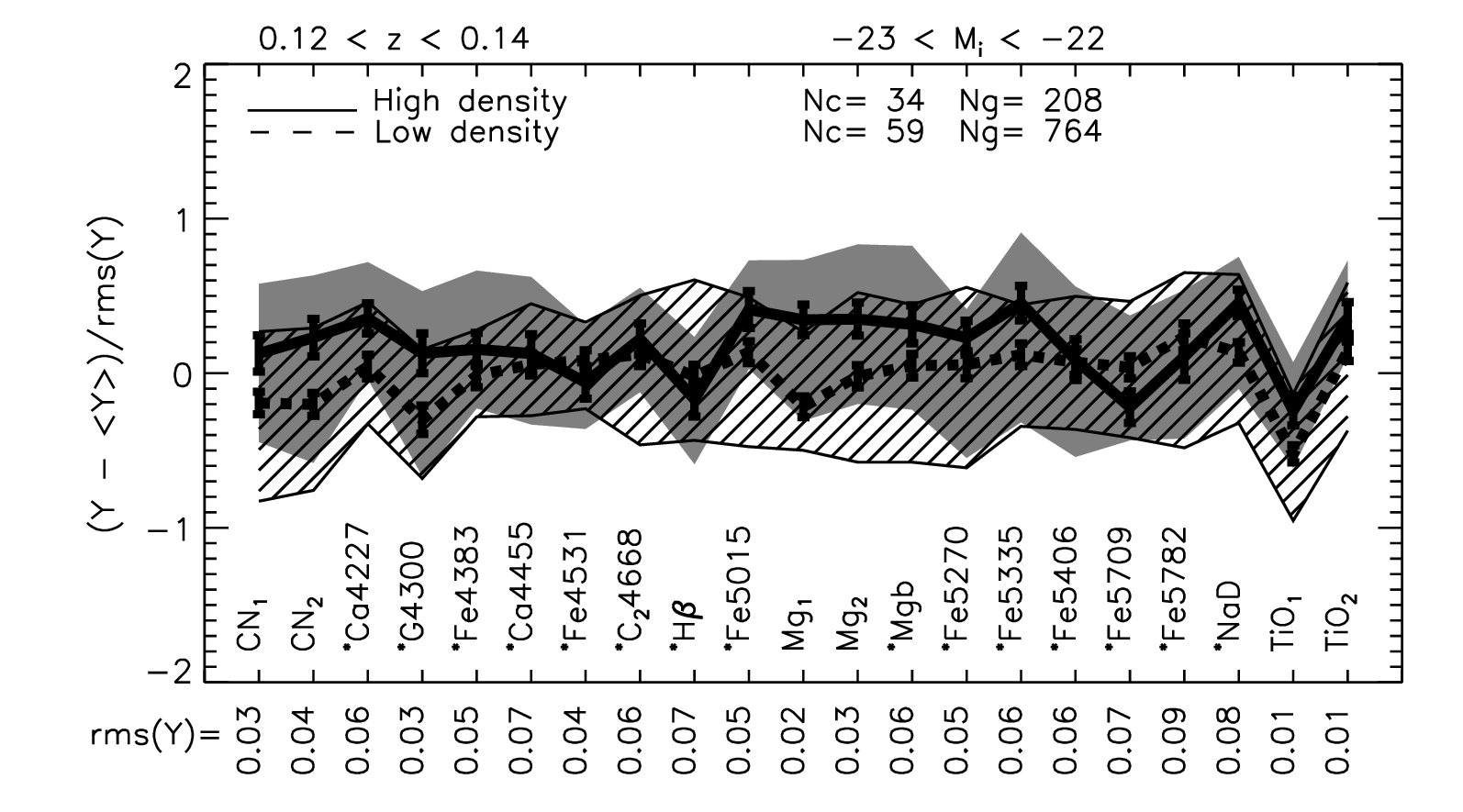}
 \caption{Continued.}
 \label{lickb}
\end{figure}

\begin{figure}
 \centering
 \plottwo{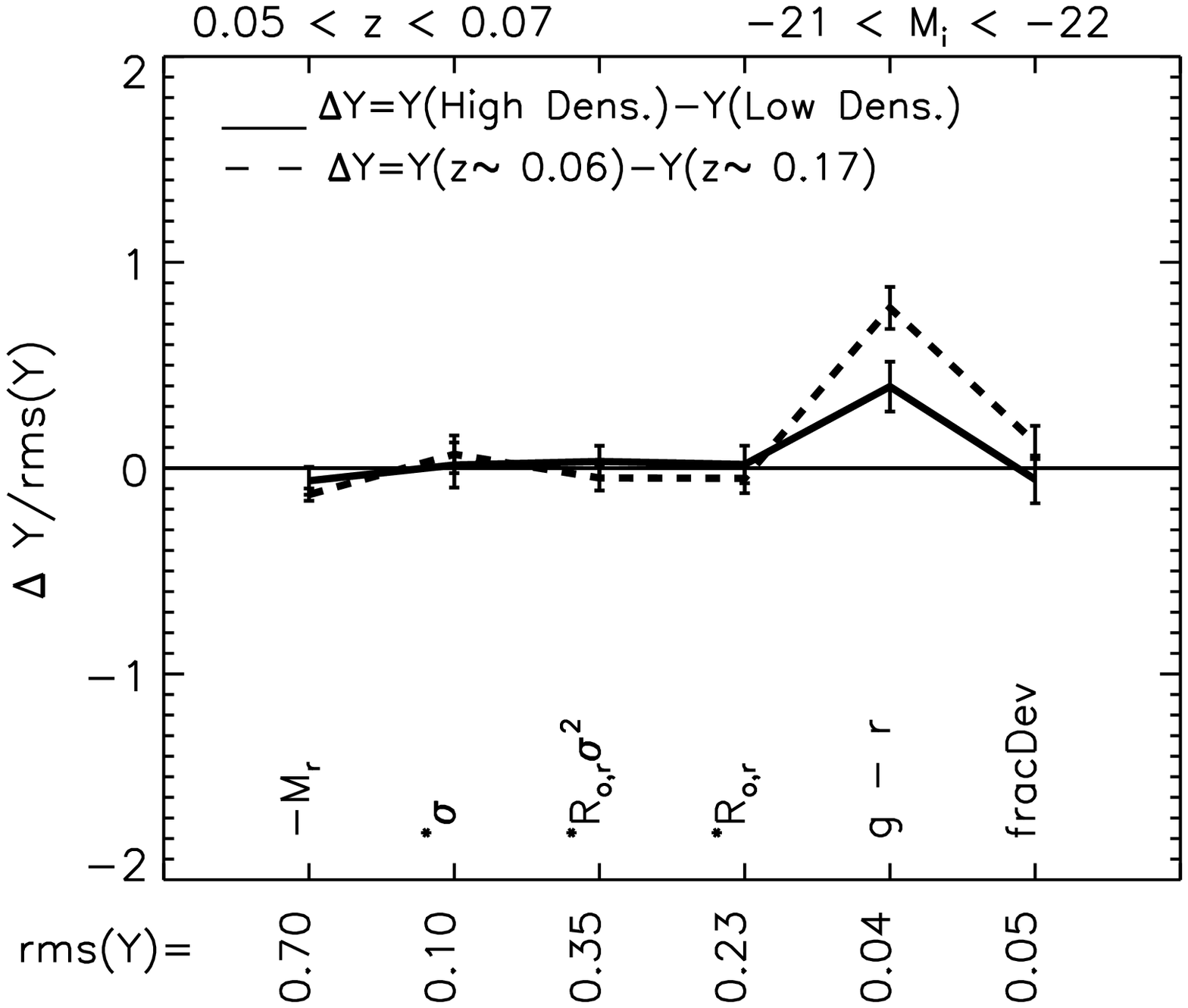}{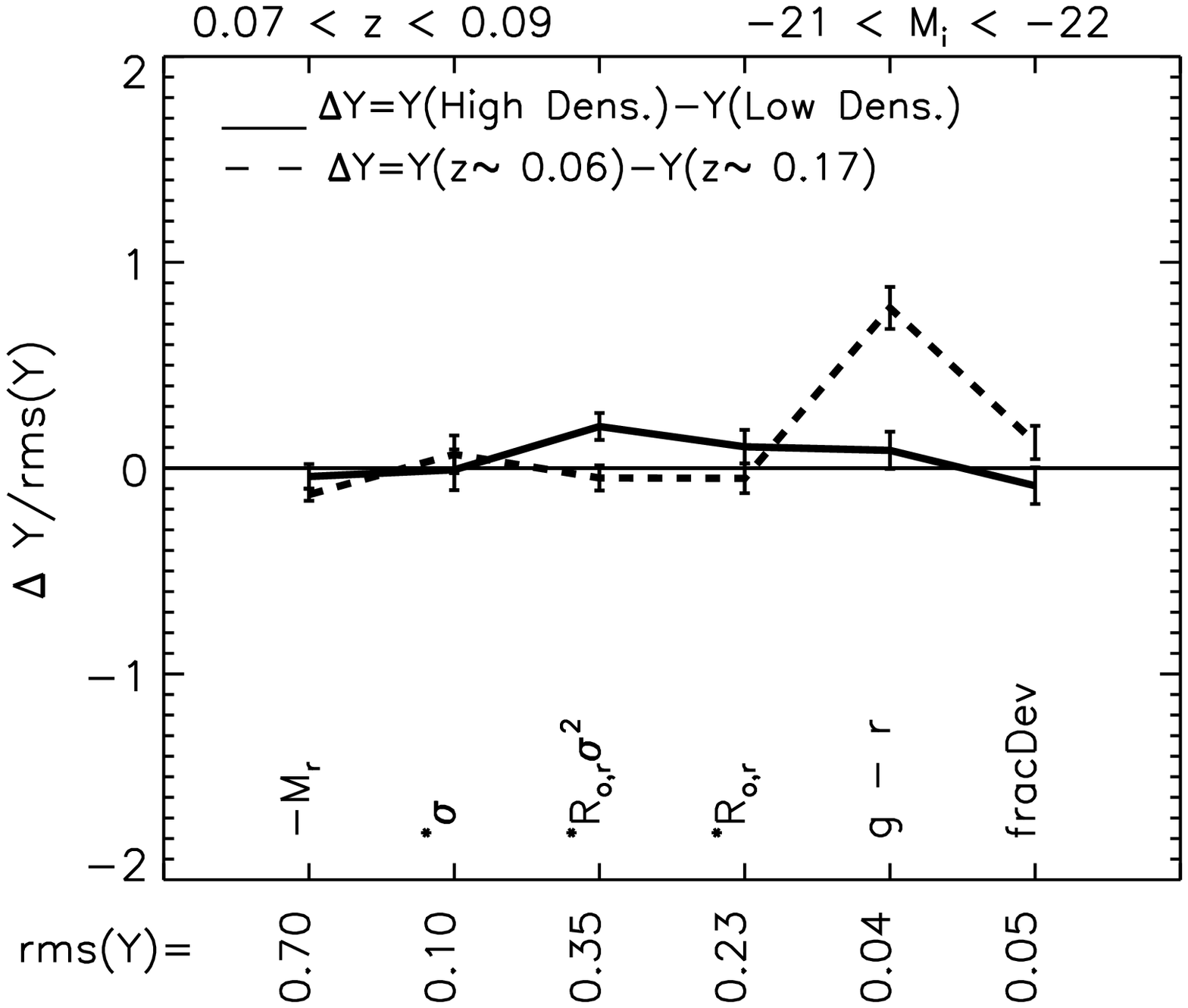}
 \plottwo{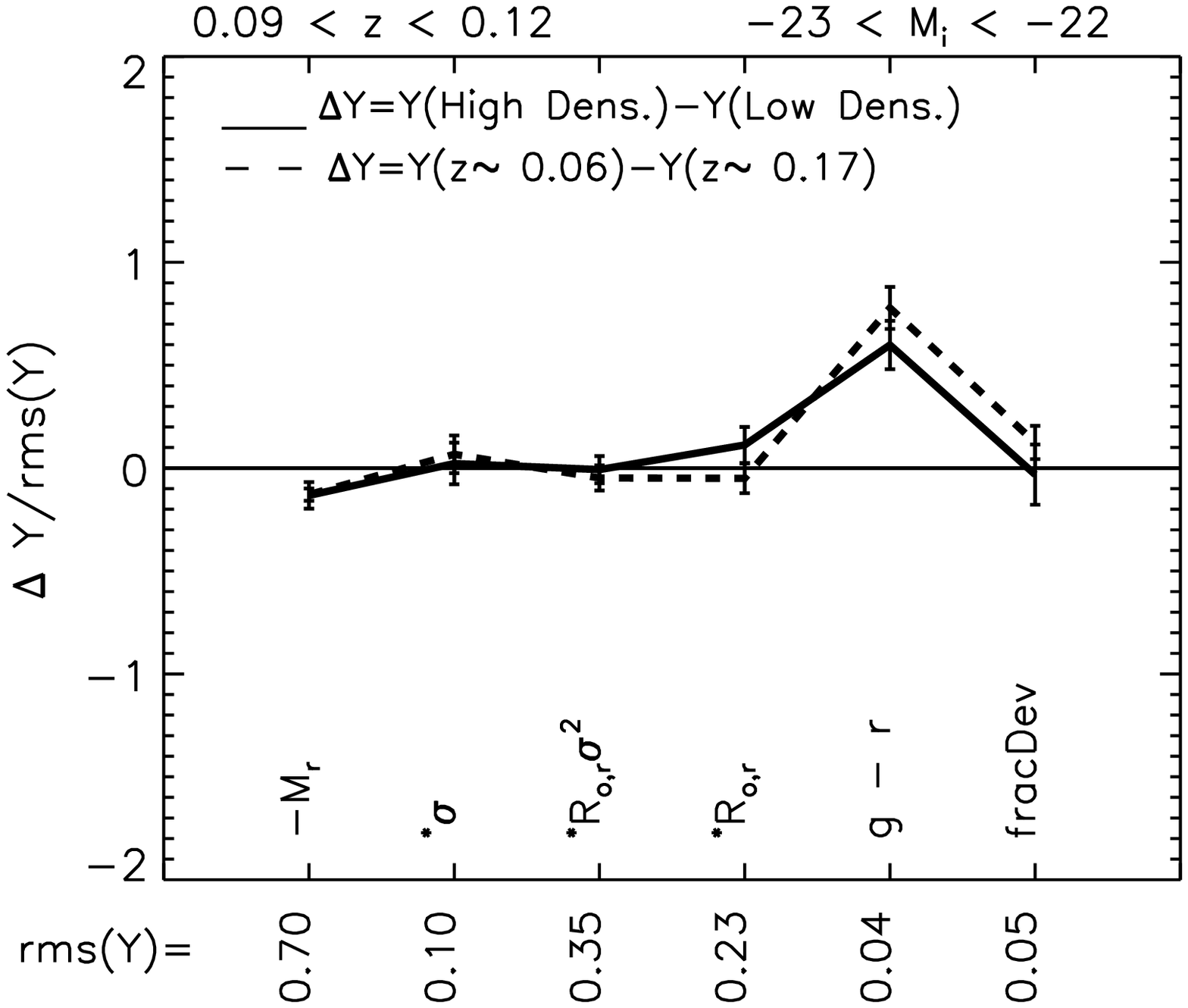}{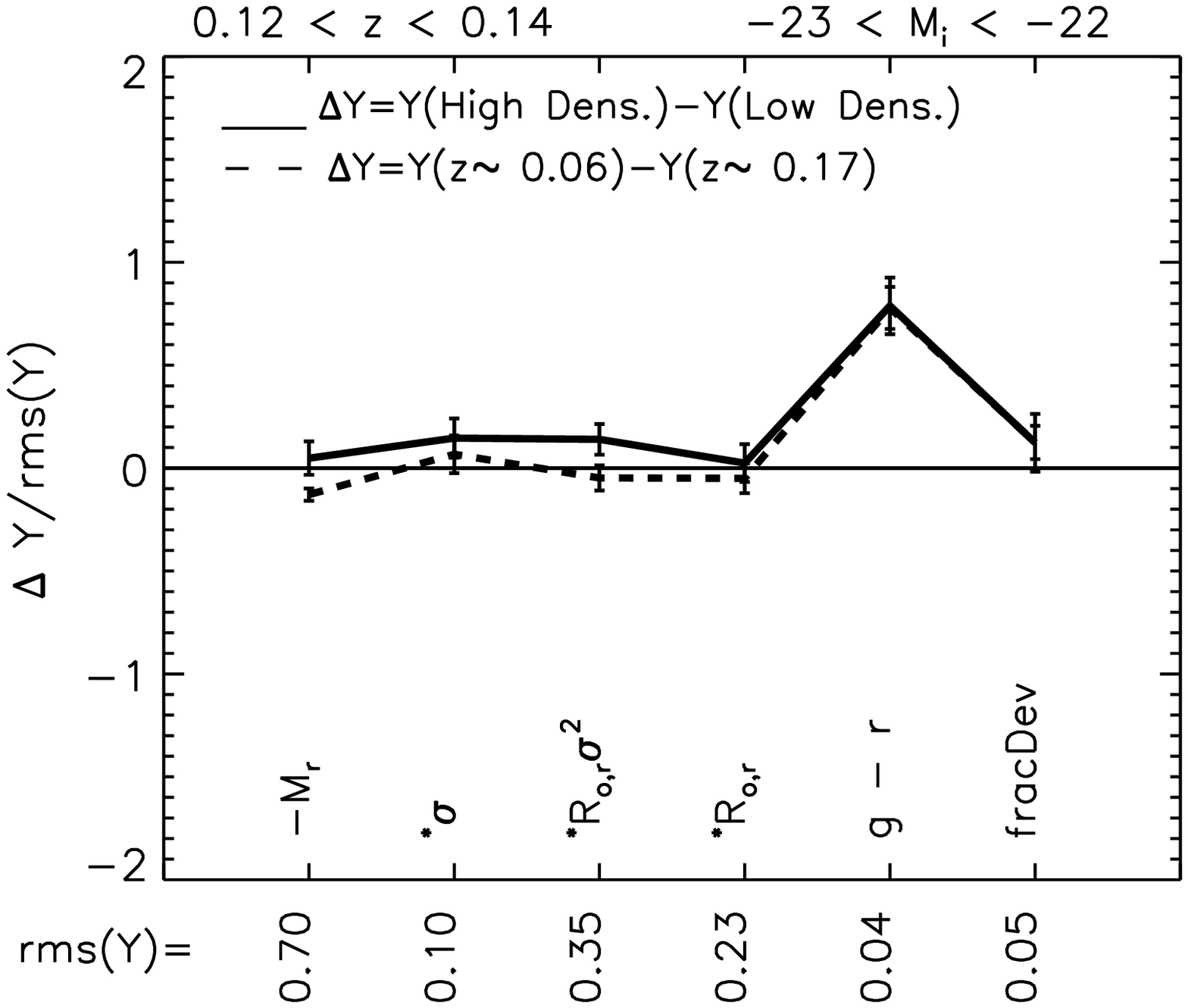}
 \caption{Evolution of observed parameters estimated using early-type
          galaxies in the entire sample which have 
          $2.35\le\log_{10}\sigma\le 2.4$ (dashed, same in all panels), 
          compared to dependence on environment (solid), for galaxies 
          which have the range of redshift and luminosity shown
          in the top left and top right of each panel, respectively.
          The time difference between $z\sim 0.17$ and $z\sim 0.06$ is 
          1.3~Gyr.  The top-right panel is different from the others; 
          it happens to contain a supercluster, and this may have 
          compromised our estimates of environment or affected the color. 
          (An asterisk signifies that the quoted rms is for $\log_{10}$ of 
          the index.)}
 \label{paramfbev}
\end{figure}

\begin{figure}
 \centering
 \plottwo{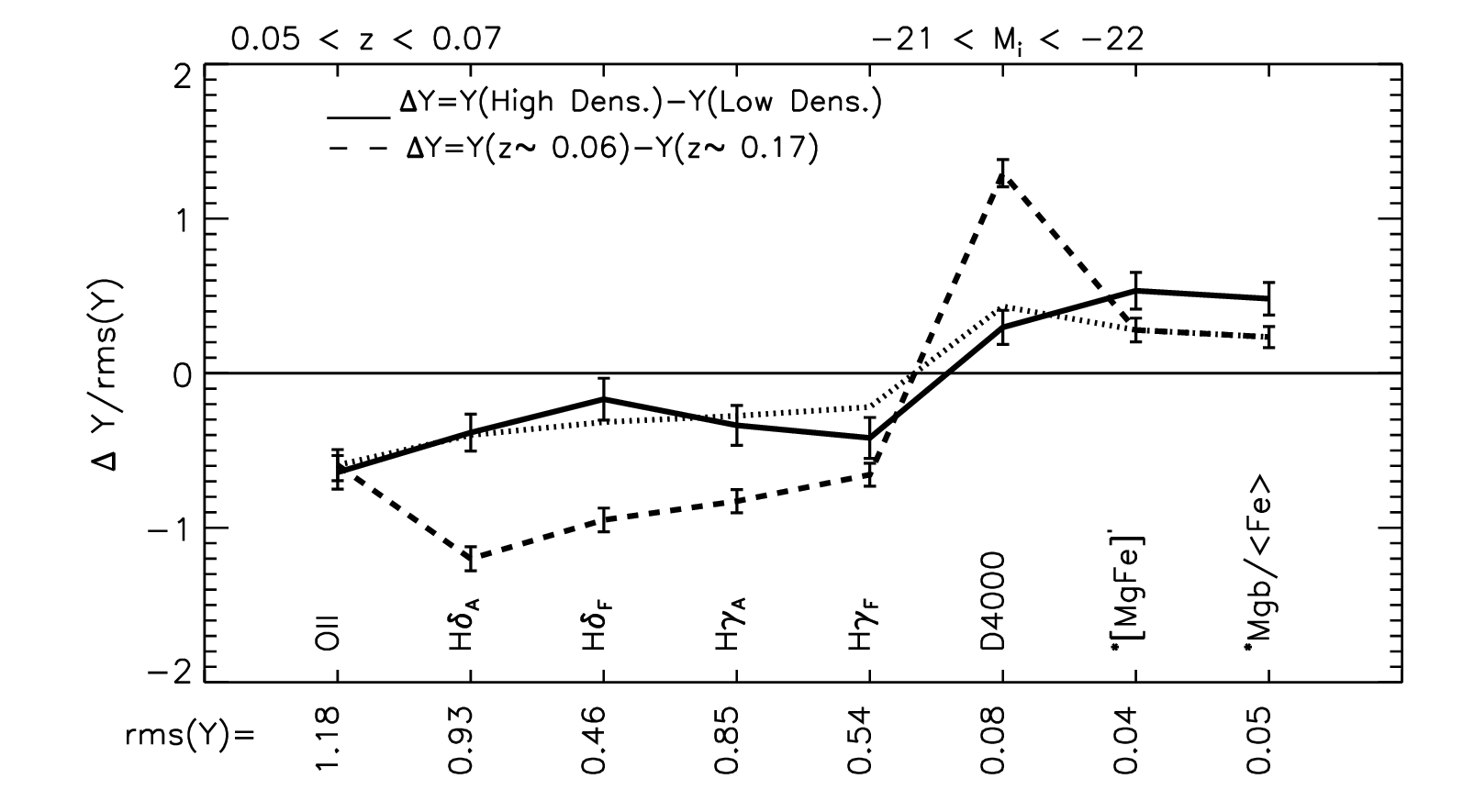}{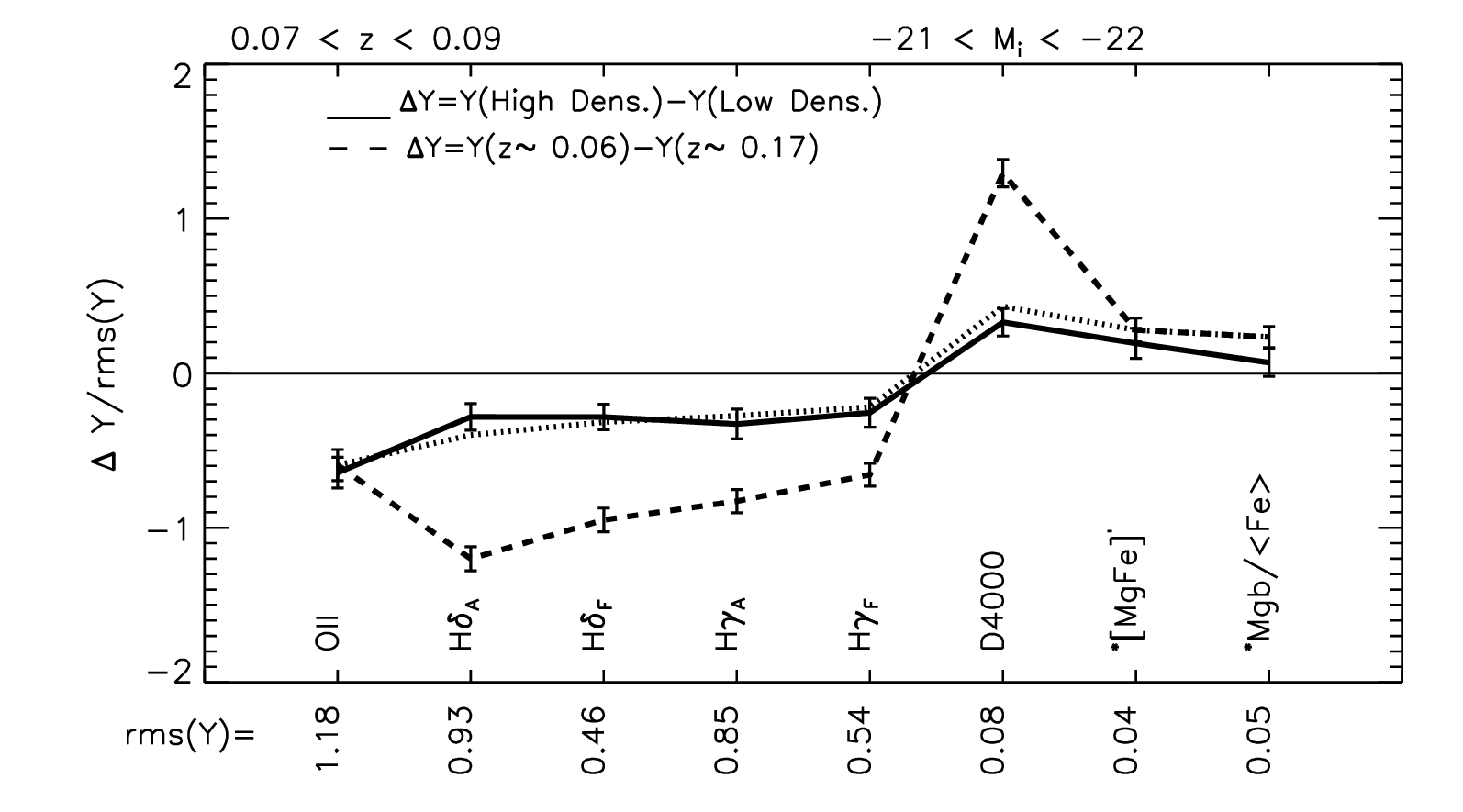}
 \plottwo{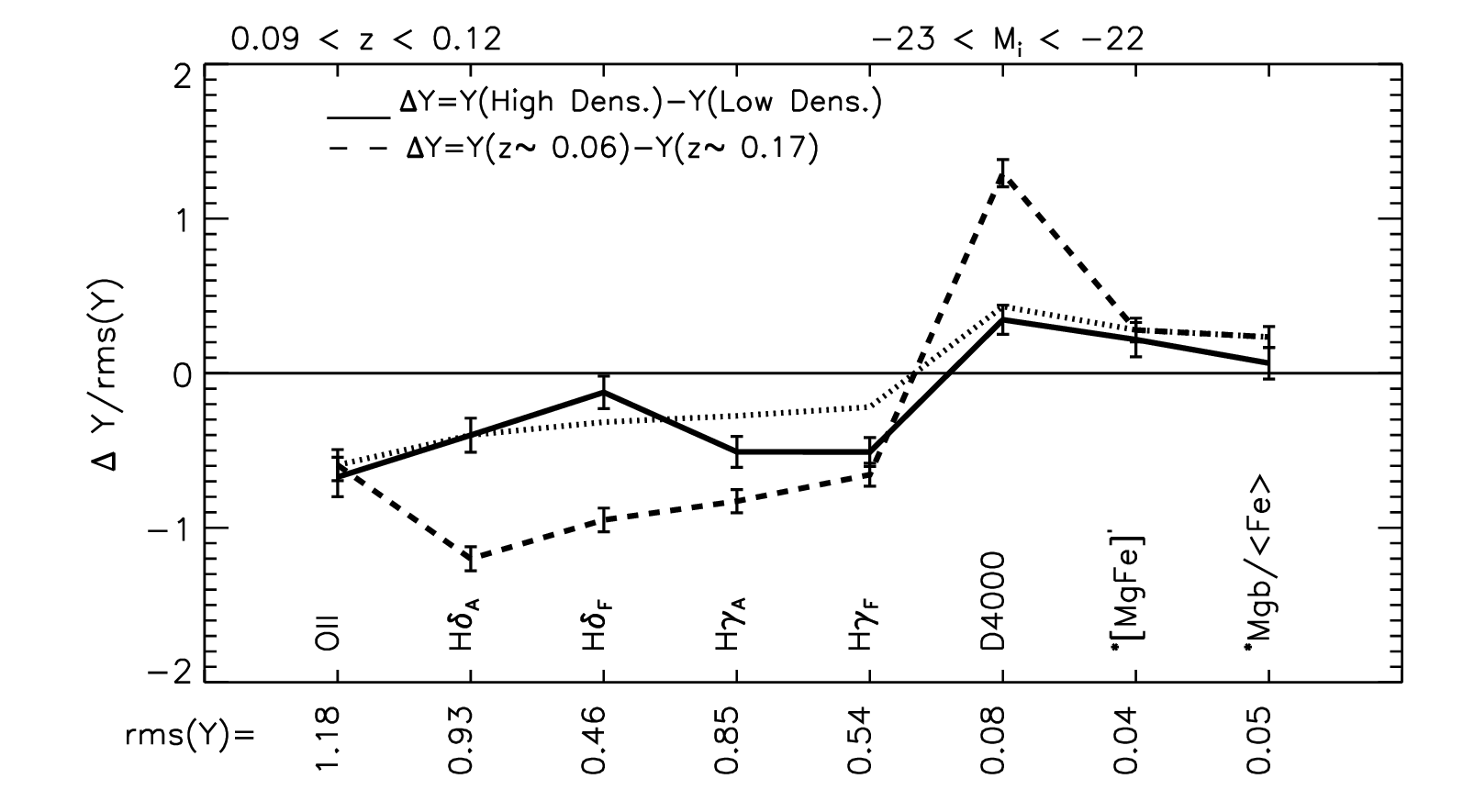}{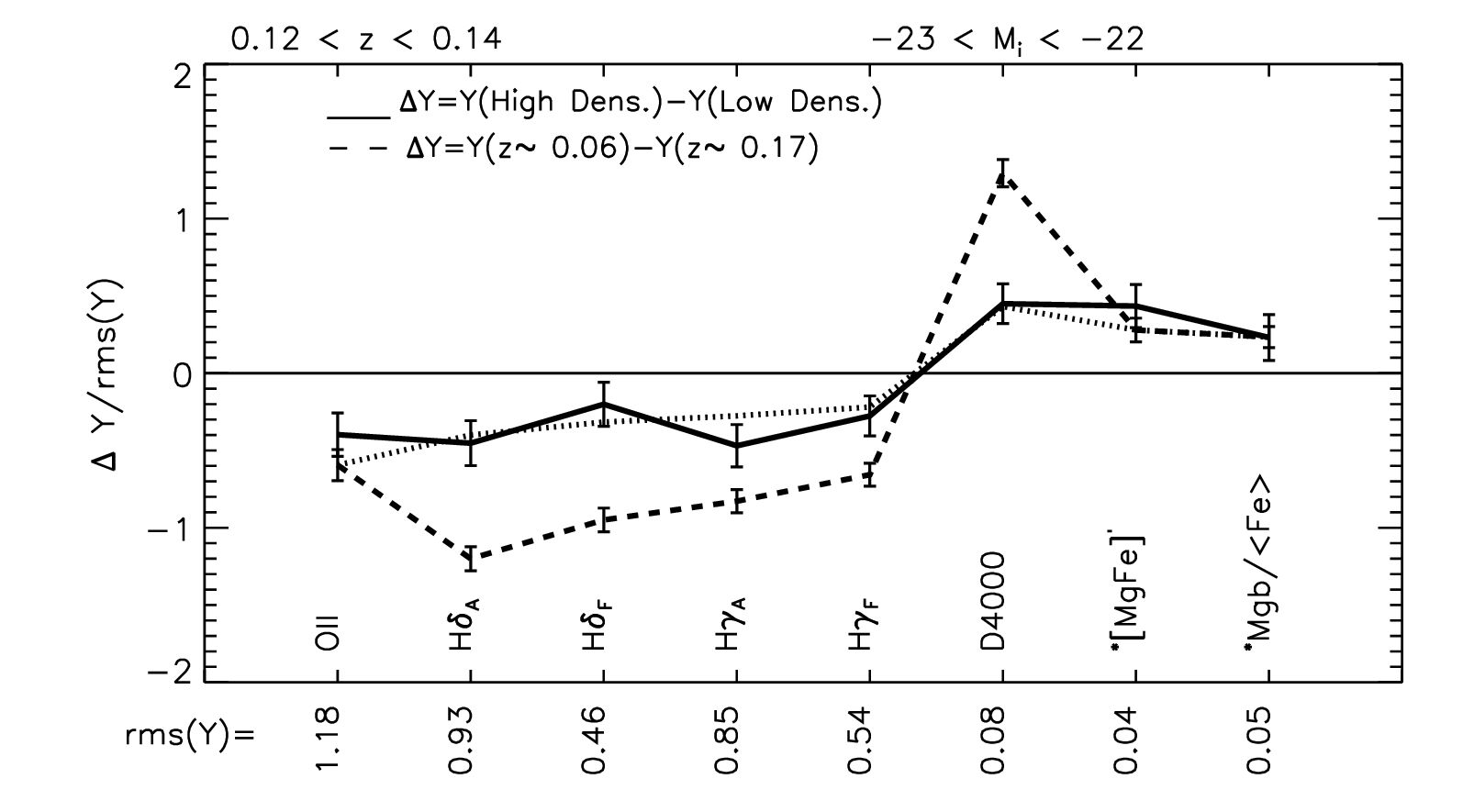}
 \caption{Evolution of observed parameter (dashed/dotted, same in all panels) 
          compared to dependence on environment (solid).  Dotted lines 
          correct for flux-calibration problems by dividing measured 
          values by a factor of three (c.f. Appendix~\ref{fluxcalib}).  
          (An asterisk signifies that the quoted rms is for $\log_{10}$ 
          of the index.)  In most cases, the difference between 
          $z\sim 0.17$ and $z\sim 0.06$ is similar to the difference 
          between cluster and less dense environments.}
 \label{linesfbev}
\end{figure}

\begin{figure}
 \centering
 \plotone{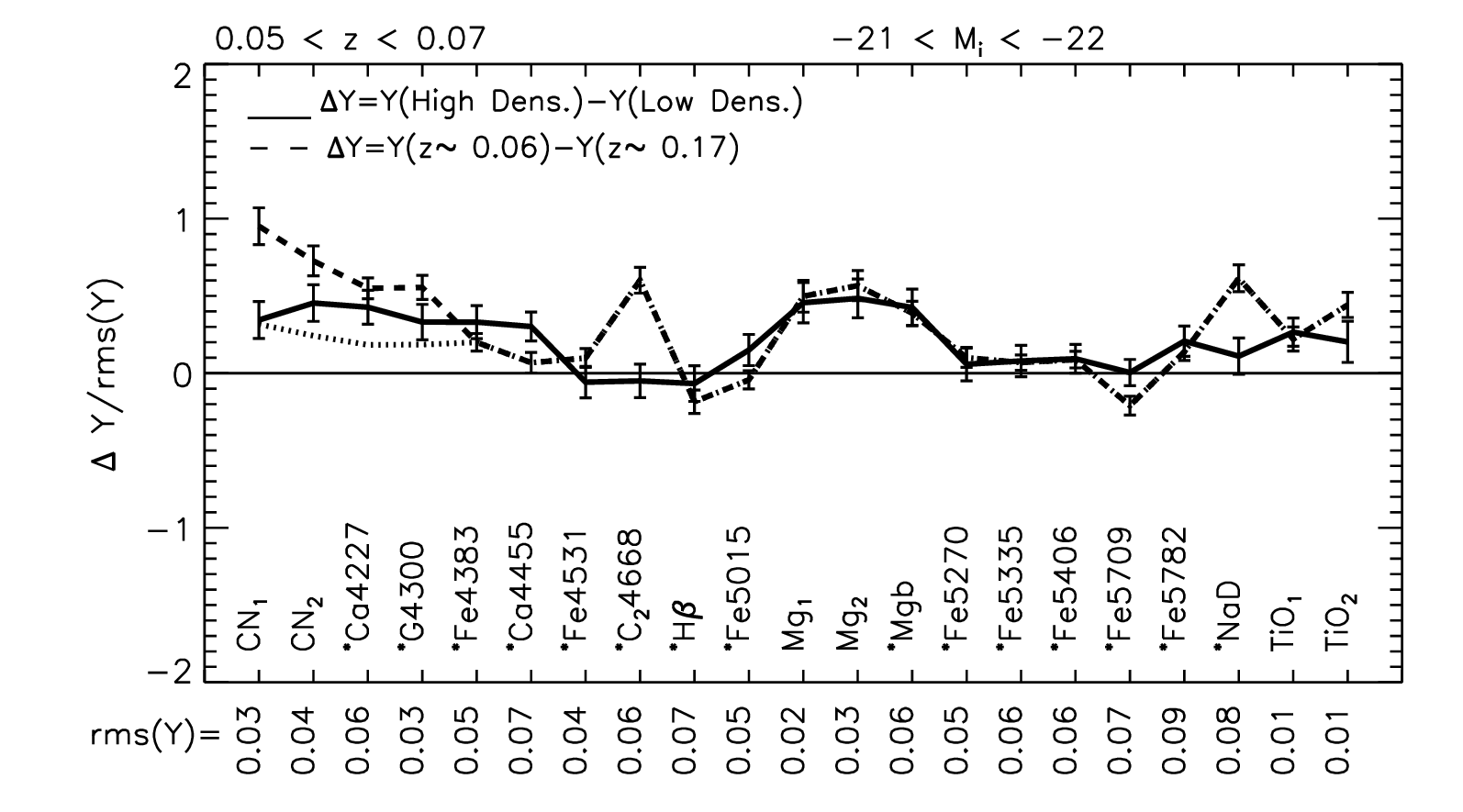}
 \plotone{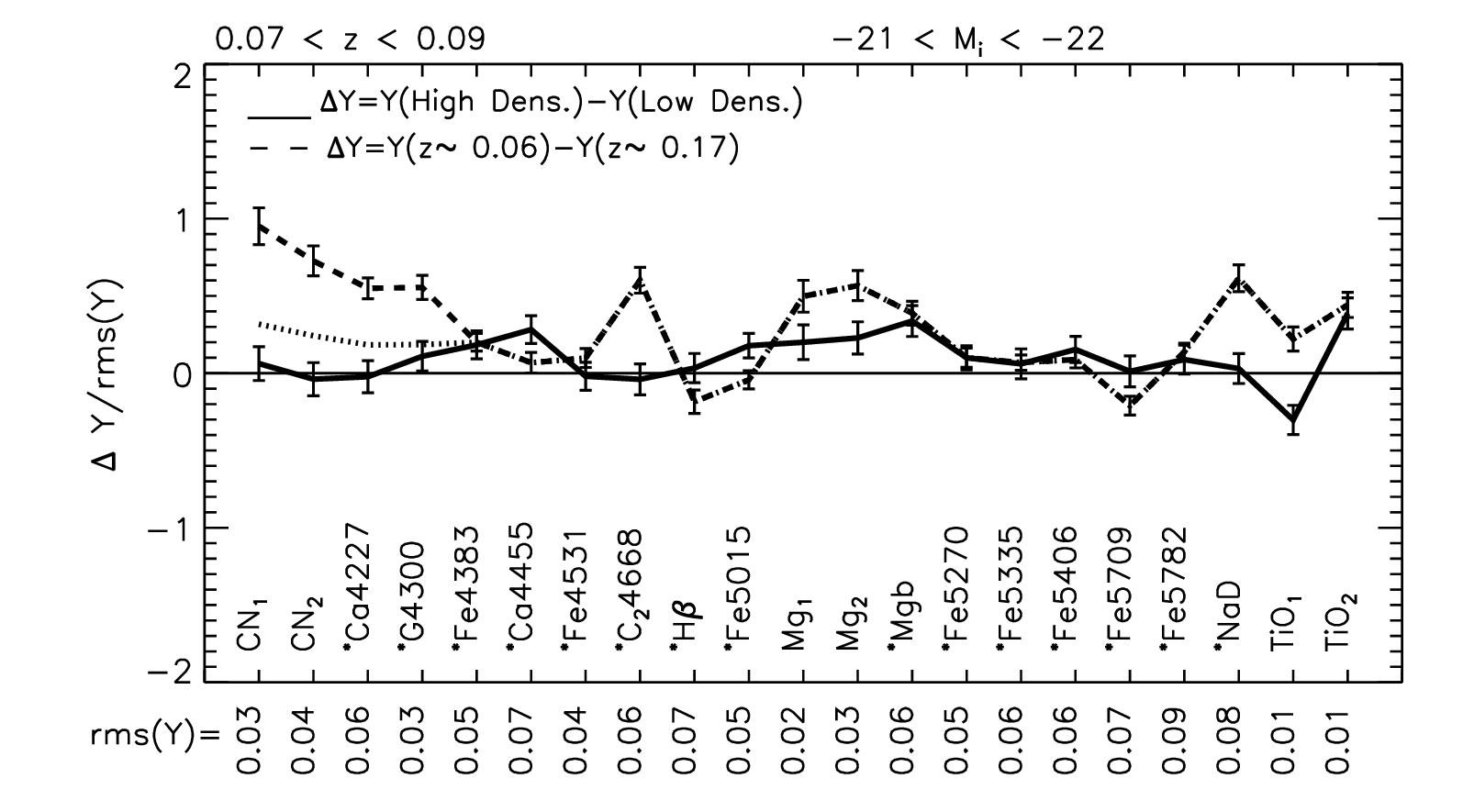}
 \caption{As for the previous figure, but now comparing the evolution 
          of the Lick indices with the dependence on environment. 
          Bottom panel contains a supercluster, 
          so environmental effects may not be accurate.
          (An asterisk signifies that the quoted rms is for $\log_{10}$ of 
          the index.)}
 \label{lickfev}
\end{figure}

\setcounter{figure}{9}

\begin{figure}
 \centering
 \plotone{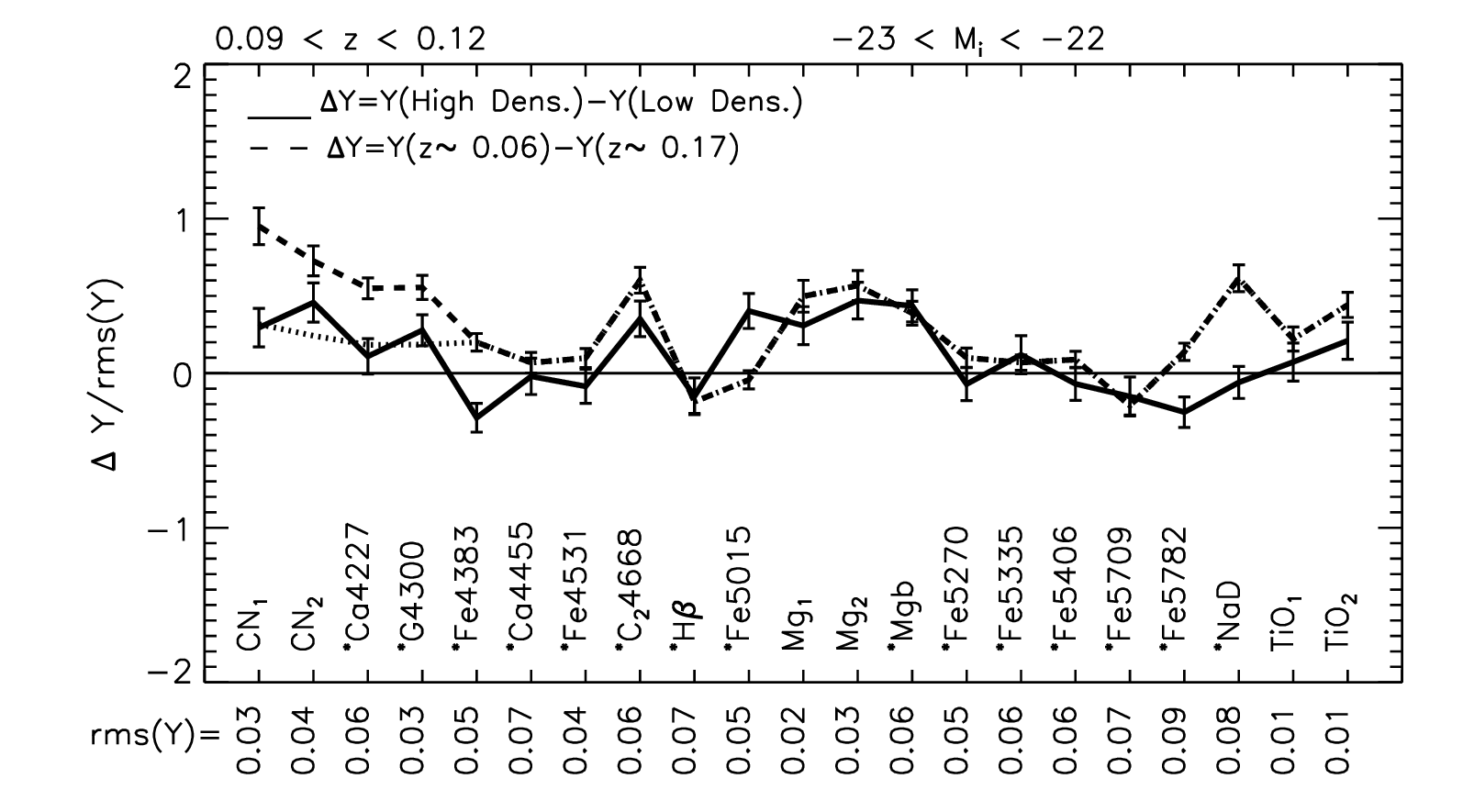}
 \plotone{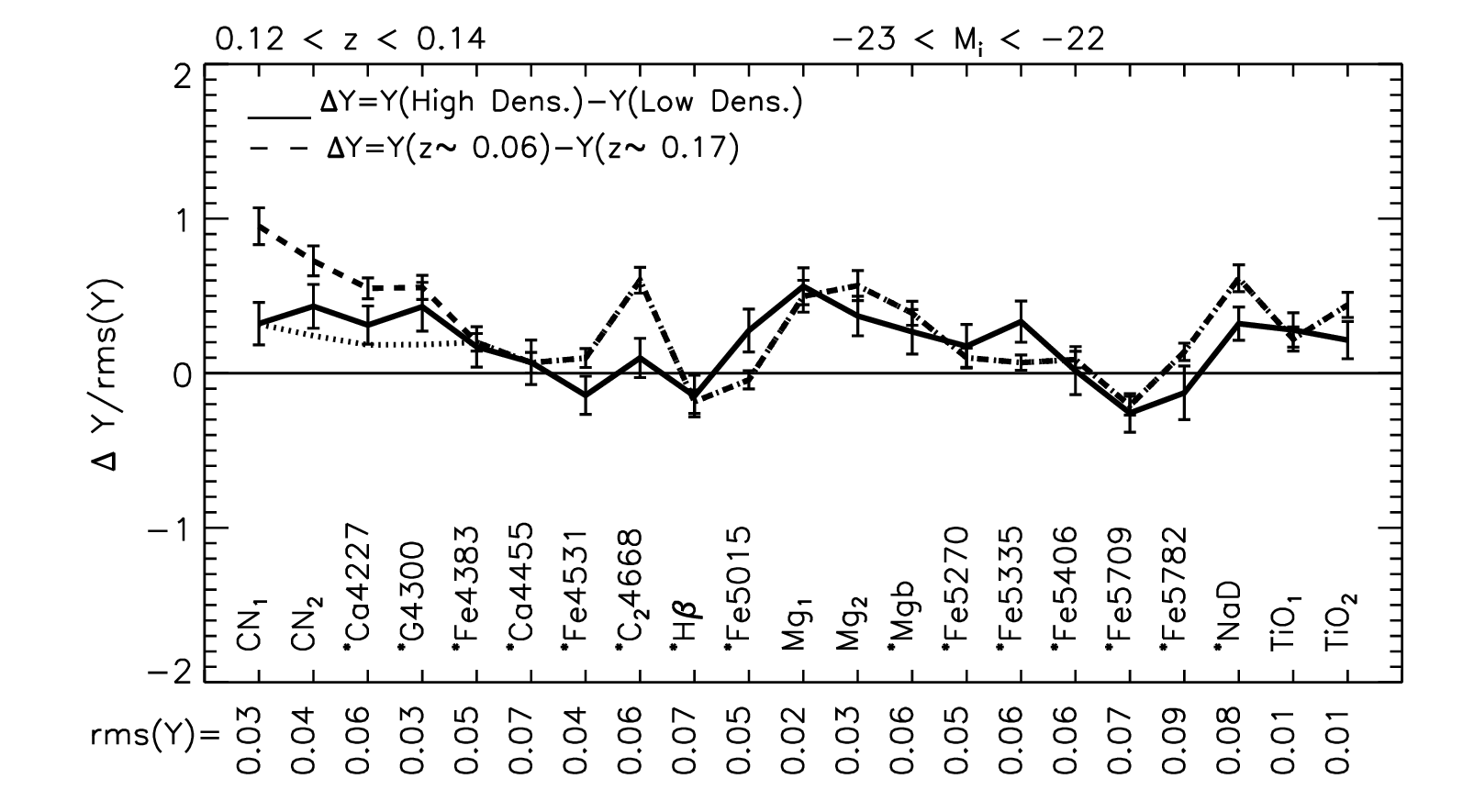}
 \caption{Continued.}
 \label{lickbev}
\end{figure}

\begin{figure}
 \centering
 \plottwo{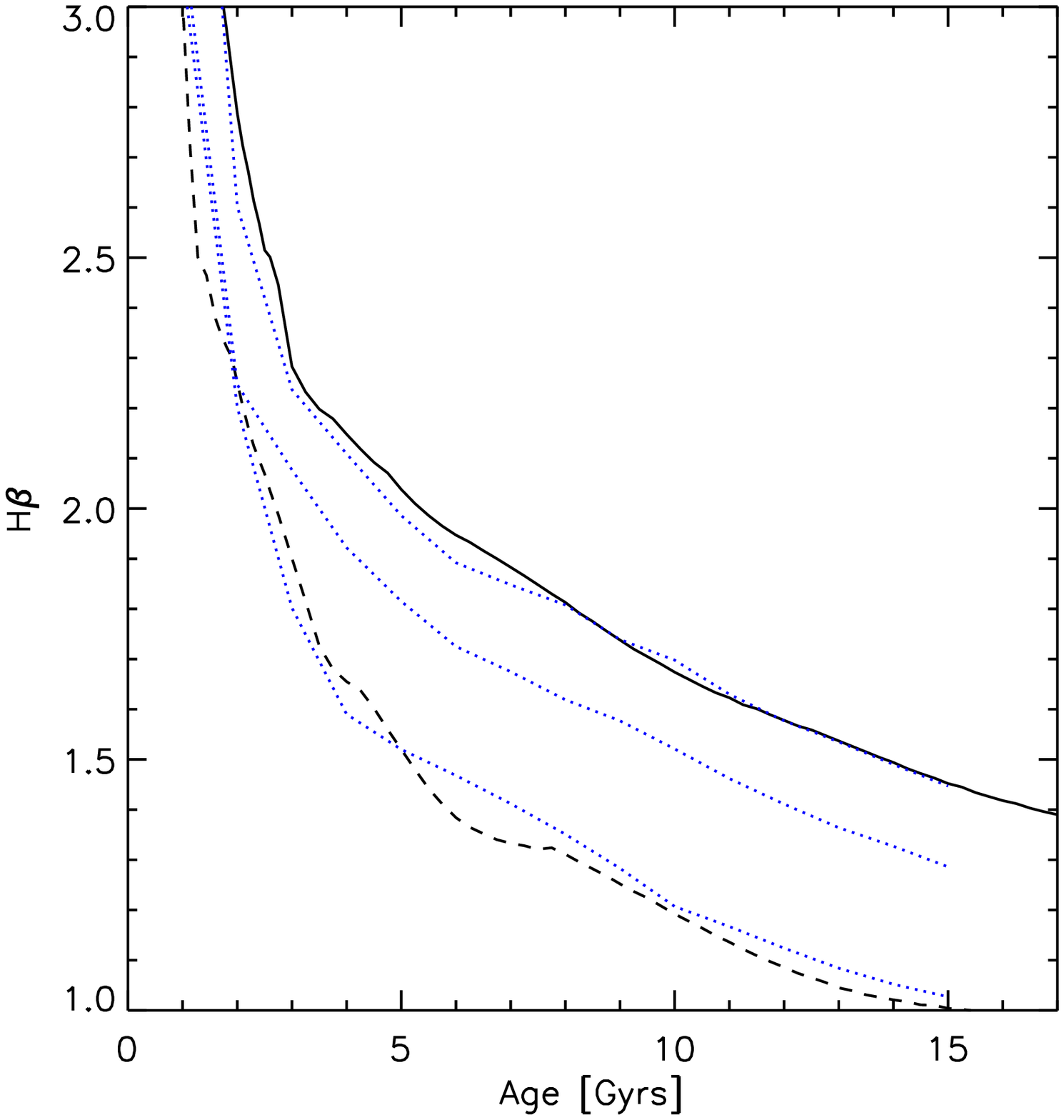}{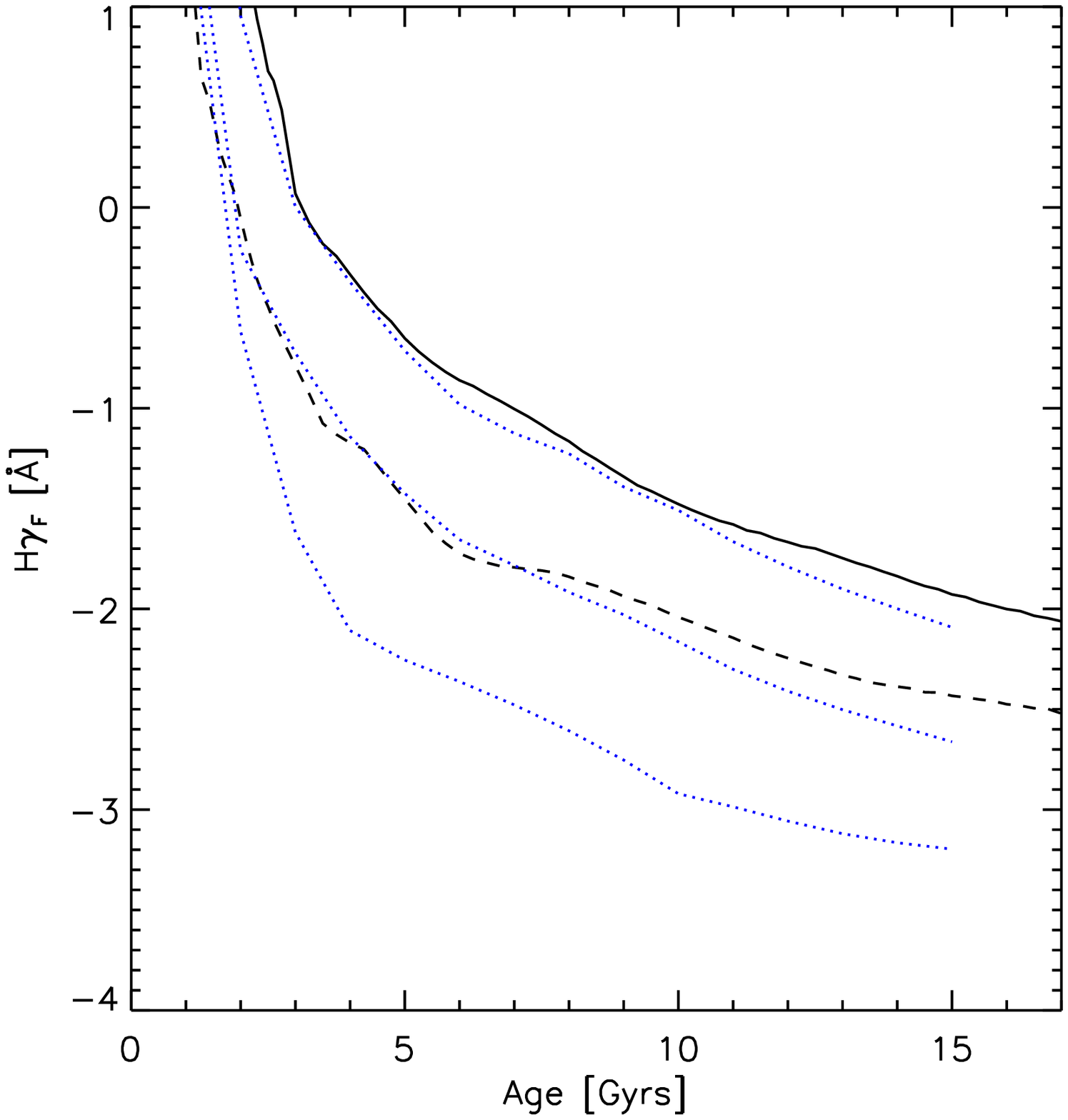}
 \plottwo{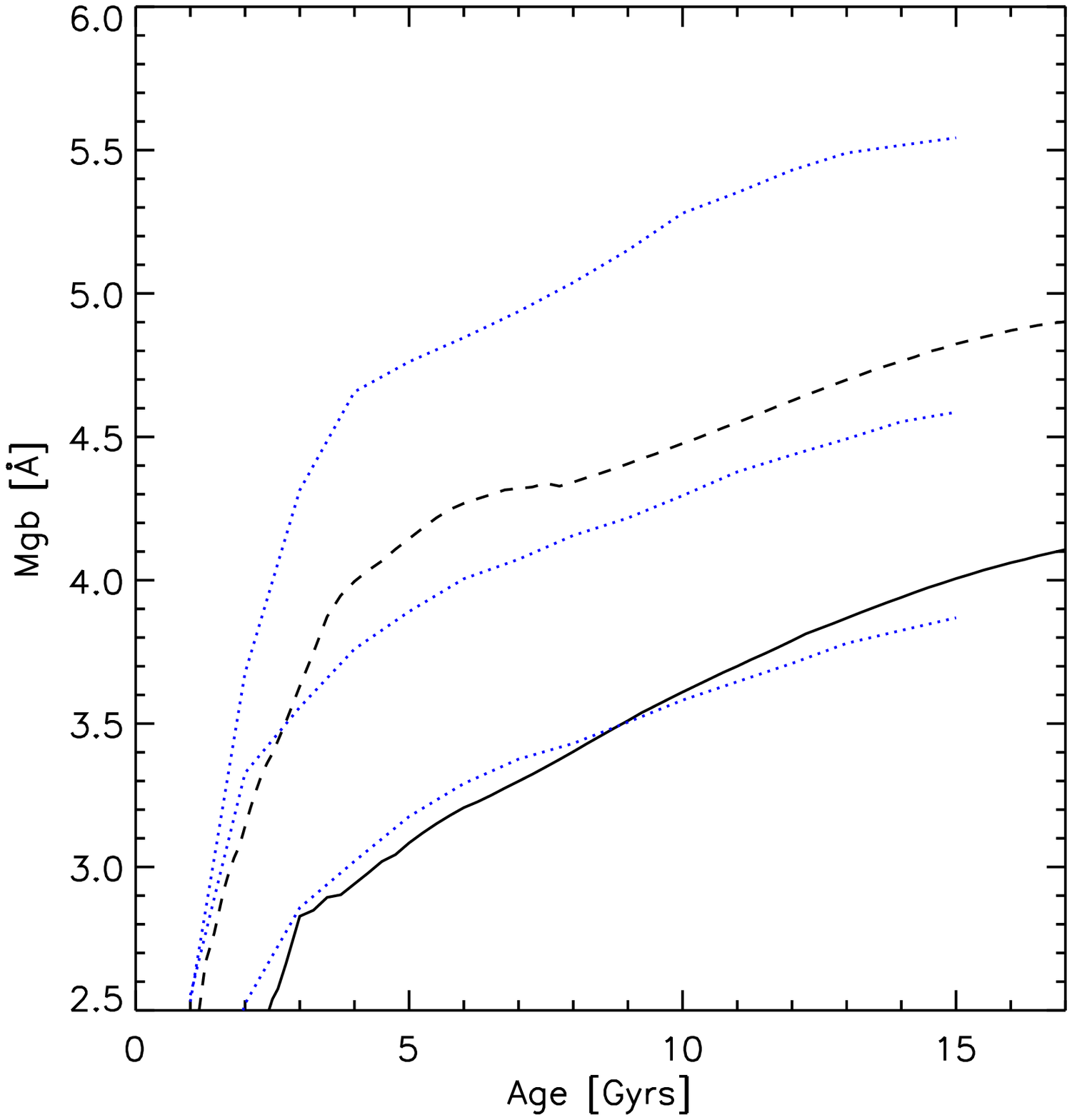}{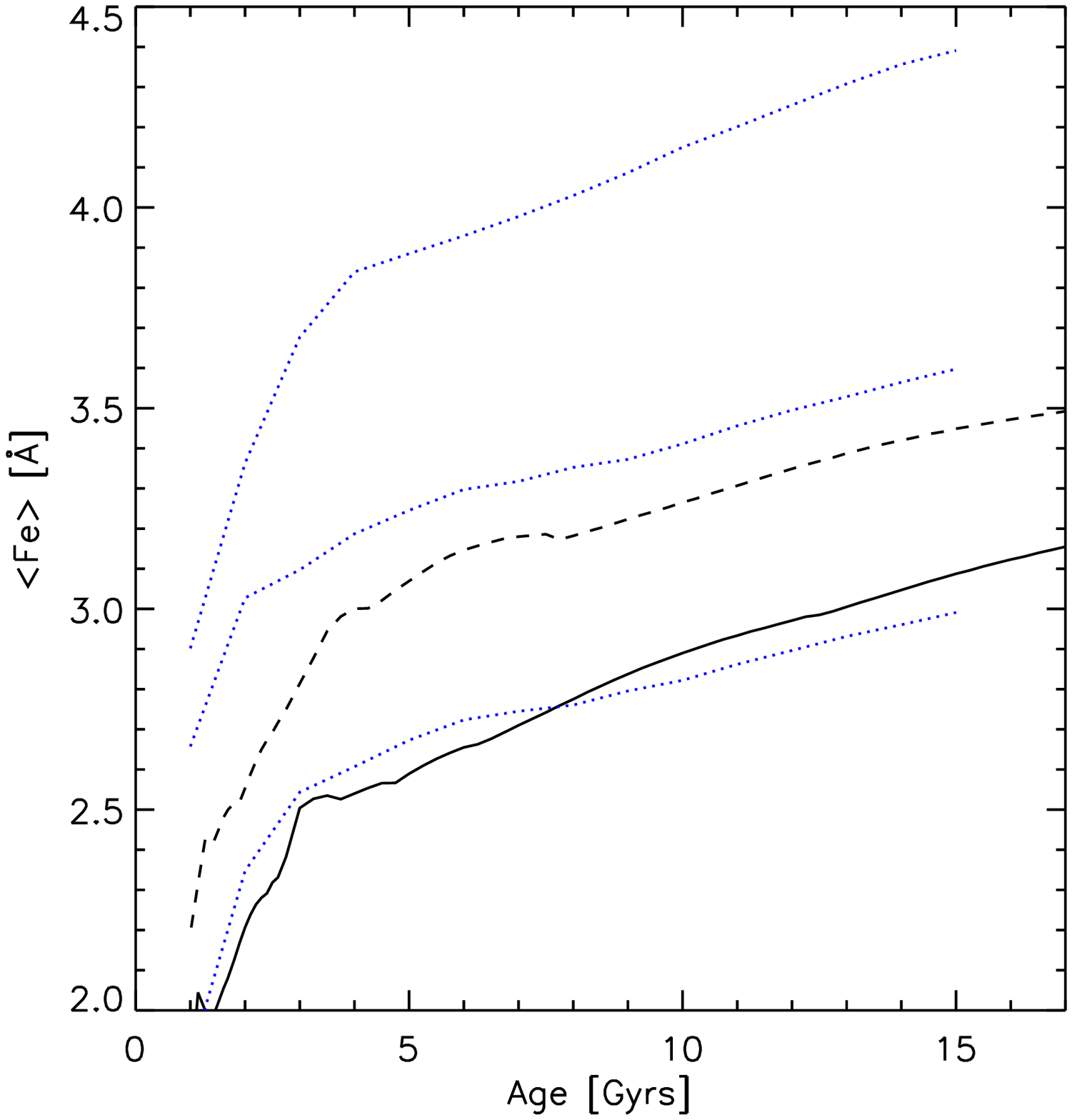}
 \caption{Evolution of various absorption line strengths in 
 single burst stellar population synthesis models at solar 
 $\alpha$-enhancement.  Solid and dashed lines show the models of 
 Bruzual \& Charlot (2003) at solar metallicity and higher.  
 Dotted lines show the models of TMB03-TMK04
 at solar metallicity and higher.  The models agree at solar metallicity 
 (bottom curves in Mg$b$ and $\langle$Fe$\rangle$, top curves in 
 H$\beta$ and H$\gamma_{\rm F}$), 
 but, in general, do not agree at the higher metallicities which 
 are expected to be characteristic of the early-type population. }
 \label{models}
\end{figure}

\begin{figure}
 \centering
 \plottwo{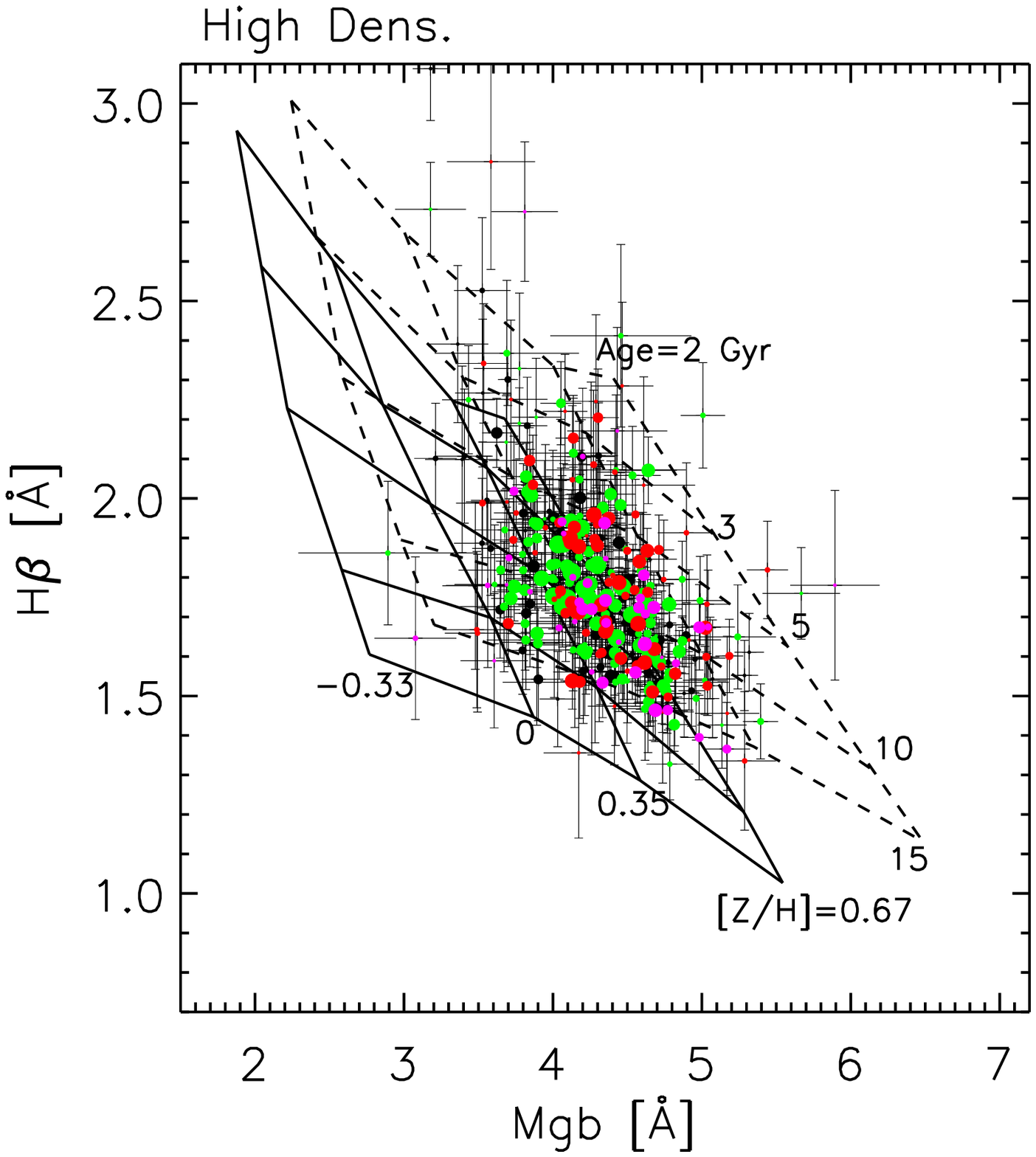}{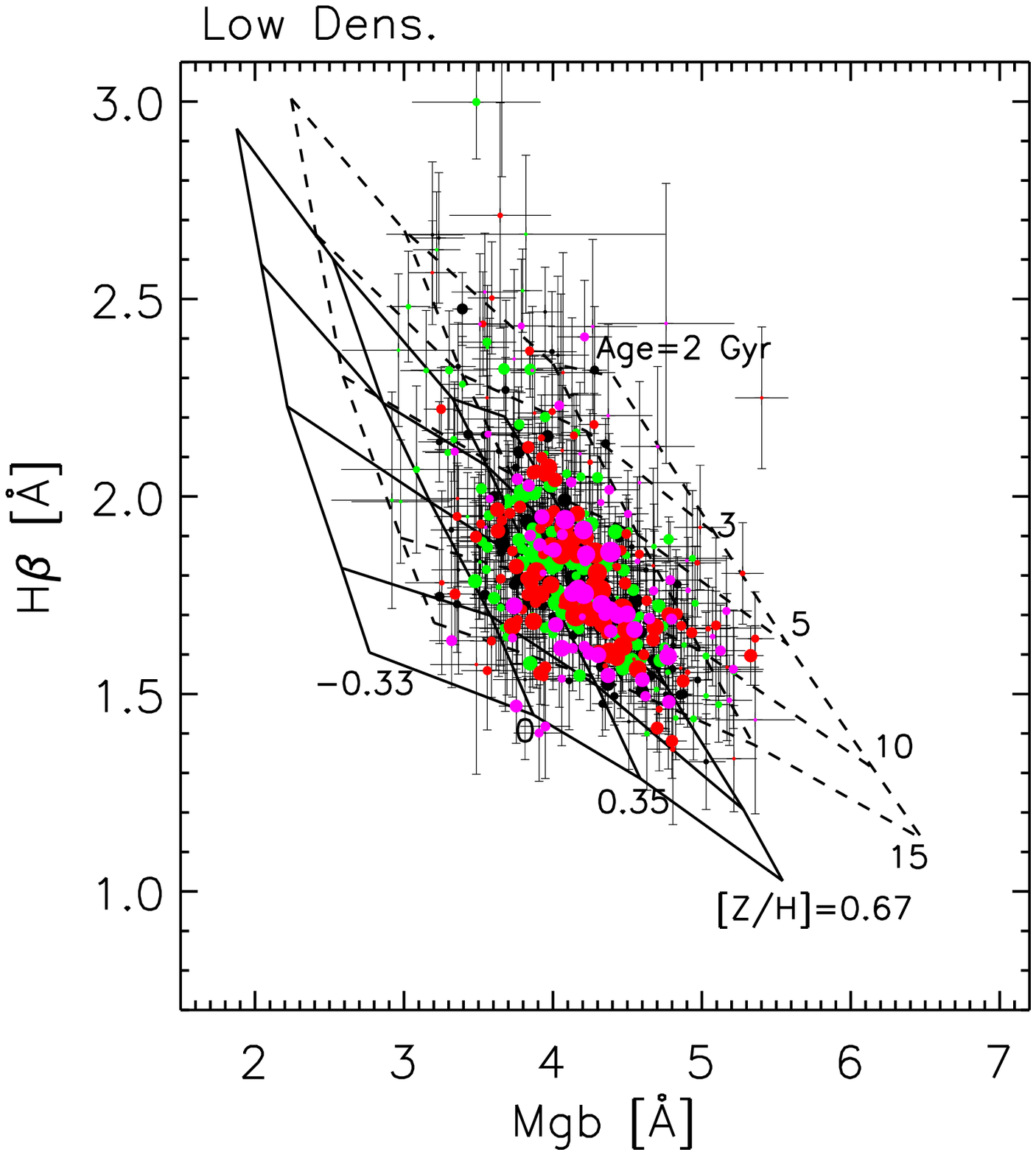}
 \plottwo{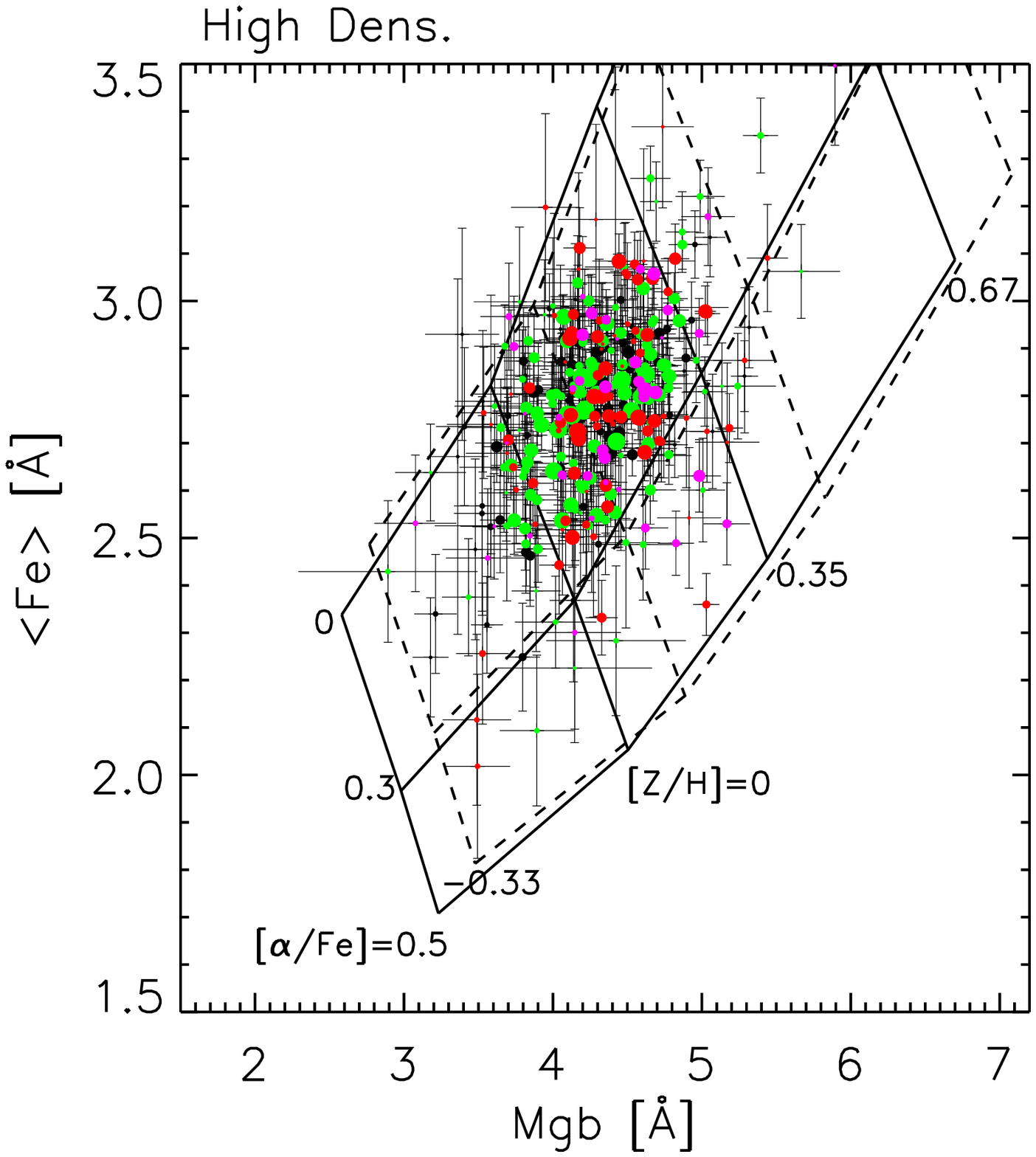}{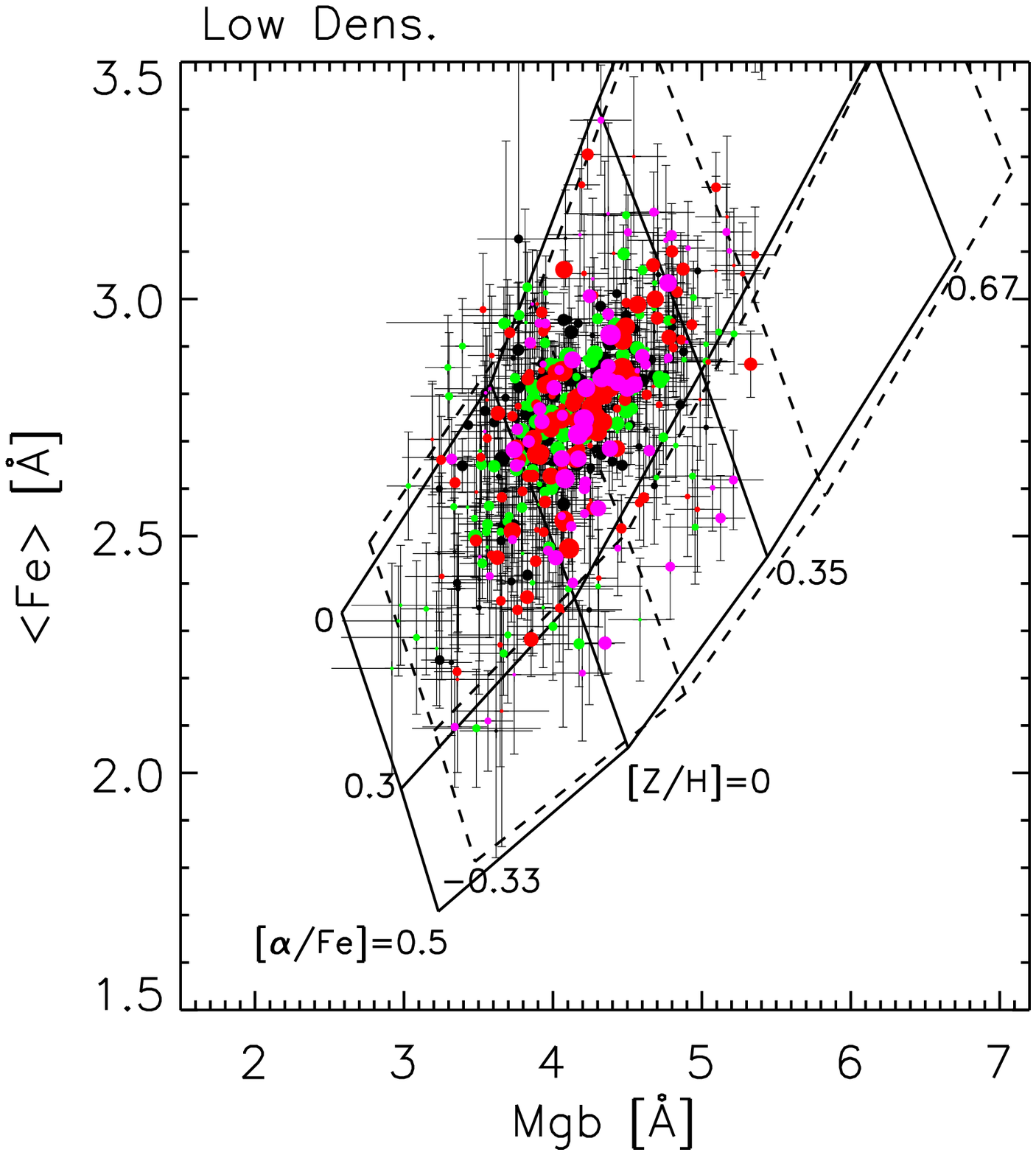}
 \plottwo{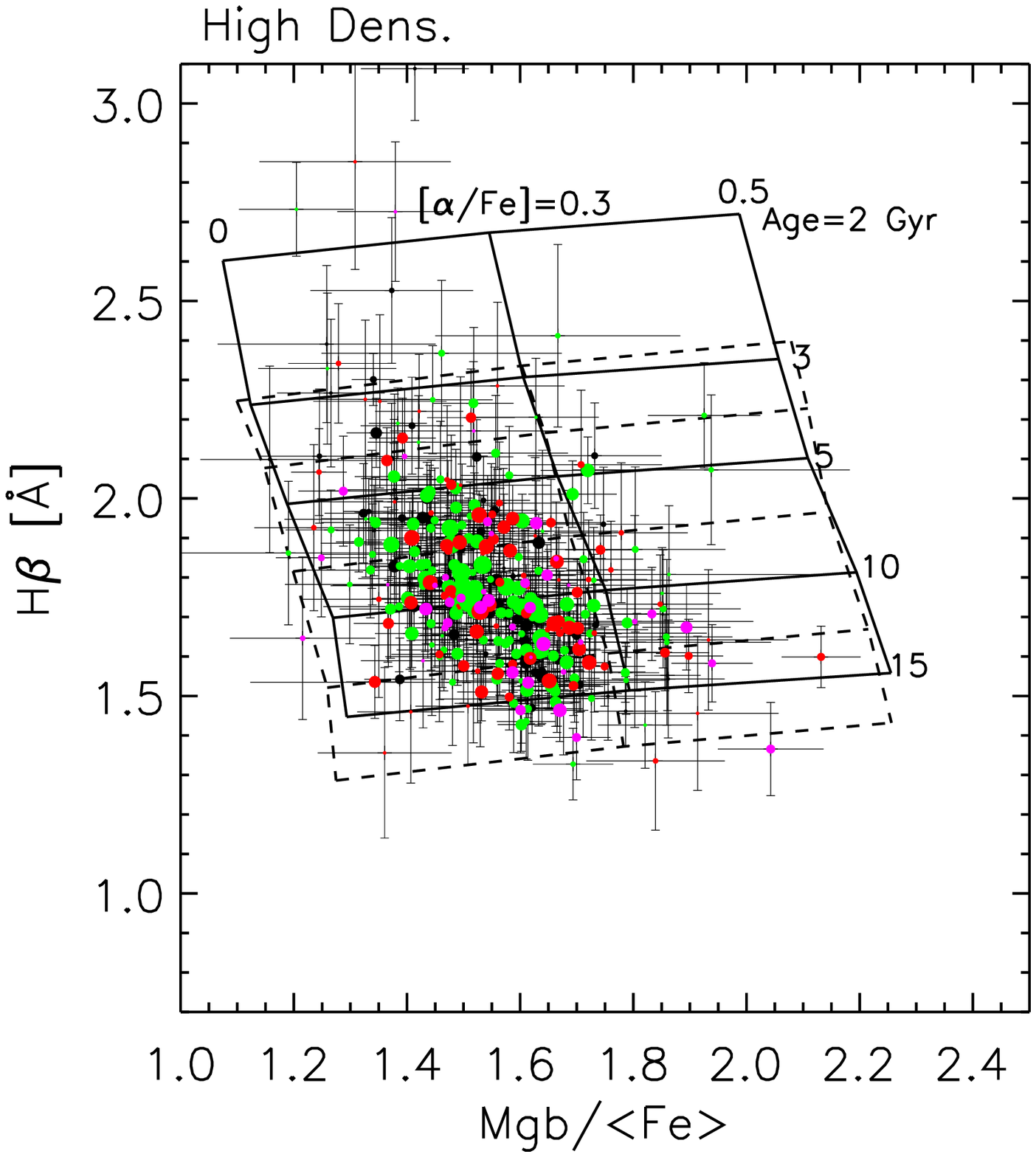}{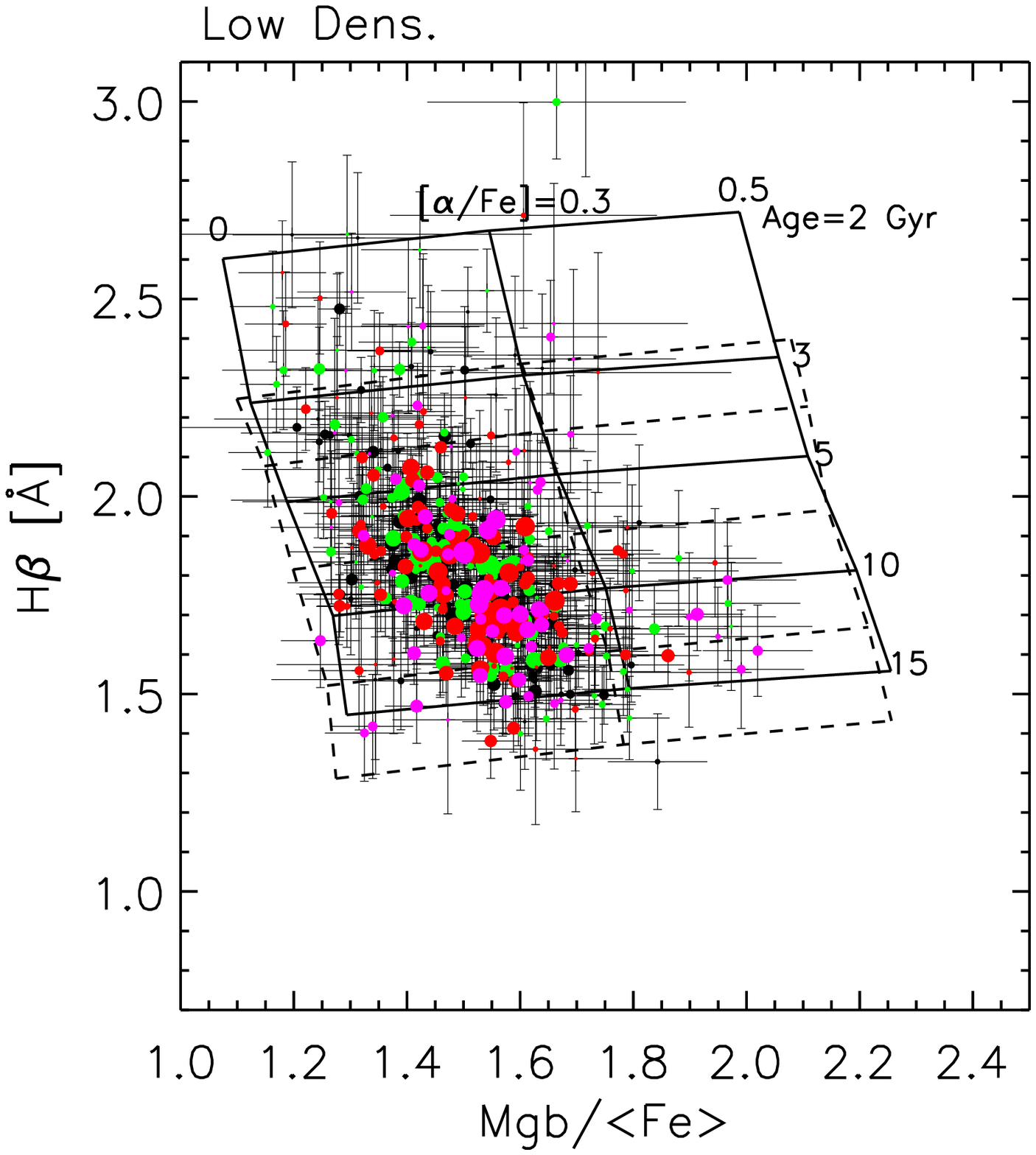}
 \caption{Distribution of H$\beta$ vs Mg (top), 
          $\langle$Fe$\rangle$ vs. Mg (middle) 
          and H$\beta$ vs Mg/$\langle$Fe$\rangle$ (bottom).  
          Symbol sizes indicate the number of galaxies in the 
          associated composite.  
          Symbol colors indicate redshift:  black, green, red and 
          magenta points represent objects at $0.05<z<0.07$, 
          $0.07\le z <0.09$, $0.09\le z<0.12$ and $0.12\le z<0.14$.  
          Grids show the models of TMB03.  
          Solid and dashed grids show age and metallicity at 
          $\alpha$-enhancements which are 0 (solar) and 0.3 (top panels), 
          metallicity and $\alpha$-enhancement for ages of 
          10 and 15~Gyrs (middle panels), and 
          ages and $\alpha$-enhancements when [Z/H] is 
          0 and 0.35 (bottom panels).  }
 \label{modelgrids}
\end{figure}

\begin{figure}
 \centering
\plottwo{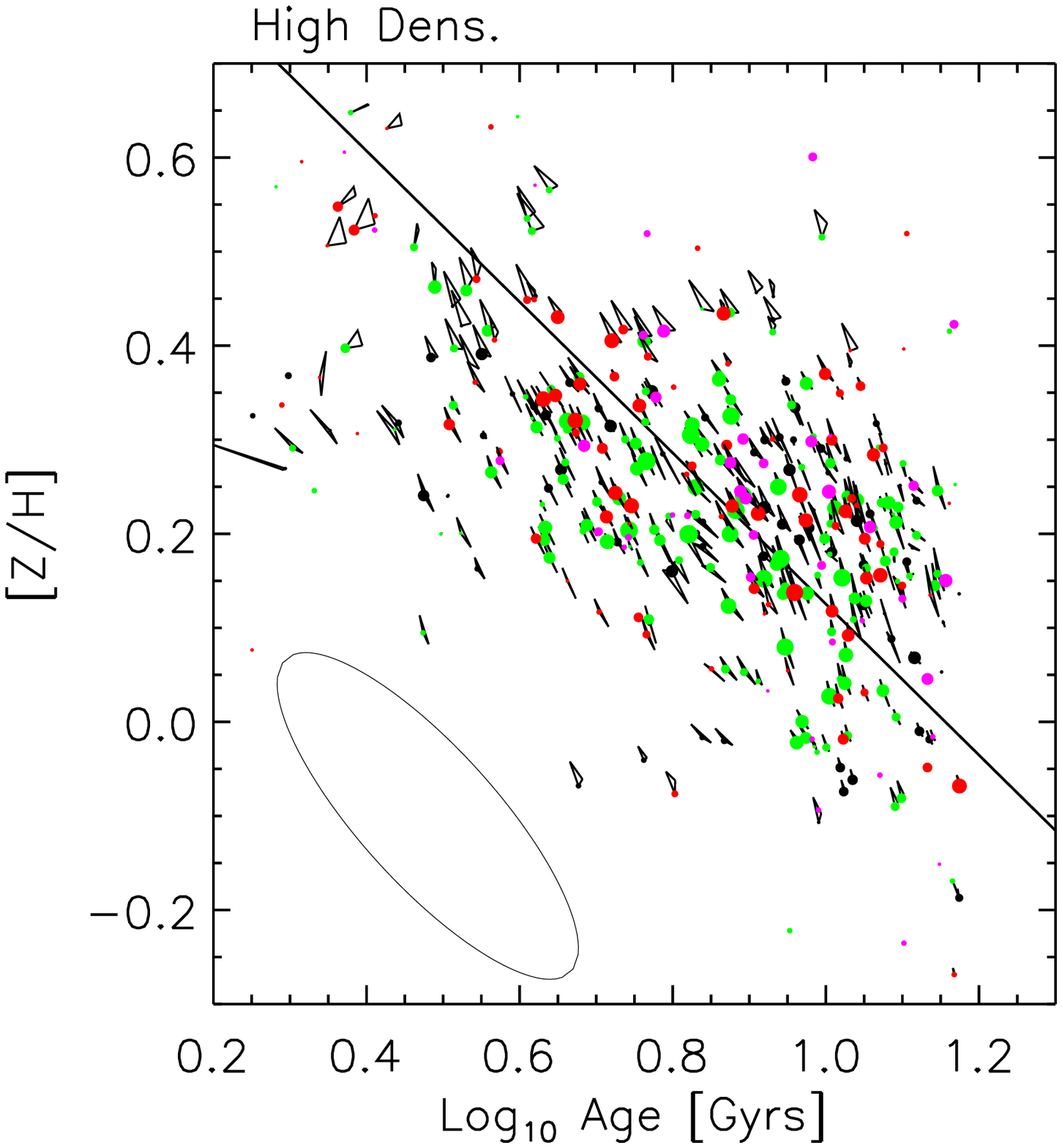}{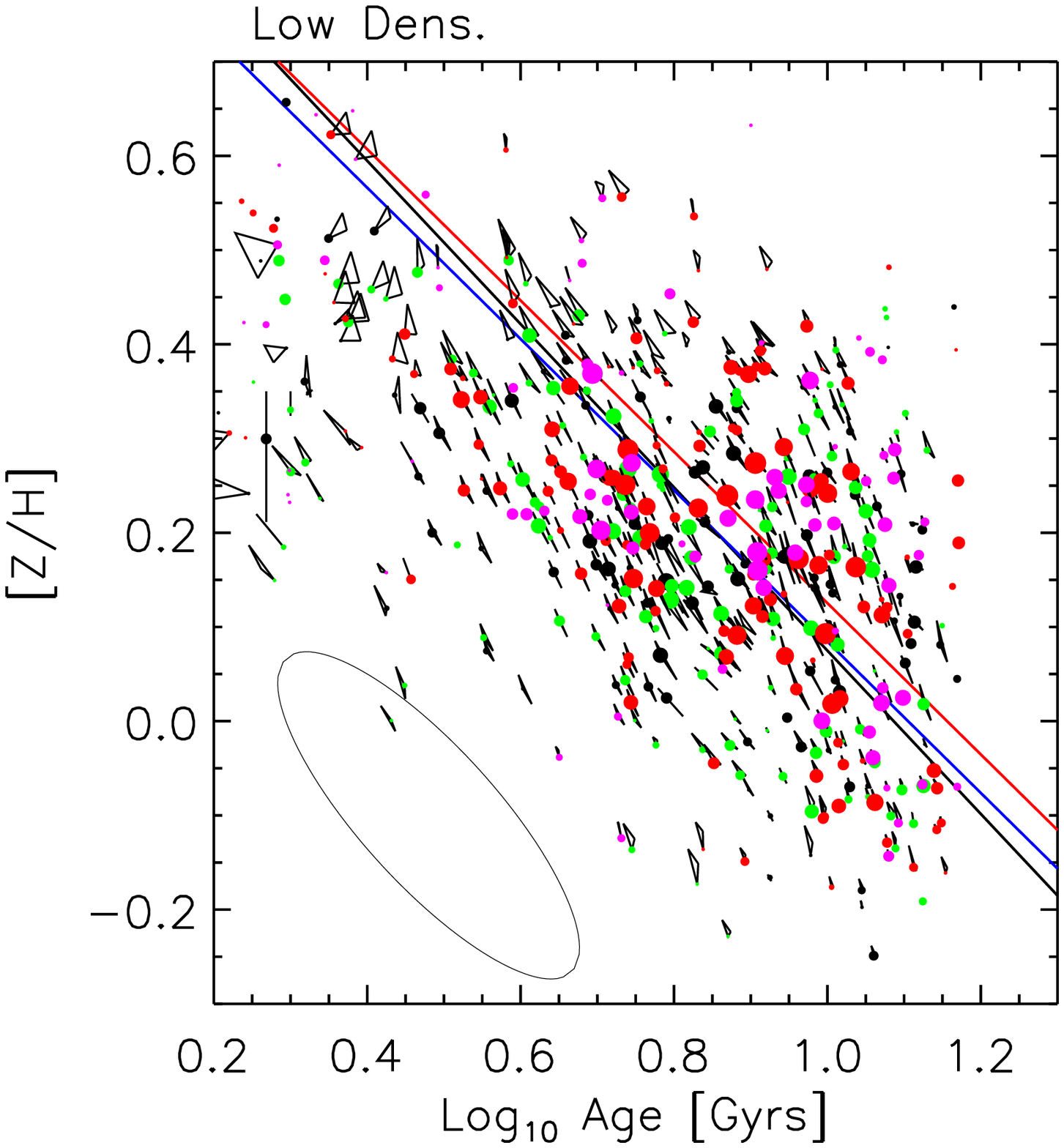}
\plottwo{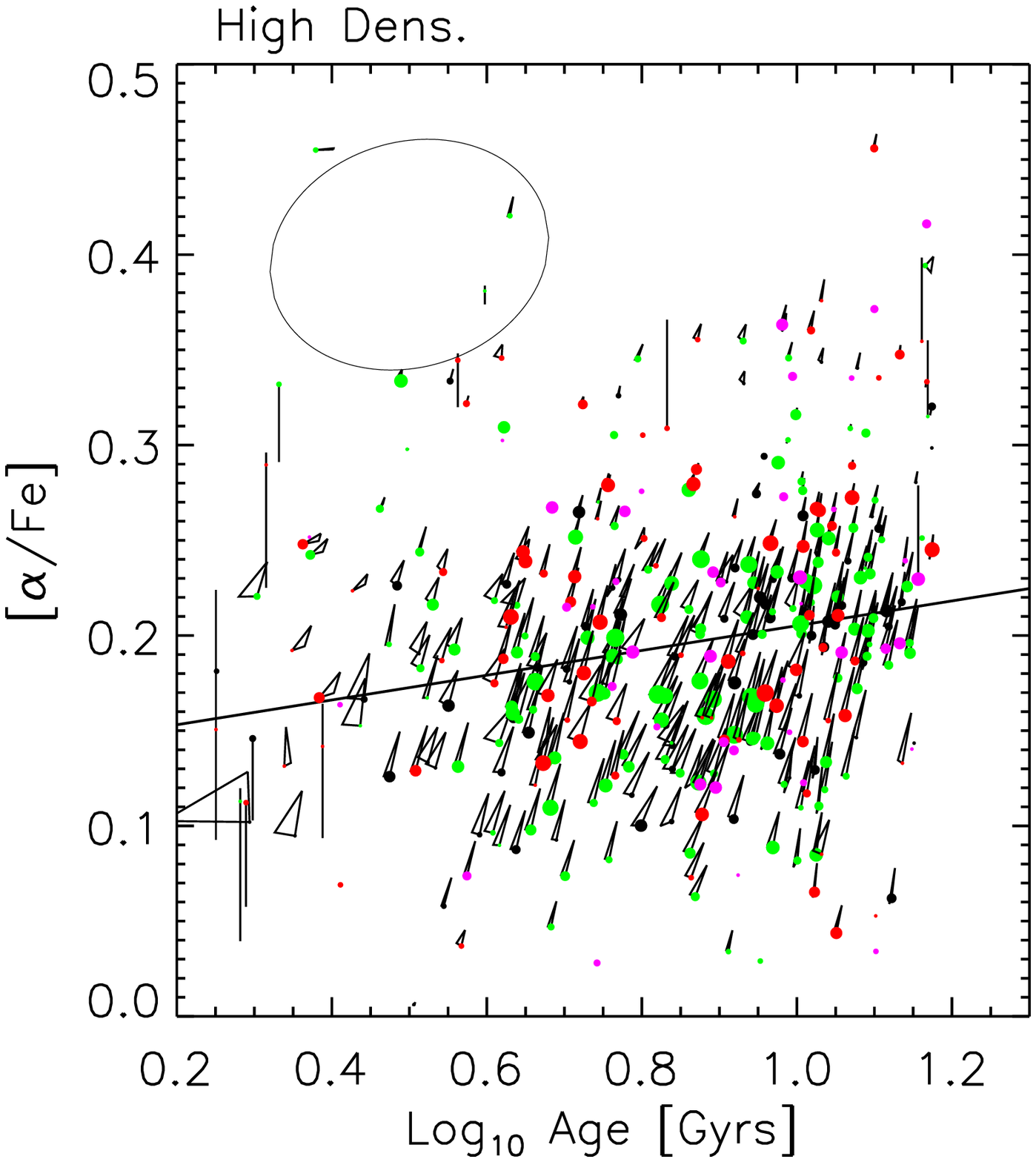}{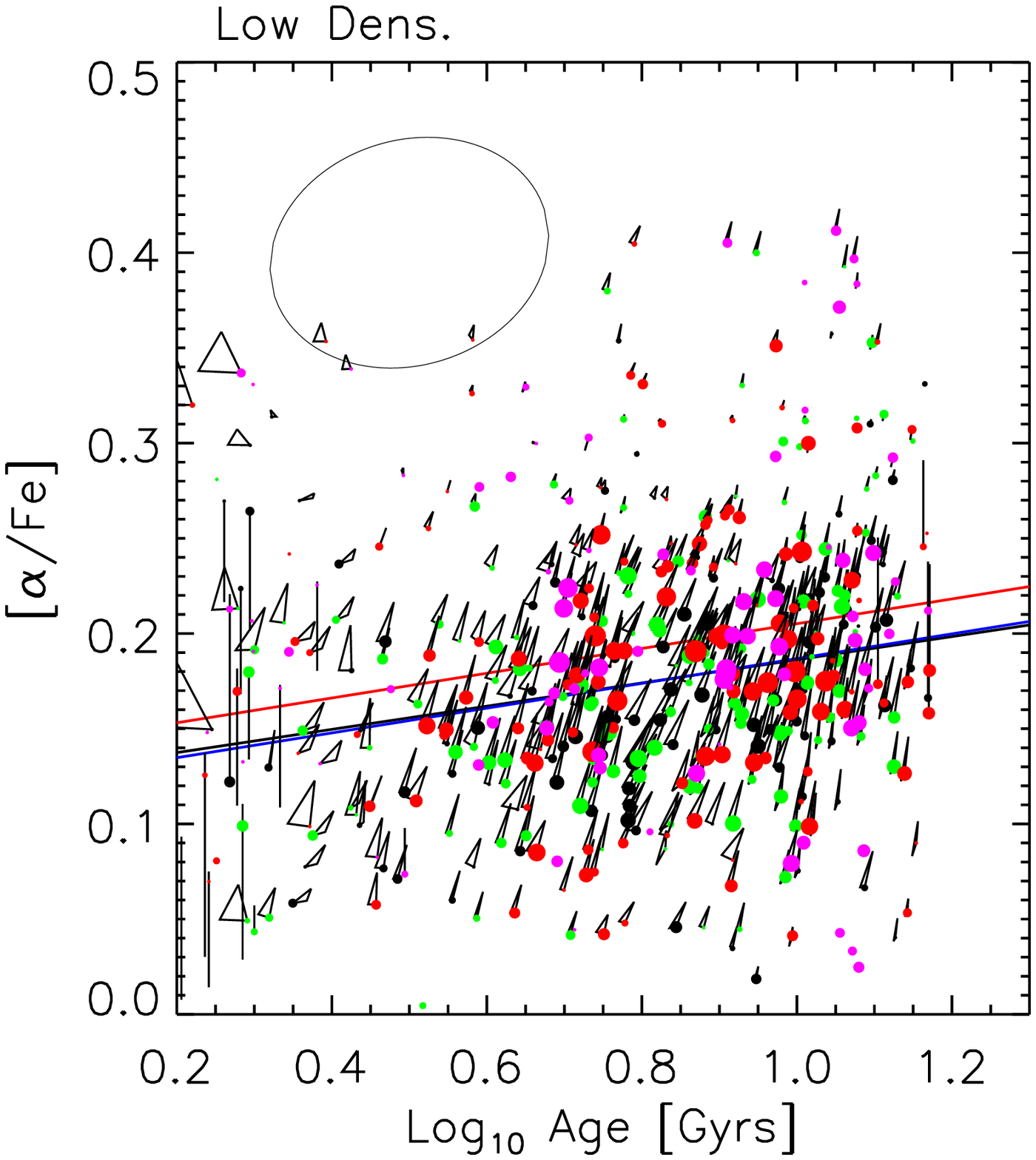}
 \caption{Pairwise correlations between derived ages, metallicities 
          and $\alpha$-enhancements.  The vertices of the triangle 
          associated with each data point show the three estimates (see 
          text) of each quantity; these provide a rough estimate of the 
          systematic uncertainty in the derived quantities.  
          The ellipse in the left-hand corner of each plot shows the 
          typical uncertainty arising from the errors associated with 
          measuring the line-indices.  
          The shape of the ellipse in the top panels reflects 
          the well-known degeneracy between age and metallicity.  
          Solid lines in left panels show bisector (top left) and 
          direct (bottom left) fits.  
          Red and blue lines in the panels on the right compare the 
          offset between high and low density environments if both 
          samples are required to have the same slope as that shown 
          in the left panel.  }
 \label{agemetalenhan}
\end{figure}

\begin{figure}
 \centering
 \plottwo{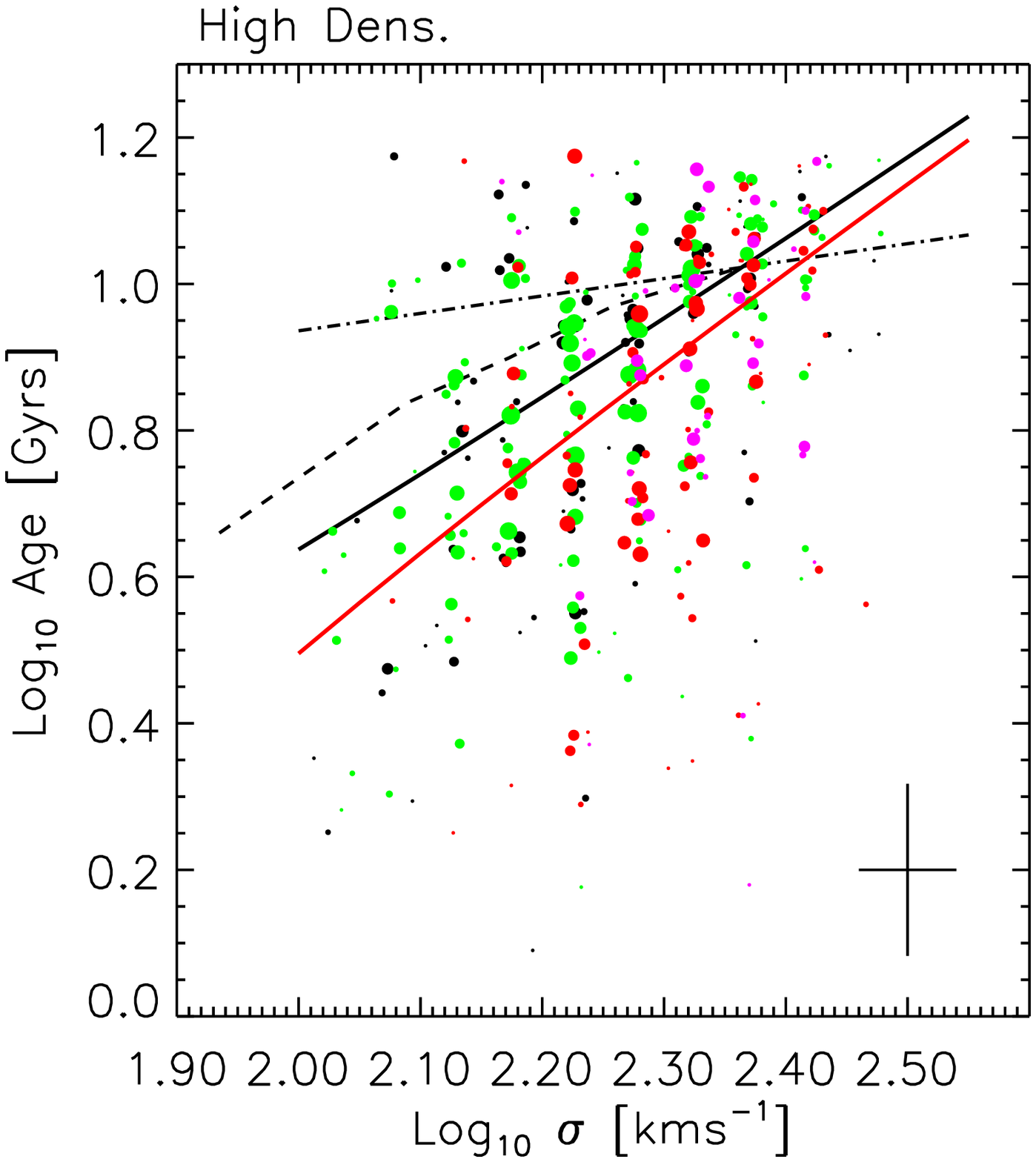}{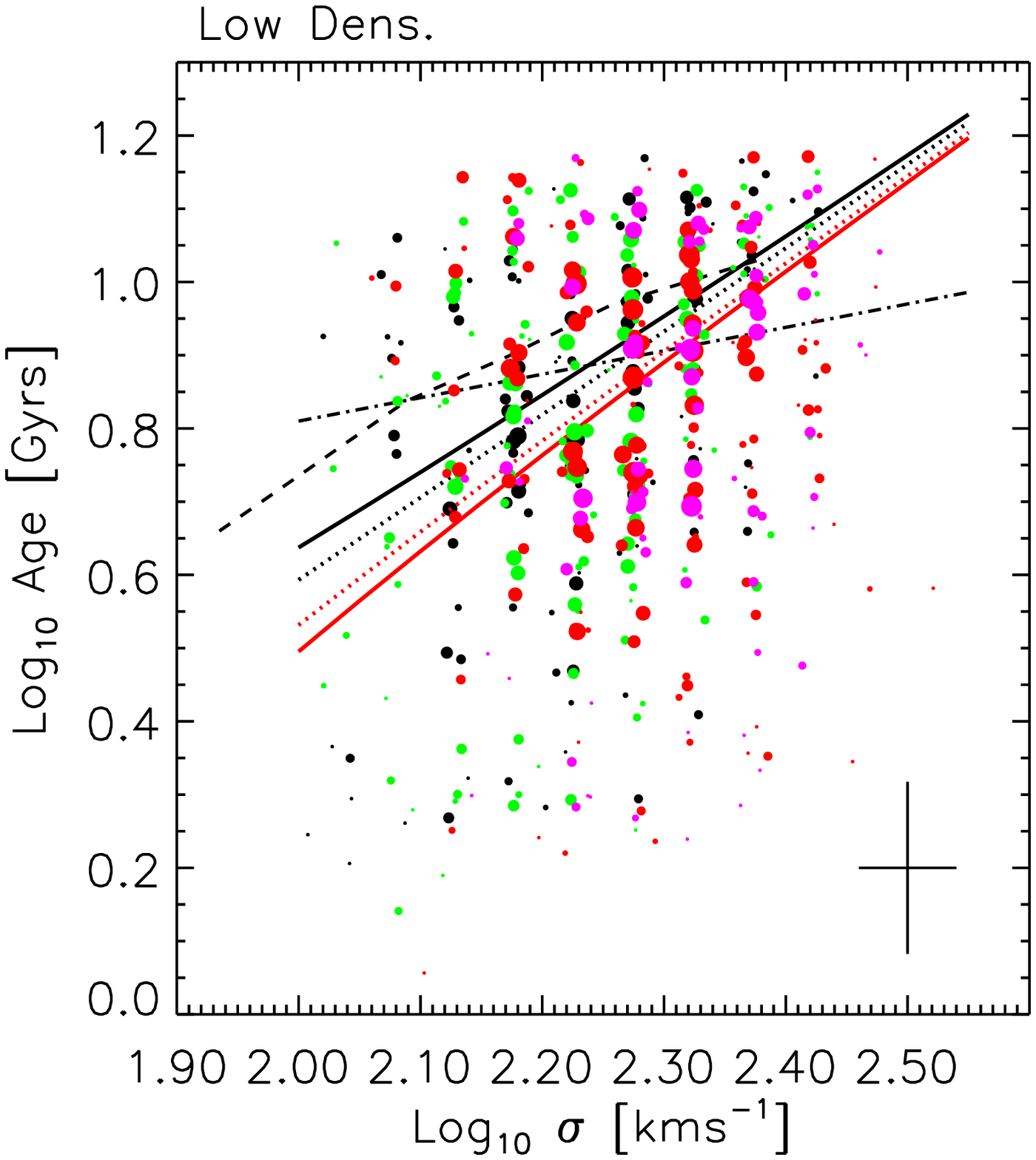}
 \plottwo{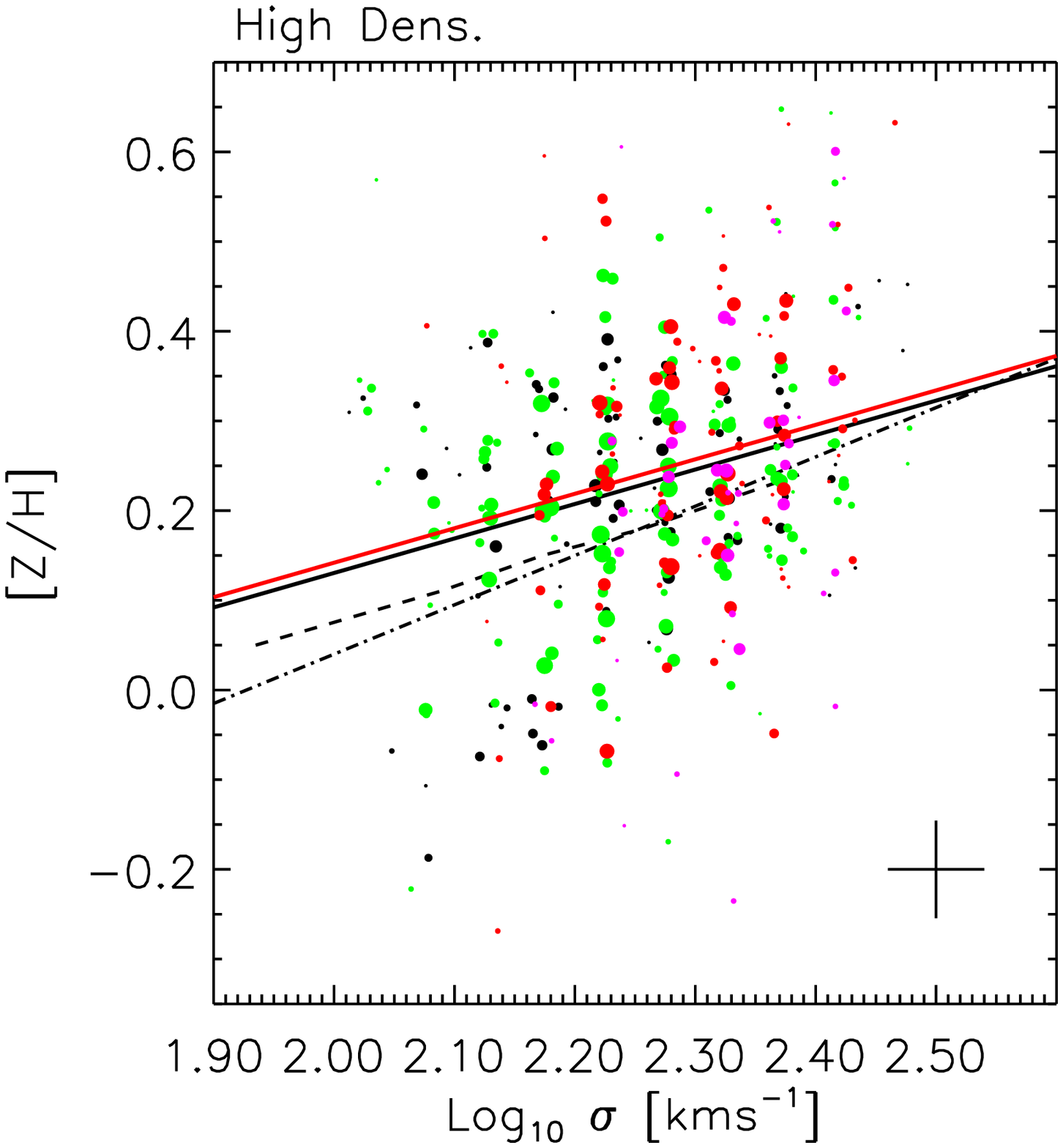}{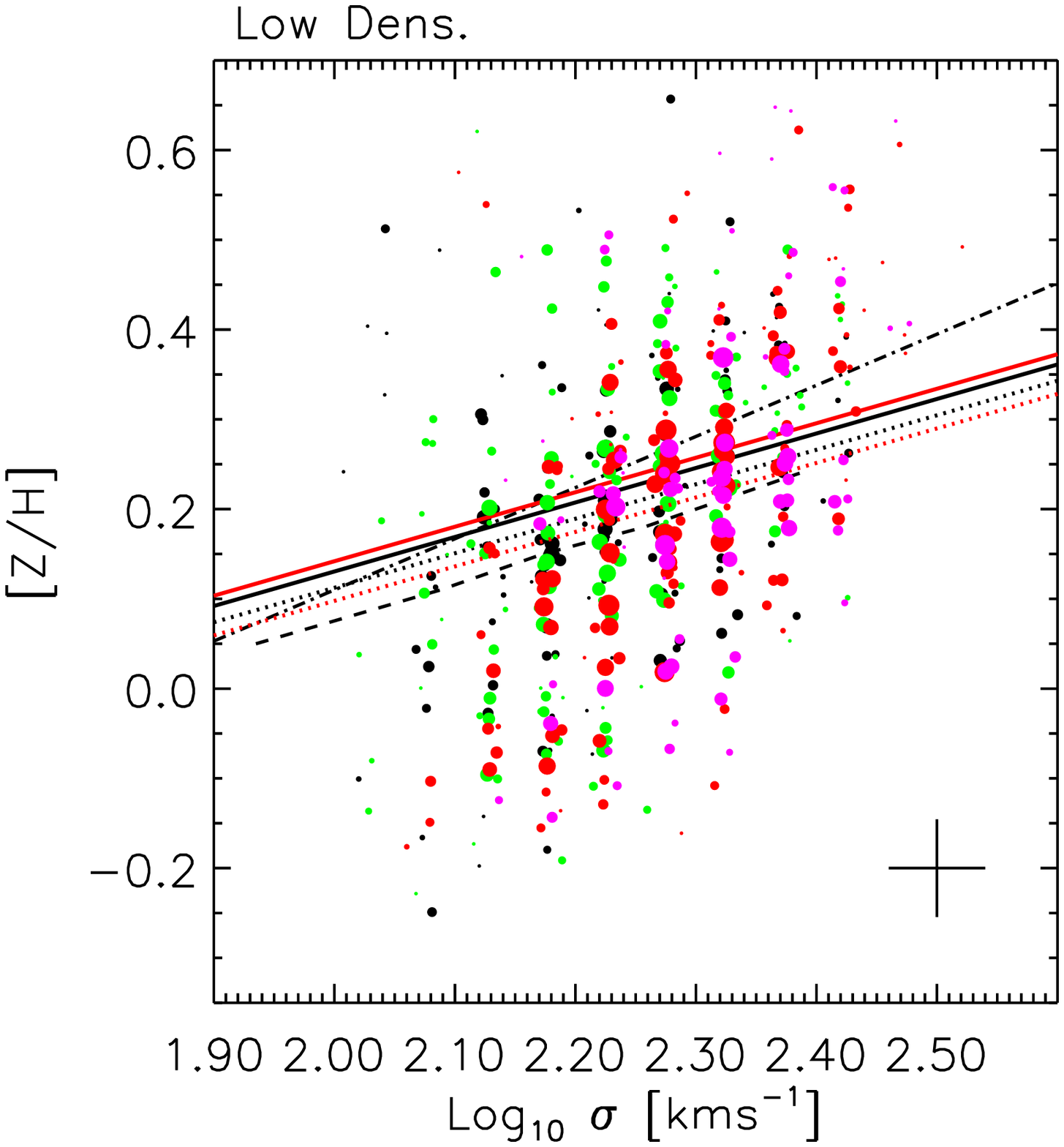}
 \plottwo{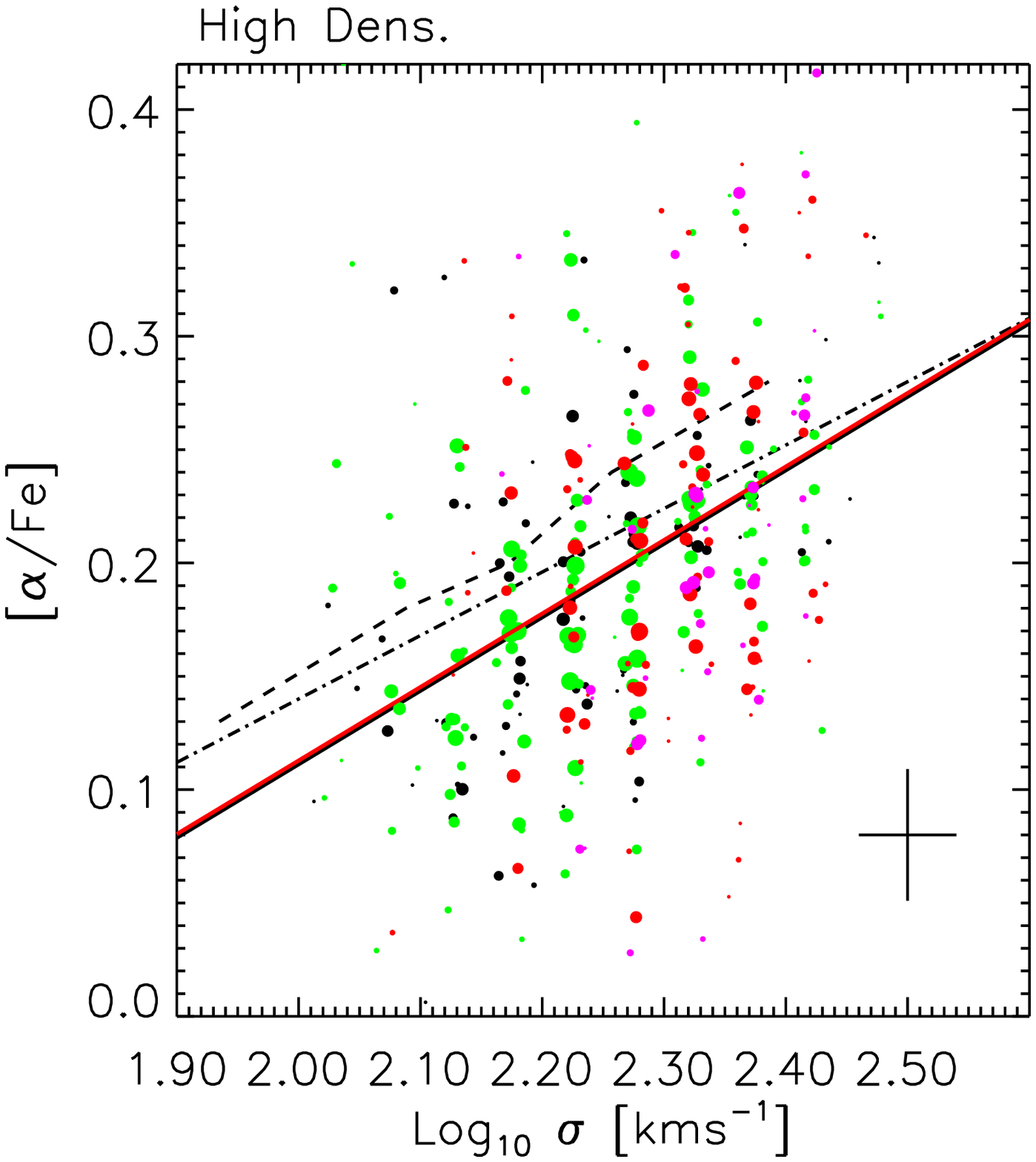}{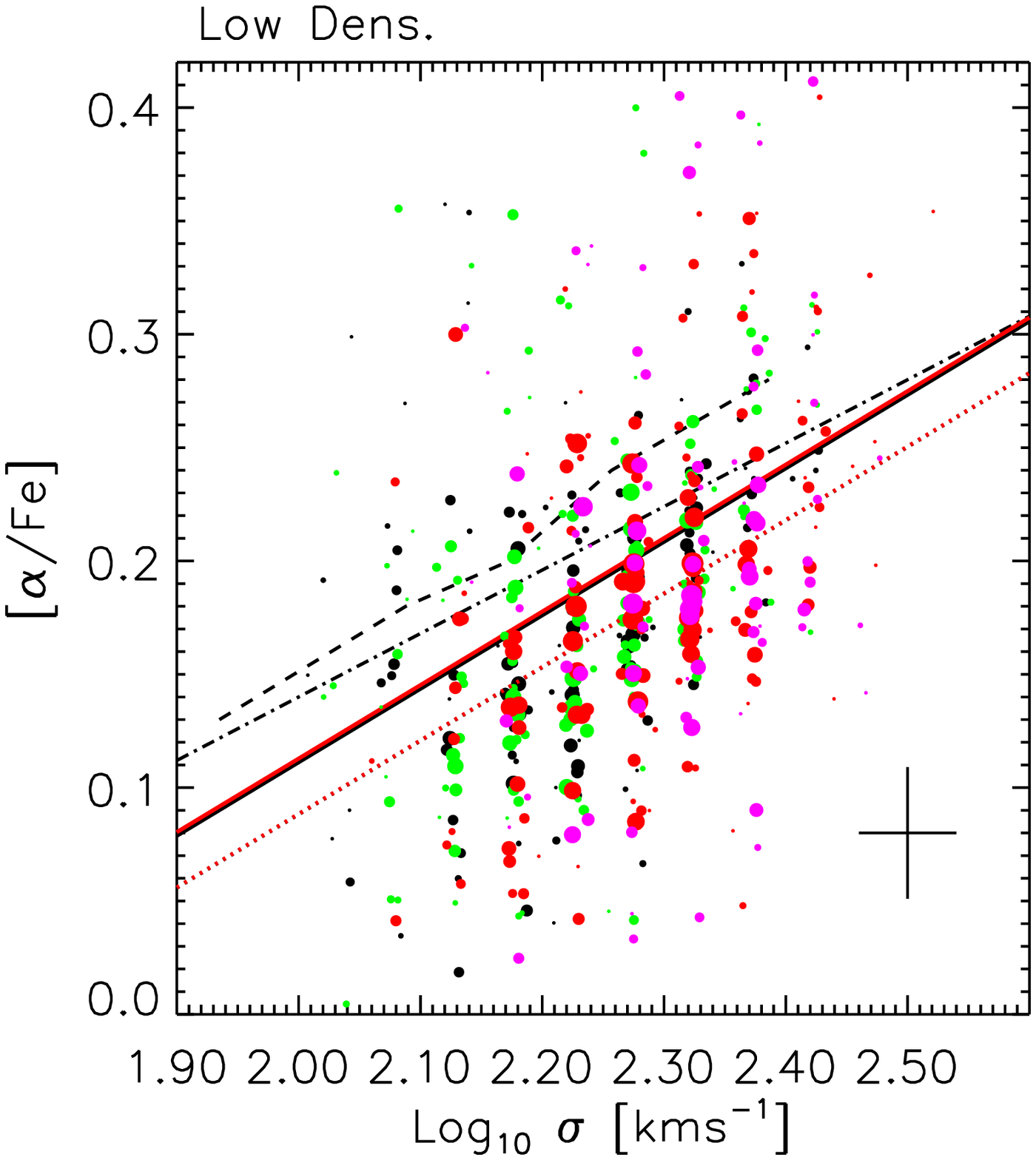}
 \caption{Objects with large $\sigma$ tend to be older (top panels), 
          have larger metallicities (middle panels), 
          and larger $\alpha$-enhancements (bottom).  
          This is true at all redshifts in our sample:  
          black lines show the mean relation at $z\sim 0.06$ and 
          red lines show the result of fitting for the shift in 
          Gyrs for the sample at $z\sim 0.11$.  Moreover, note that 
          there is clear evolution in the age-$\sigma$ relation, but 
          not in [Z/H]-$\sigma$ or [$\alpha$/Fe]-$\sigma$.  
          In addition, at fixed $\sigma$, objects in dense regions 
          (solid lines) tend to have similar ages and metallicities 
          as their counterparts in less dense regions (dotted lines 
          in panels on right), 
          and slightly larger $\alpha$-enhancements.  Cross in the bottom 
          right corner of each panel shows the typical uncertainty. 
          Dashed and dot-dashed lines show the relations obtained by 
          Nelan et al. 2005 and Thomas et al. (2005) in their analysis of other 
          datasets.}
 \label{modelsigma}
\end{figure}

\begin{figure}
 \centering
 \plottwo{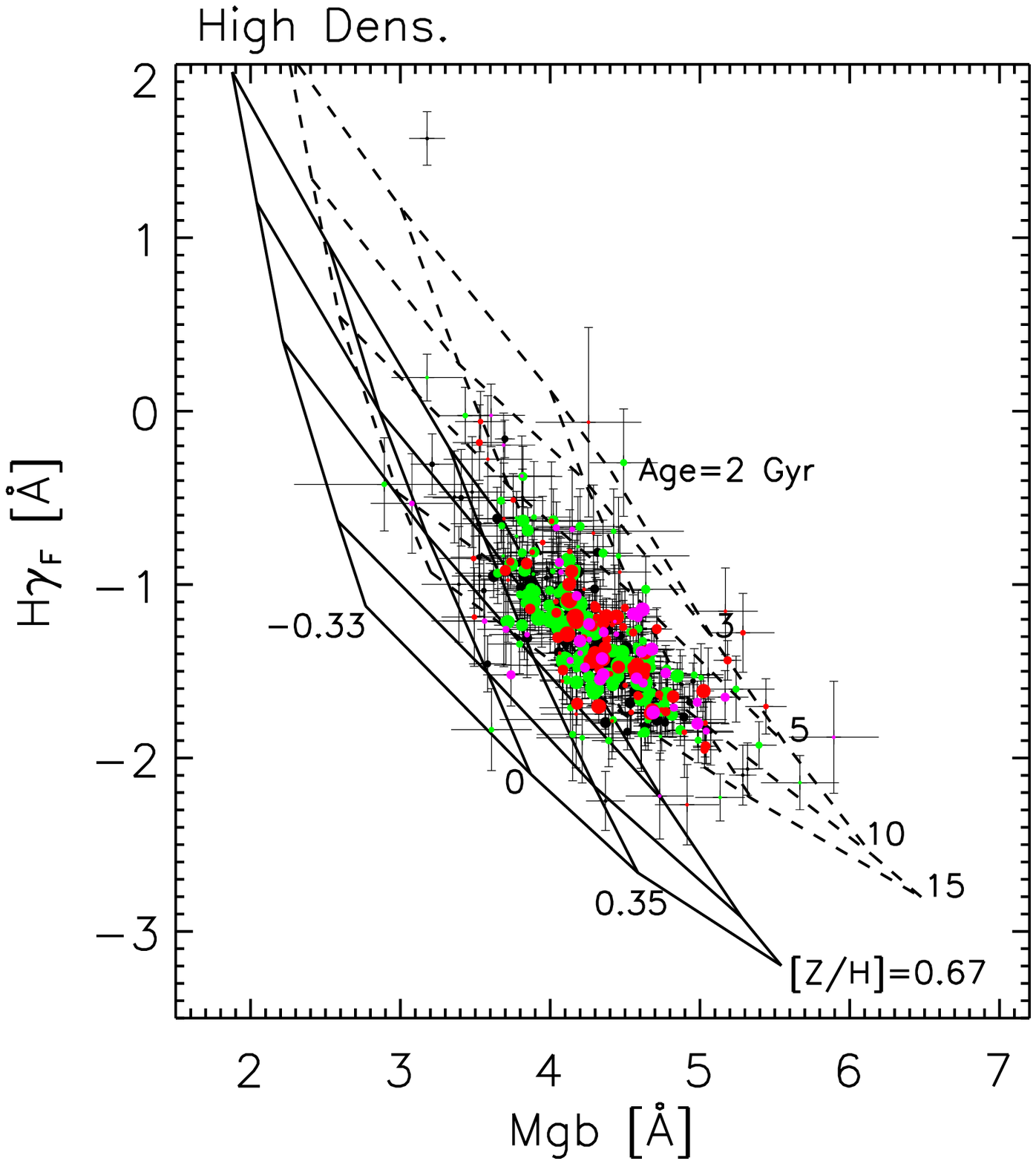}{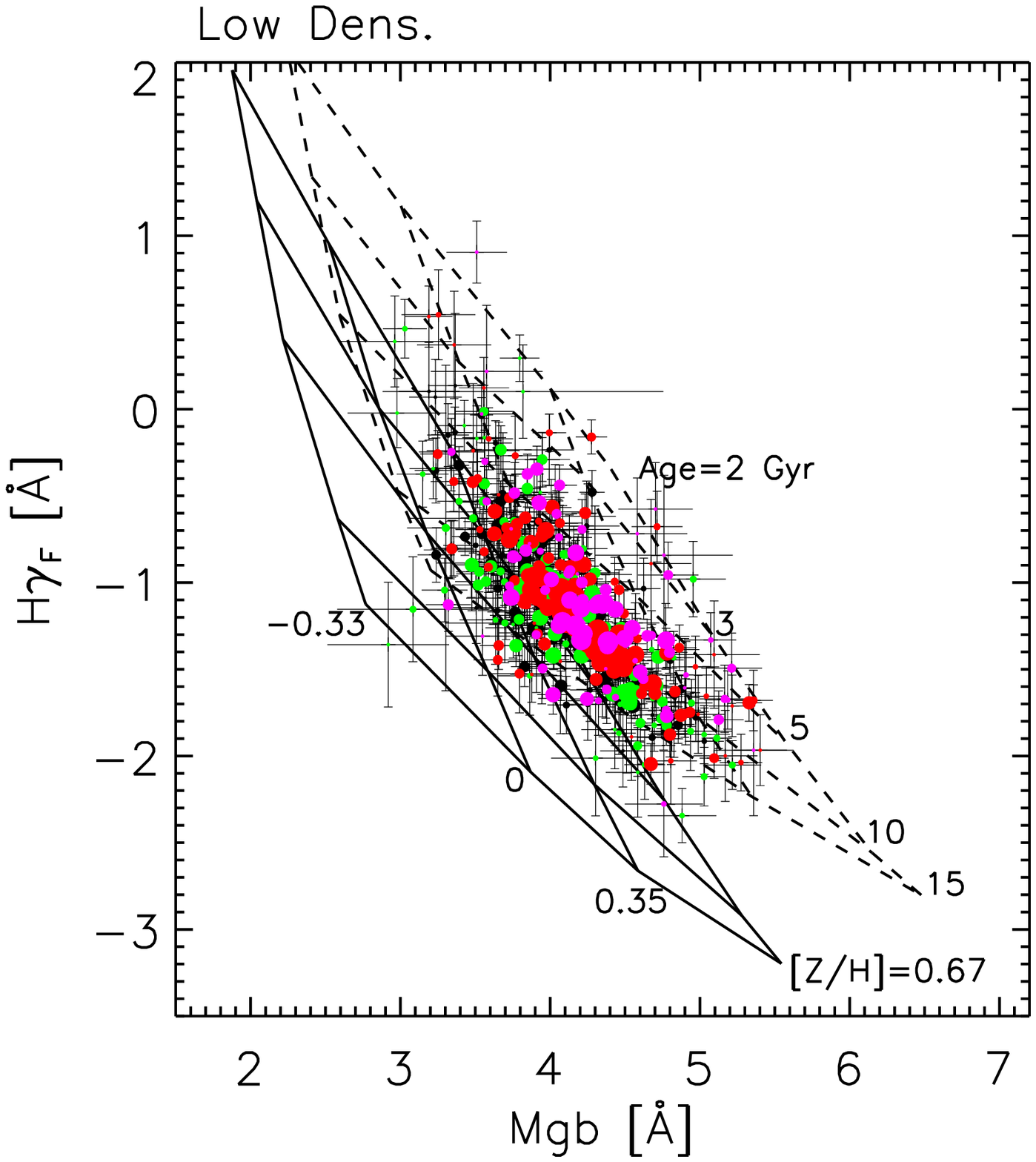}
 \plottwo{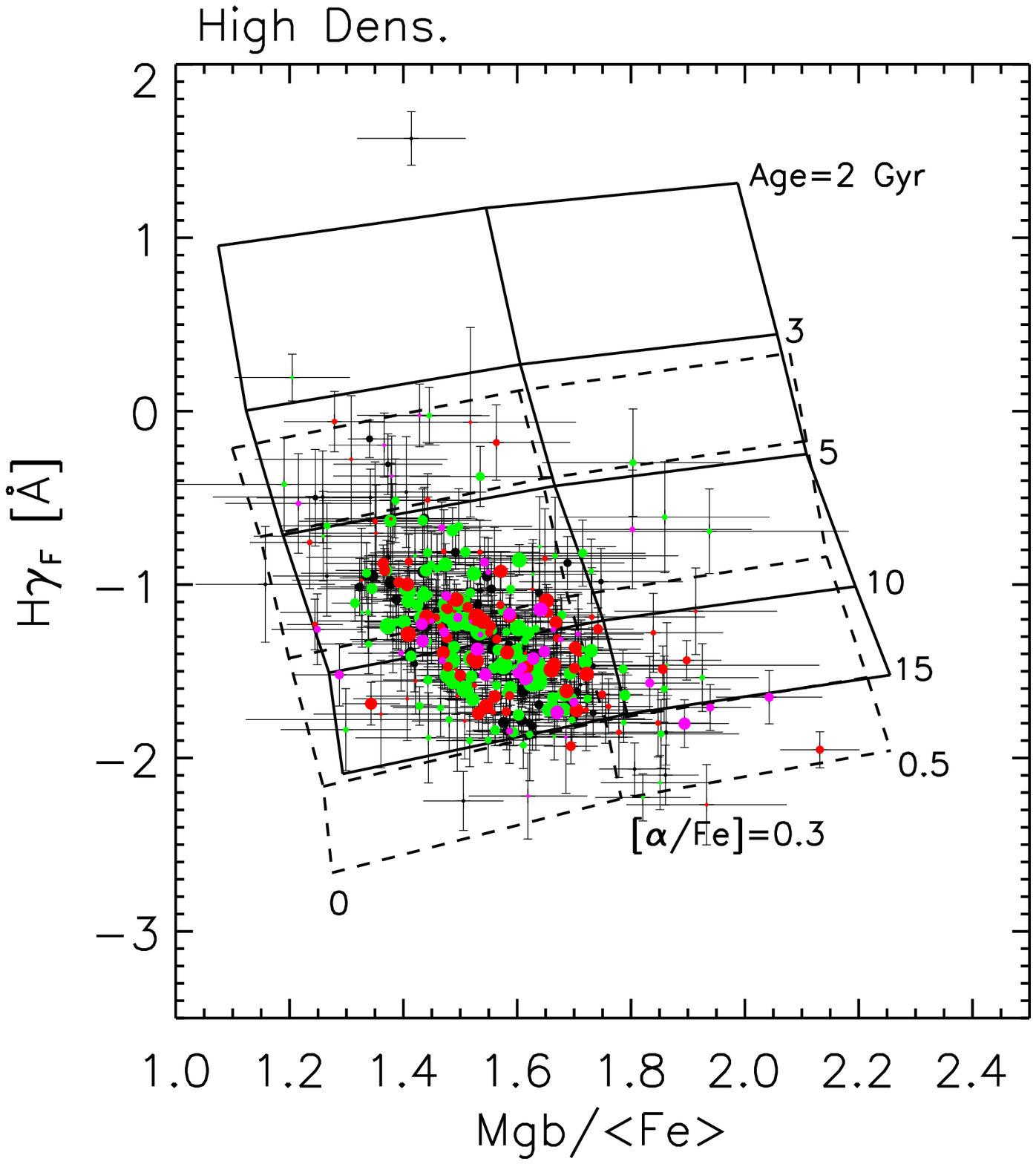}{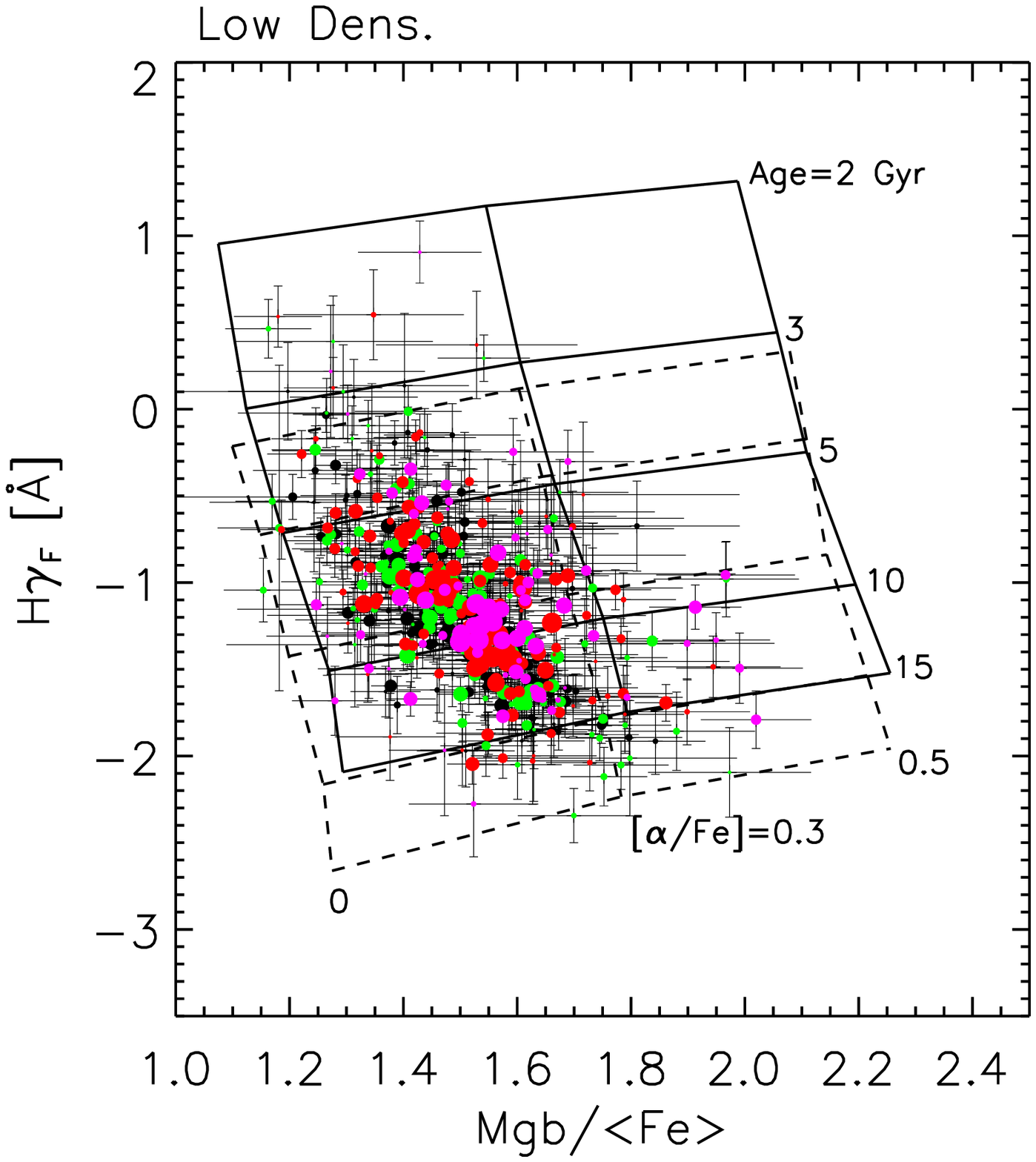}
 \caption{Same as Figure~\ref{modelgrids}, but with 
          H$\gamma_{\rm F}$ instead of H$\beta$. 
          The top panel shows clearly that these objects have 
          non-solar $\alpha$-element abundance ratios.  (But see 
          Appendix~\ref{fluxcalib} for a discussion of whether flux 
          calibration problems compromise the measurements in the 
          lowest redshift bin.)  }
 \label{modelHdHg}
\end{figure}

\begin{figure}
 \centering
 \plottwo{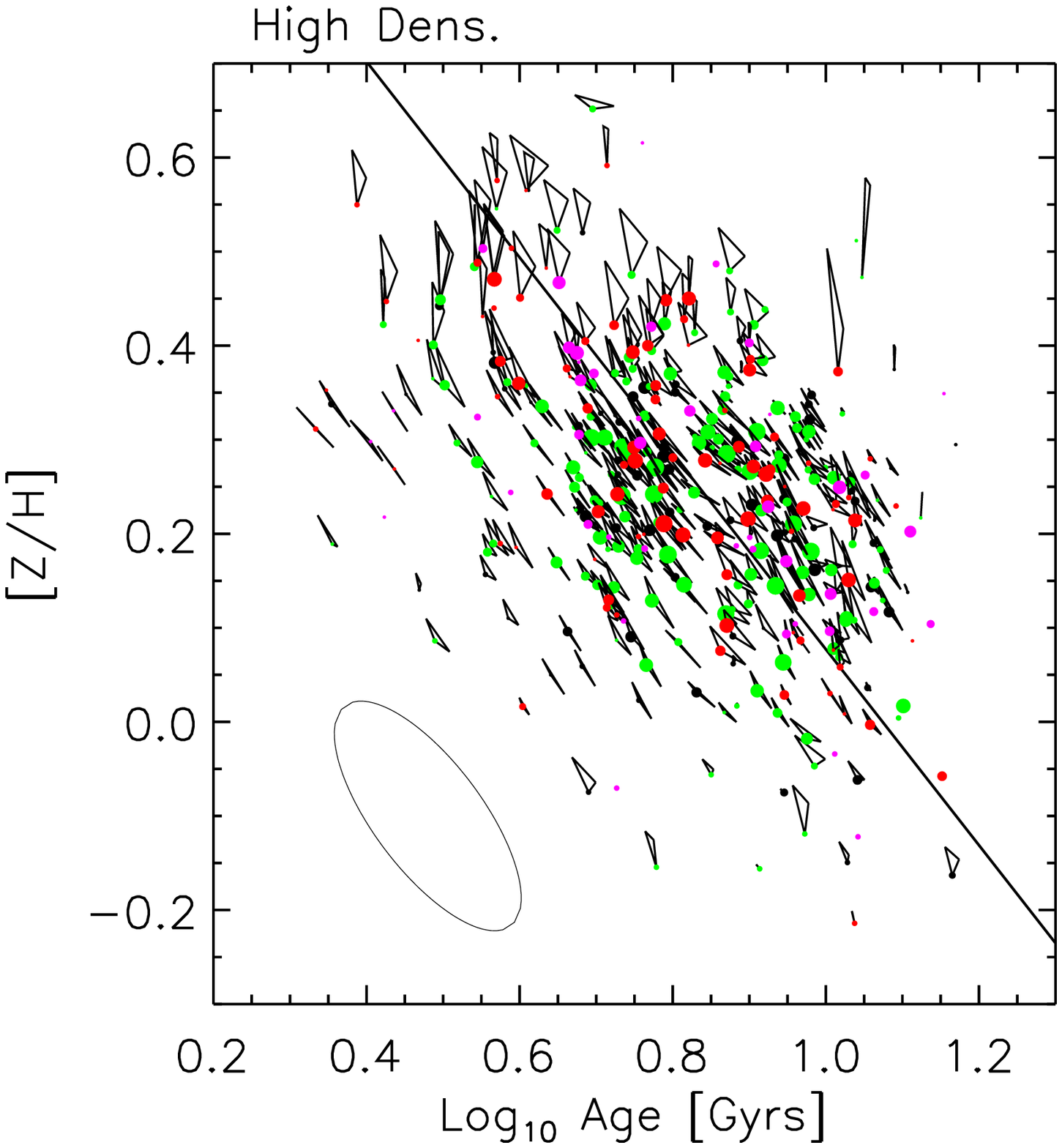}{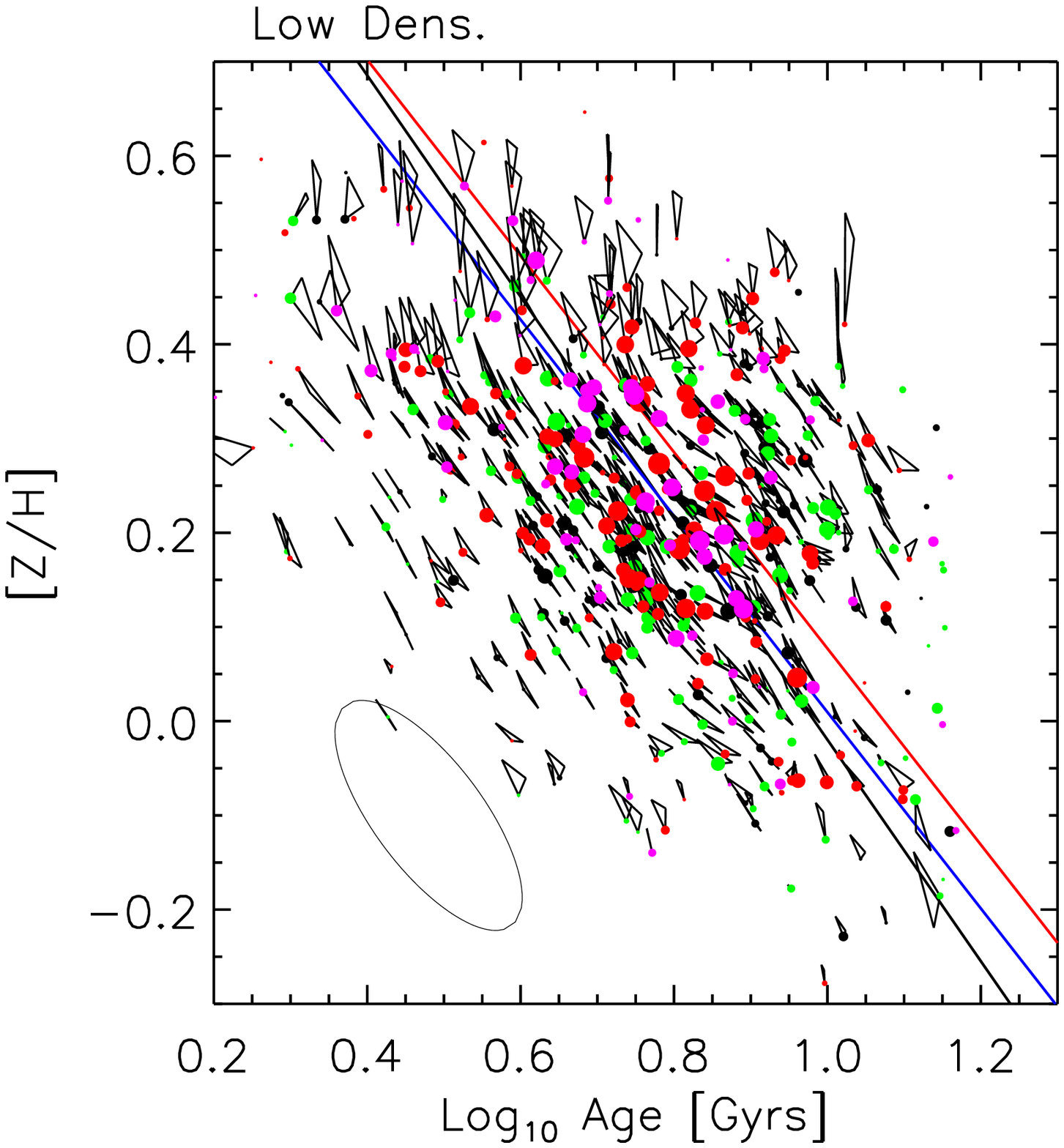}
 \plottwo{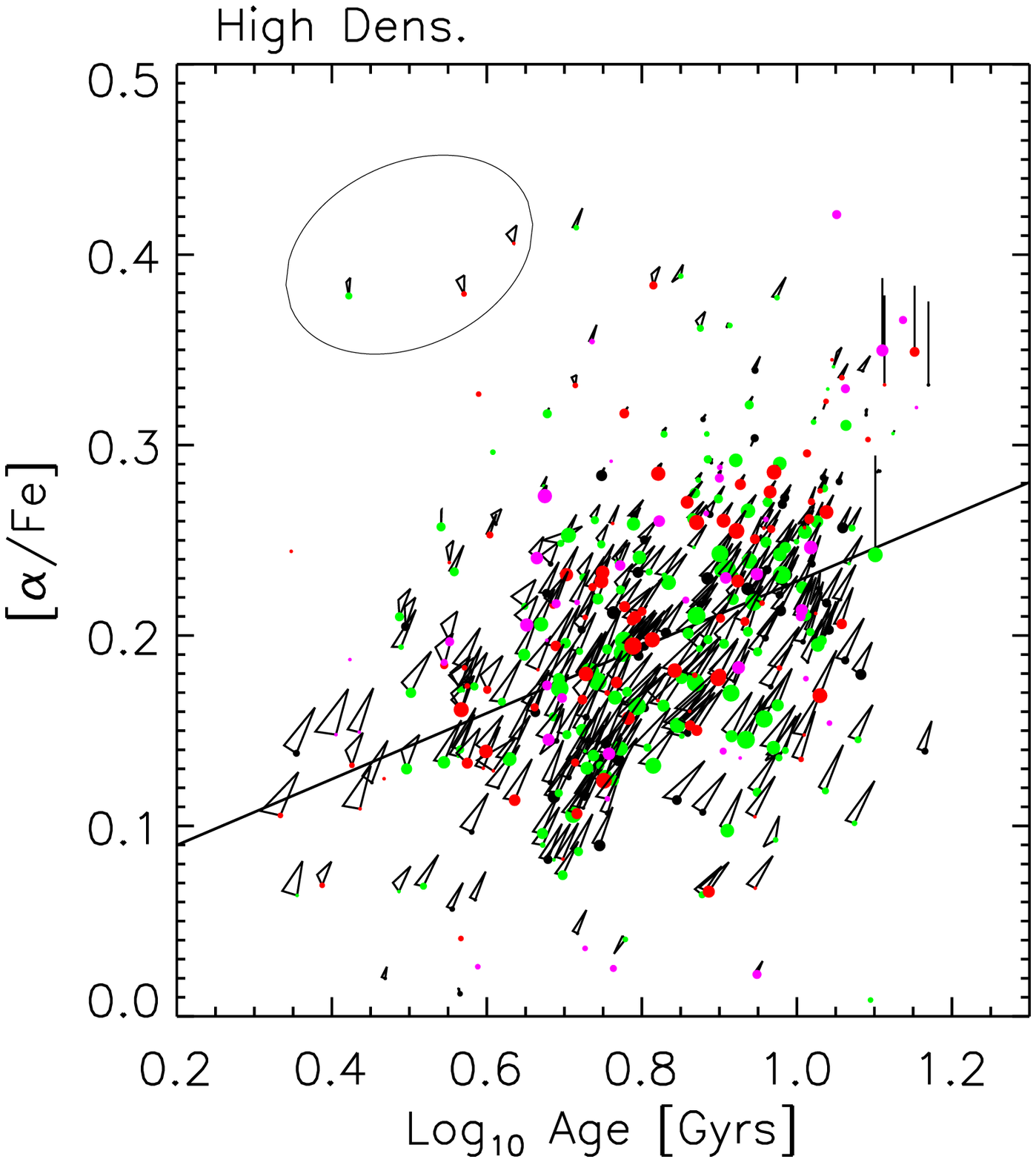}{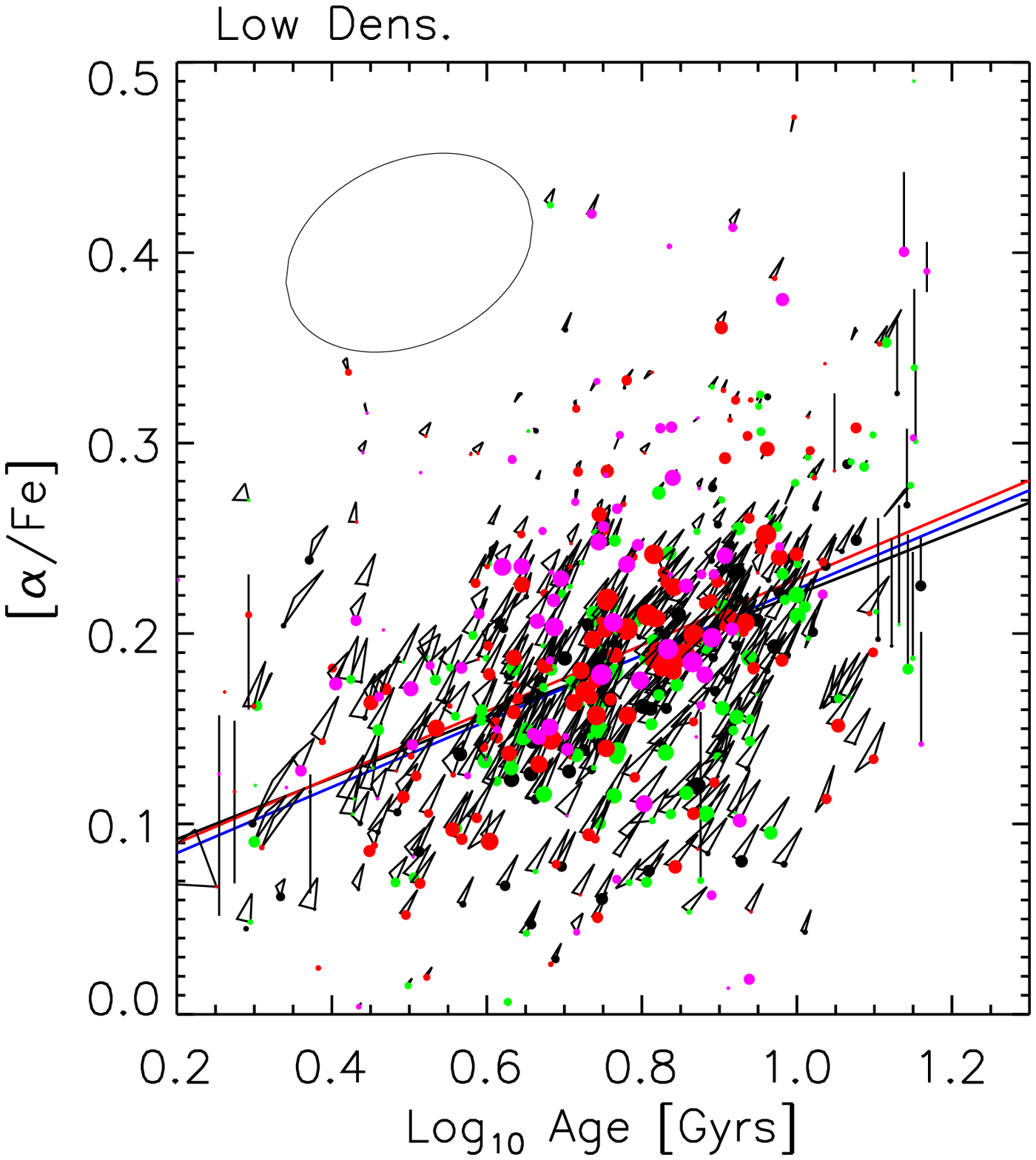}
 \caption{Same as Figure~\ref{agemetalenhan}, but with 
          H$\gamma_{\rm F}$ instead of H$\beta$.  
          Here, the anti-correlation between [Z/H] and age is slightly 
          steeper than before, and the correlation between [$\alpha$/Fe] 
          and age is steeper and tighter than before.  
          (See Appendix~\ref{fluxcalib} for a discussion of whether 
          flux calibration problems compromise the results in the 
          lowest redshift bin.) }
 \label{agemetalenhanHdHg}
\end{figure}

\clearpage

\begin{figure}
 \centering
 \plottwo{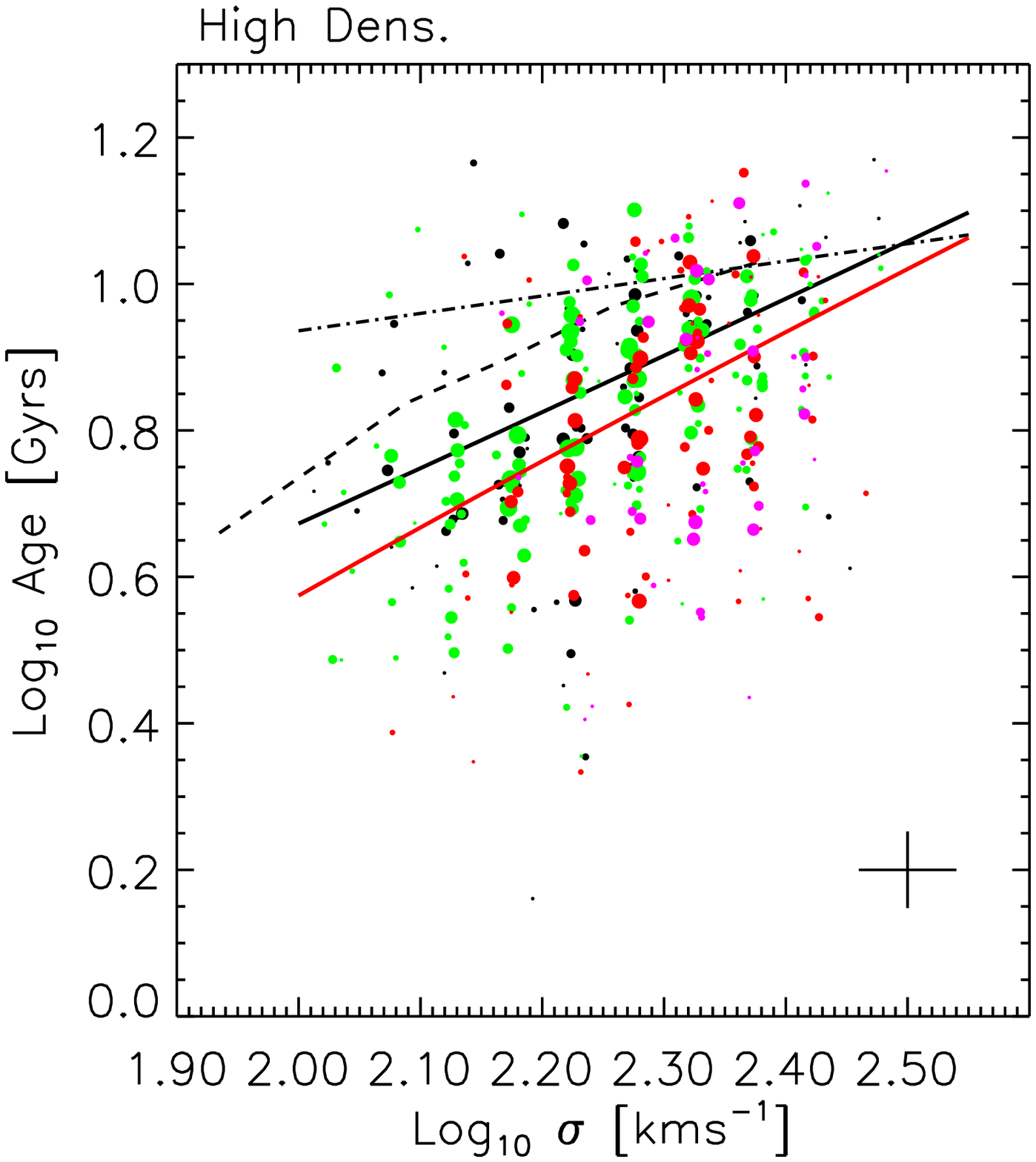}{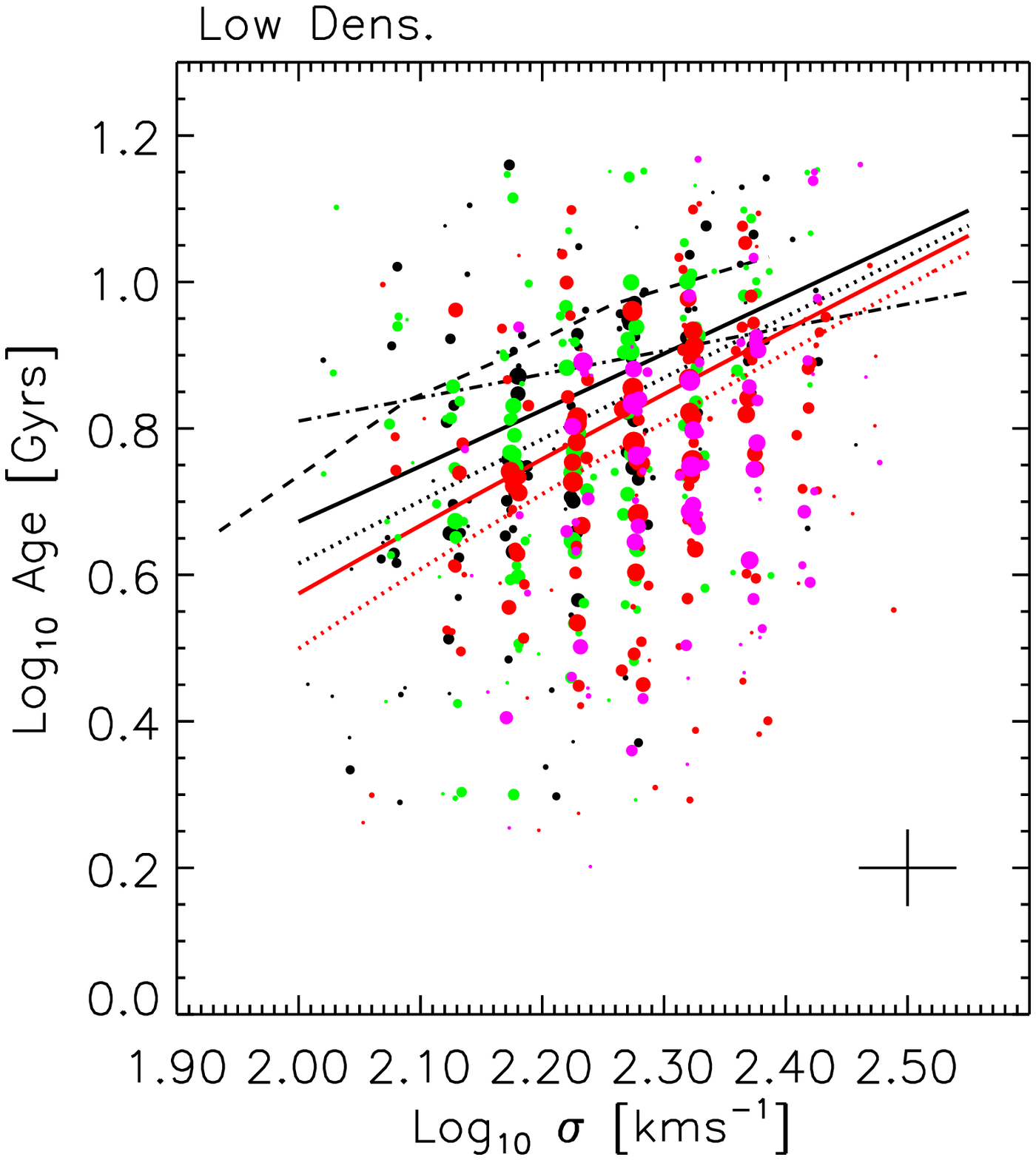}
 \plottwo{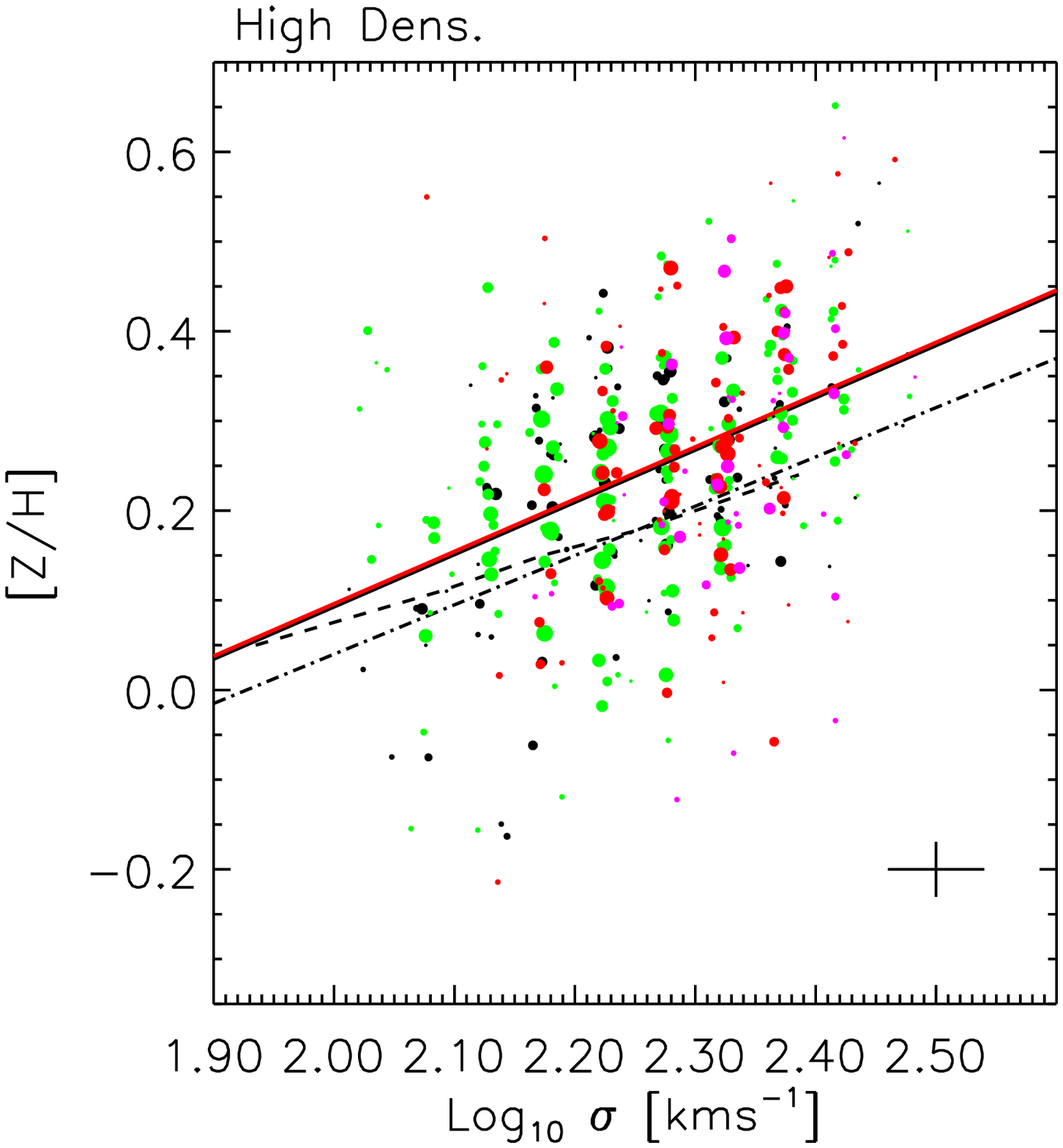}{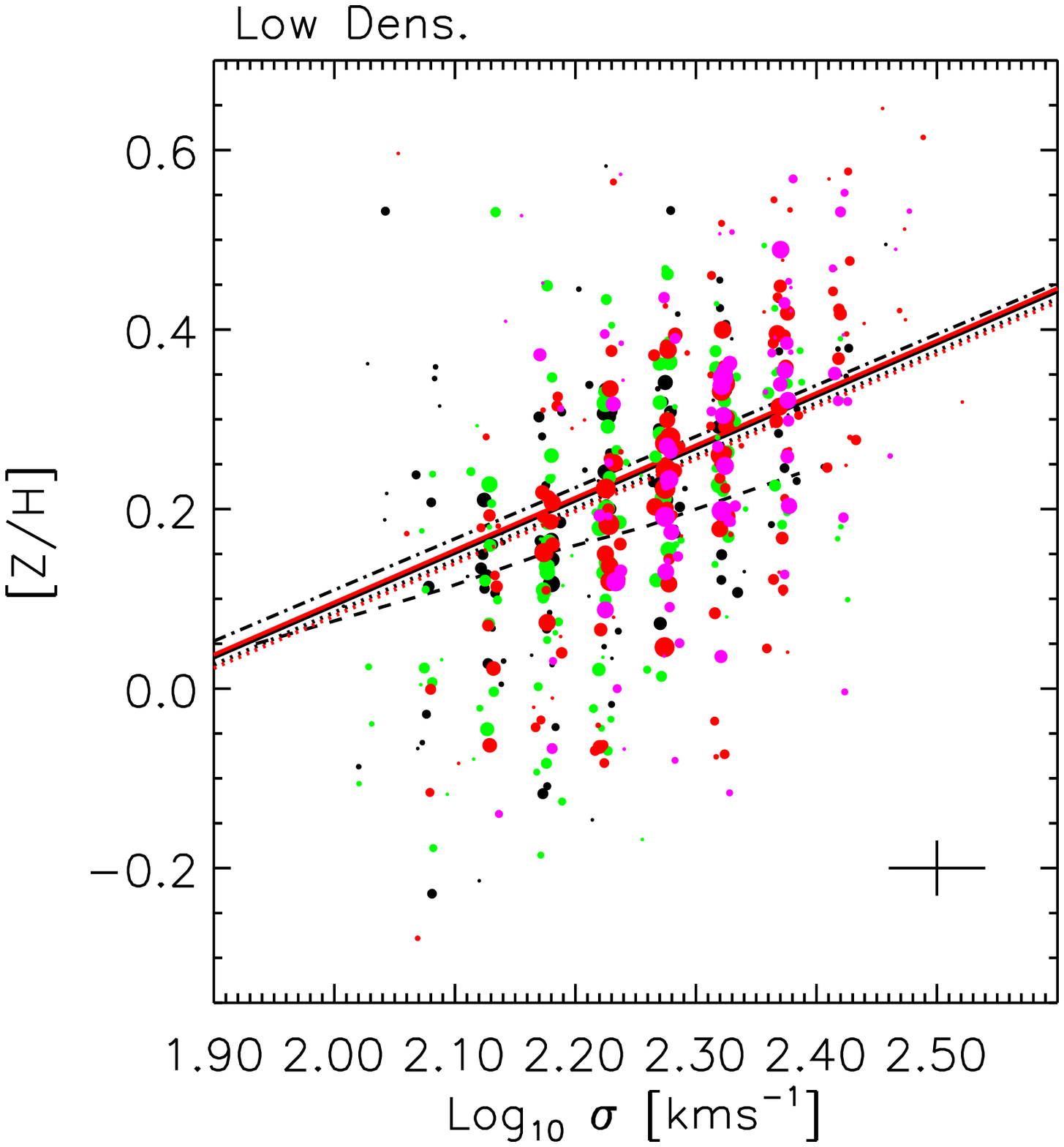}
 \plottwo{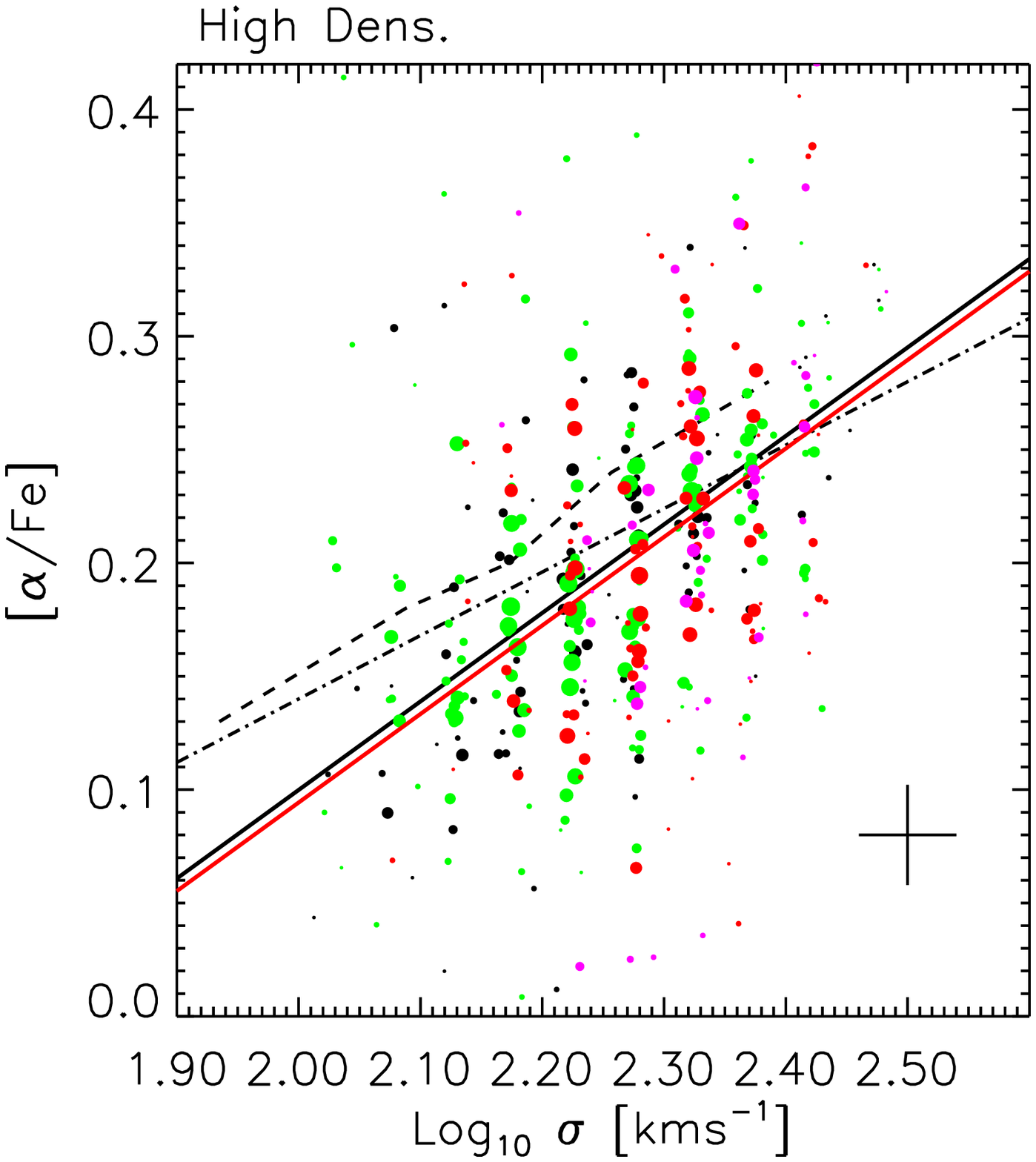}{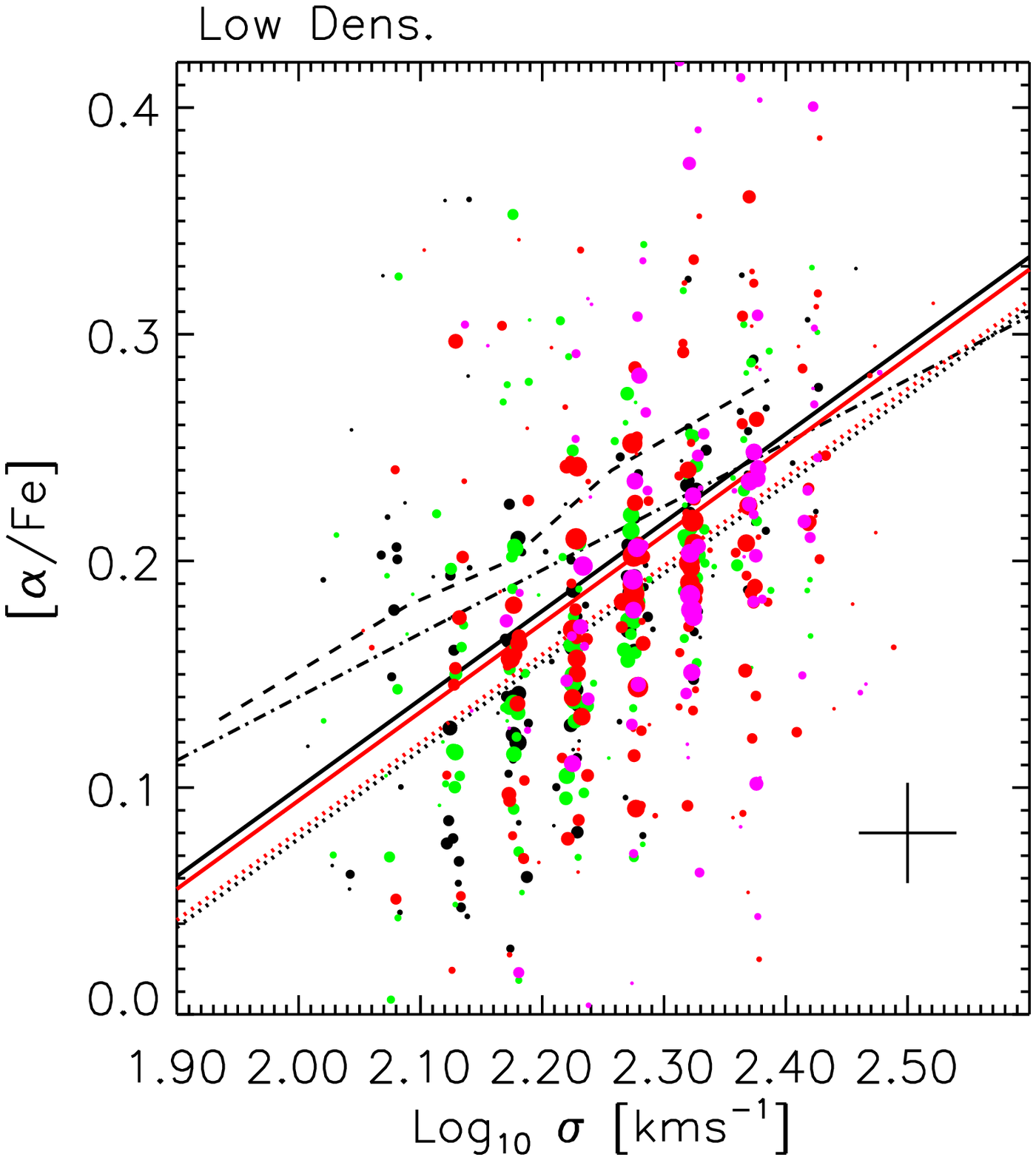}
 \caption{Same as Figure~\ref{modelsigma}, but with 
          H$\gamma_{\rm F}$ instead of H$\beta$.  
          These panels show that the objects in the high 
          redshift population (red lines) are younger than their 
          counterparts of the same $\sigma$ at lower redshift (black 
          lines).  But there is no such trend for [Z/H] or [$\alpha$/Fe].  
          (But see Appendix~\ref{fluxcalib} for a discussion of whether 
          flux calibration problems compromise the results in the 
          lowest redshift bin.)   In addition, objects in lower density 
          regions are slightly younger and less $\alpha$-enhanced than 
          their counterparts in denser regions.}
 \label{modelsigmaHdHg}
\end{figure}

\begin{figure}
 \centering
 \plottwo{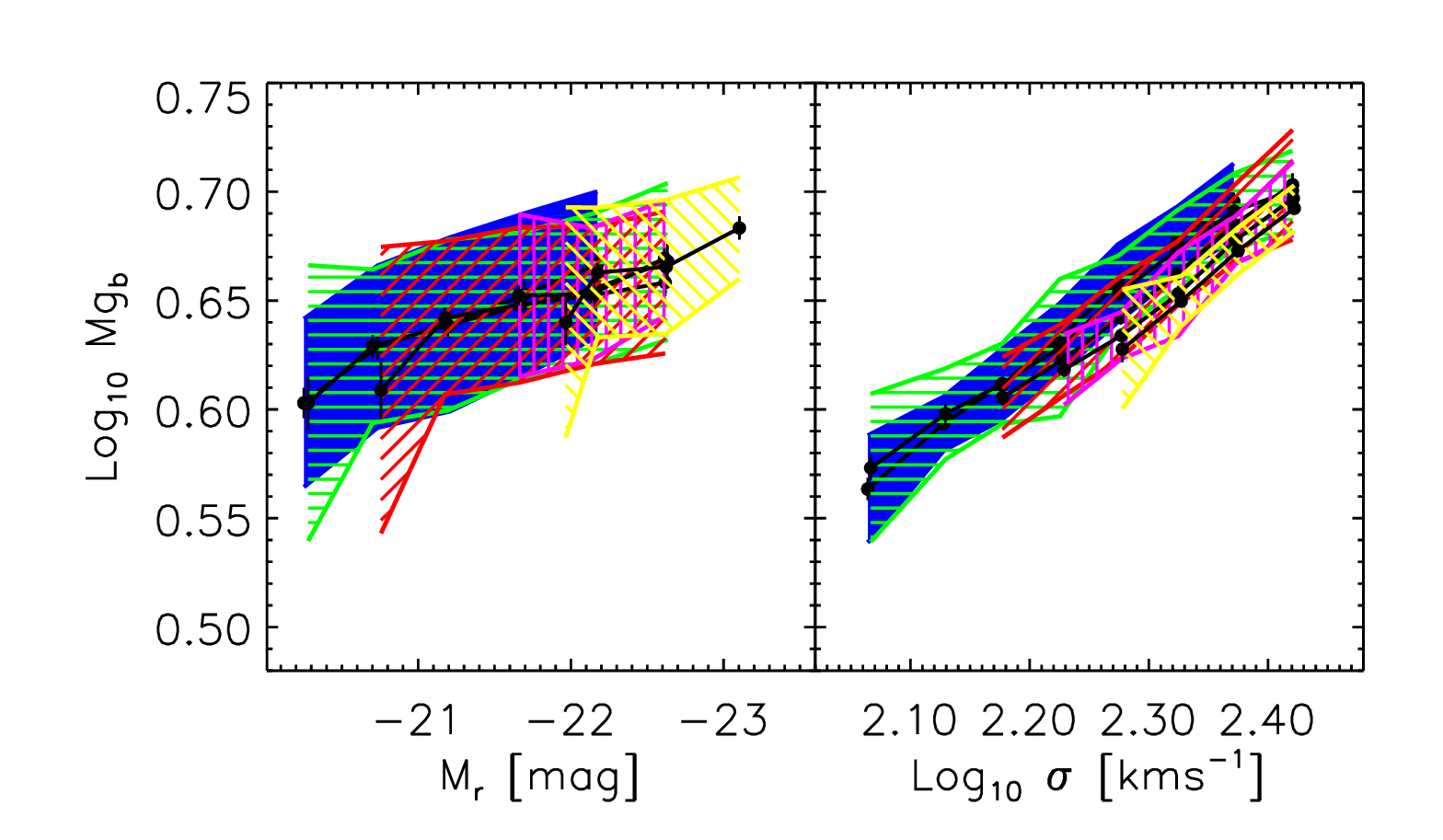}{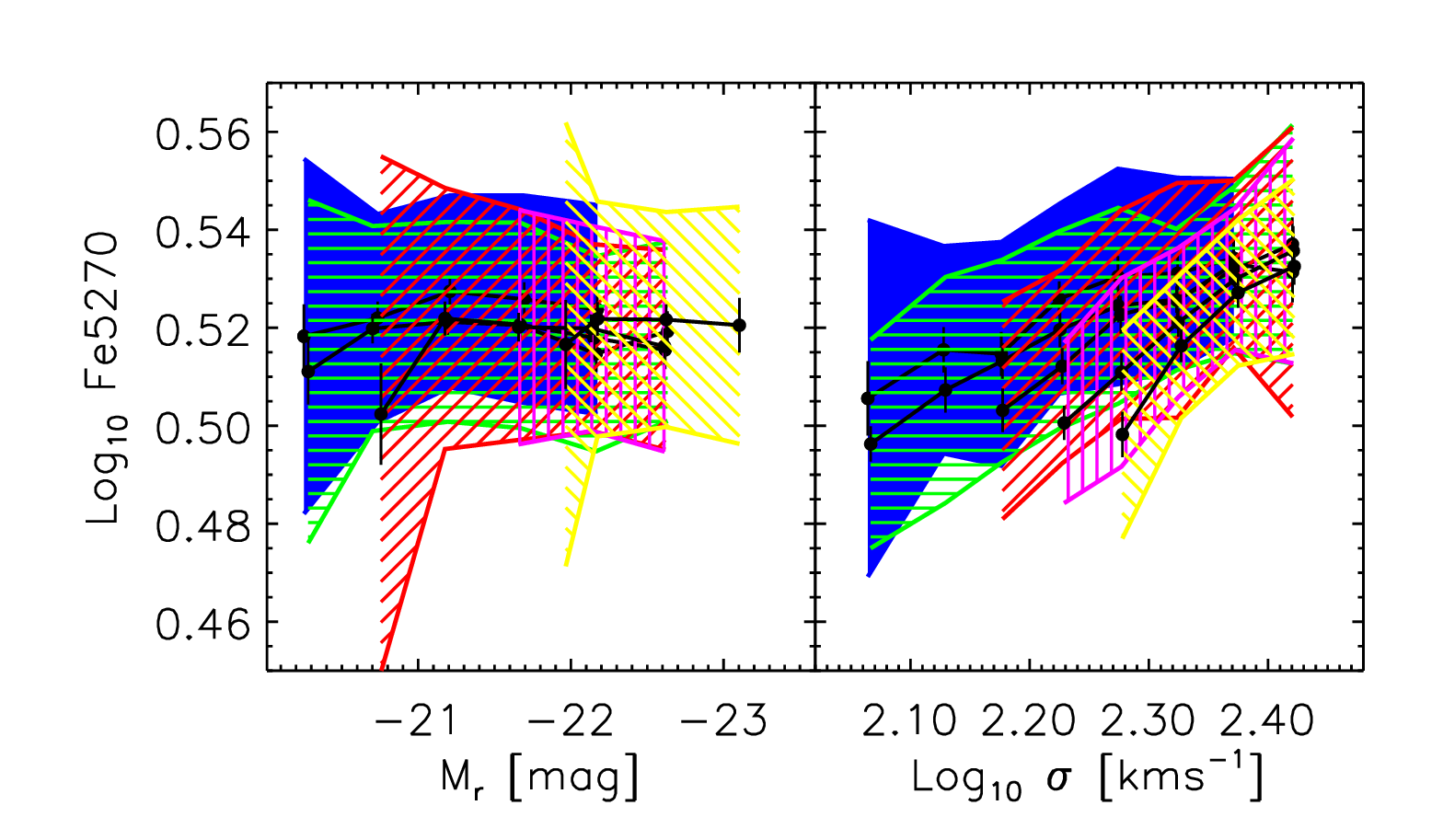}
 \caption{Top: correlation between Mg$_{\rm b}$ and magnitude (left
               panel) and velocity dispersion (right panel).
               Hashed regions show different redshift bins.
          Bottom: correlation between Fe5270 and magnitude (left
               panel) and velocity dispersion (right panel). The
               differential evolution see in the Fe-$\sigma$ relation 
               is due to selection effects (see text for details).}  
 \label{MgFesigma}
\end{figure}

\begin{figure}
 \centering
 \plotone{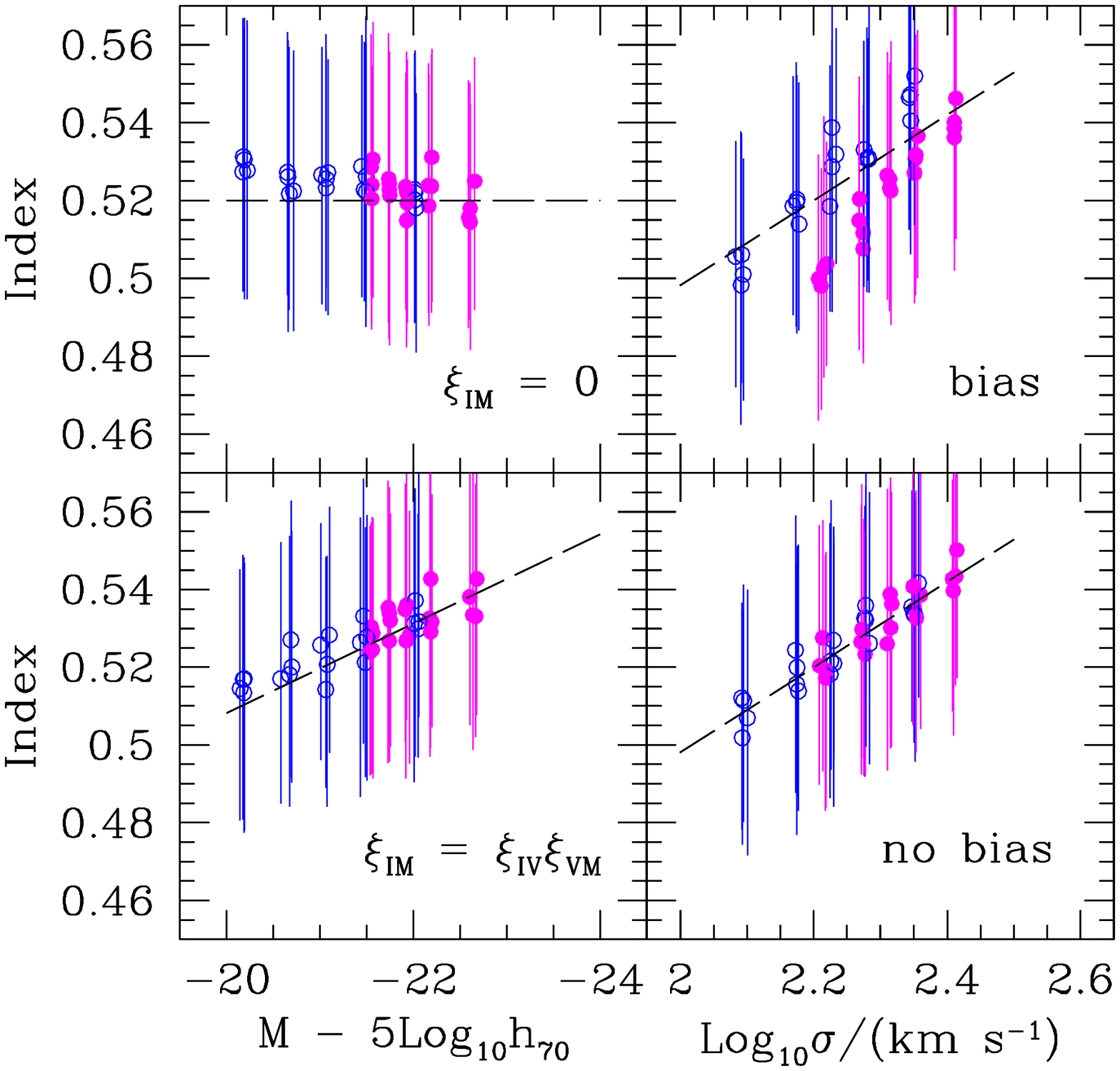}
\caption{Index-luminosity and index-$\sigma$ relations measured 
          from a magnitude limited catalog in low (open circles) 
          and high-$z$ bins (filled circles).  Error bars show 
          the values at the 25th and 75th percentile.  Dashed 
          lines show the true relations.  Top panel shows a model 
          in which index strength does not correlate with magnitude, 
          and bottom panel shows a model in which the index-magnitude 
          correlation is entirely due to the index-$\sigma$ and 
          magnitude-$\sigma$ correlations.  Note the bias in the 
          top right panel, which appears to indicate differential 
          evolution. } 
 \label{biasev}
\end{figure}

\begin{figure}
 \centering
 \plotone{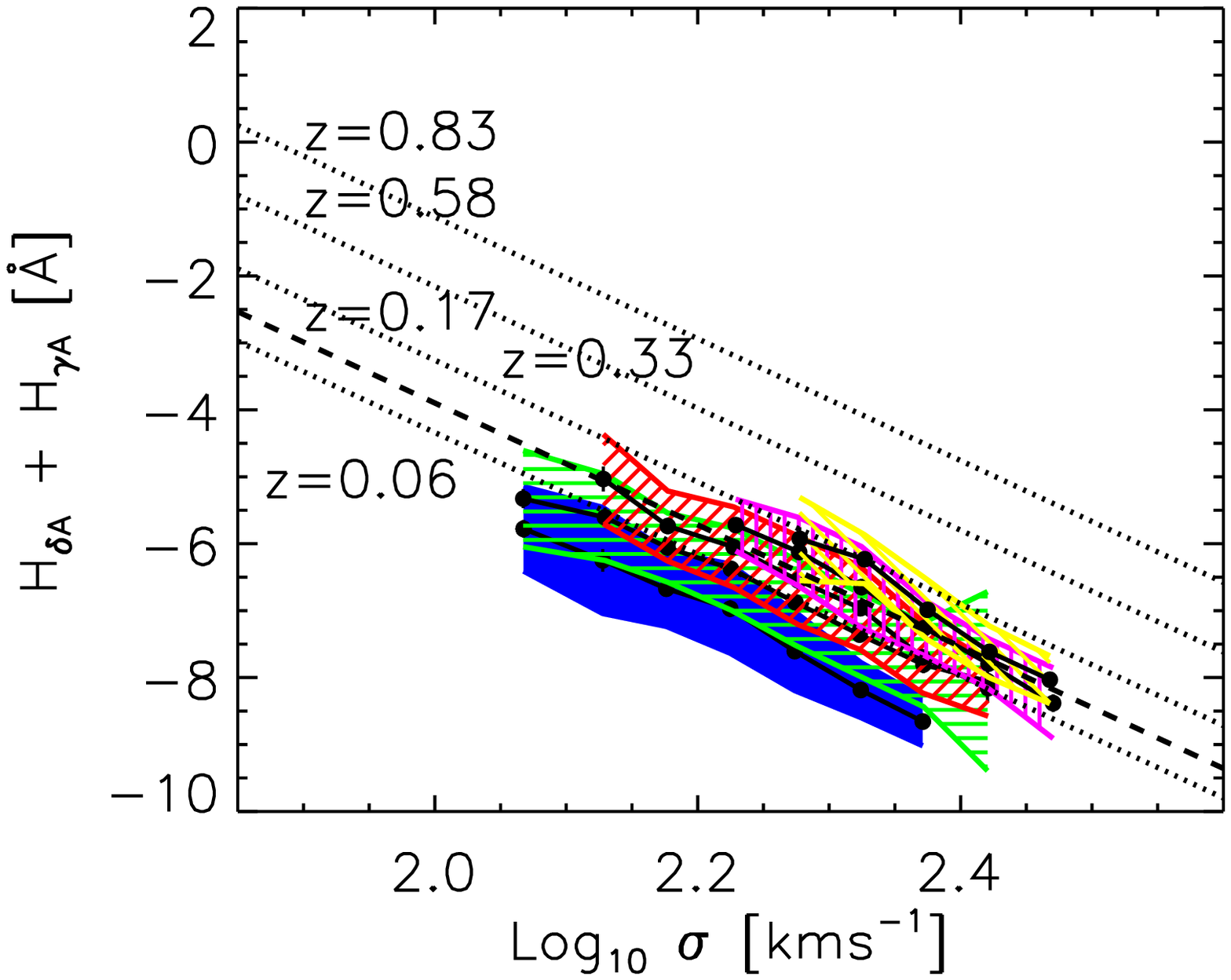}
 \caption{Comparison of the correlation between 
          (H$\delta_{\rm A}+$H$\gamma_{\rm A}$) and velocity dispersion 
          in our sample (shaded regions show different redshift bins) 
          with the relation measured using galaxies in clusters at 
          redshifts $z=0.06$, 0.33, 0.58 and 0.83 (dotted lines, 
          from Kelson et al. 2001).  Dashed line shows the relation 
          obtained by interpolating between the 0.06 and 0.33 lines 
          to $z=0.17$ for ease of comparison with our data.  
          While the SDSS data have the same slope, they have a smaller 
          zero-point at low redshift.  Thus, the apparent evolution 
          in the SDSS sample is about a factor of three larger than 
          indicated by the dotted lines.}
 \label{Hsigma}
\end{figure}

\begin{figure}
 \centering
 \plottwo{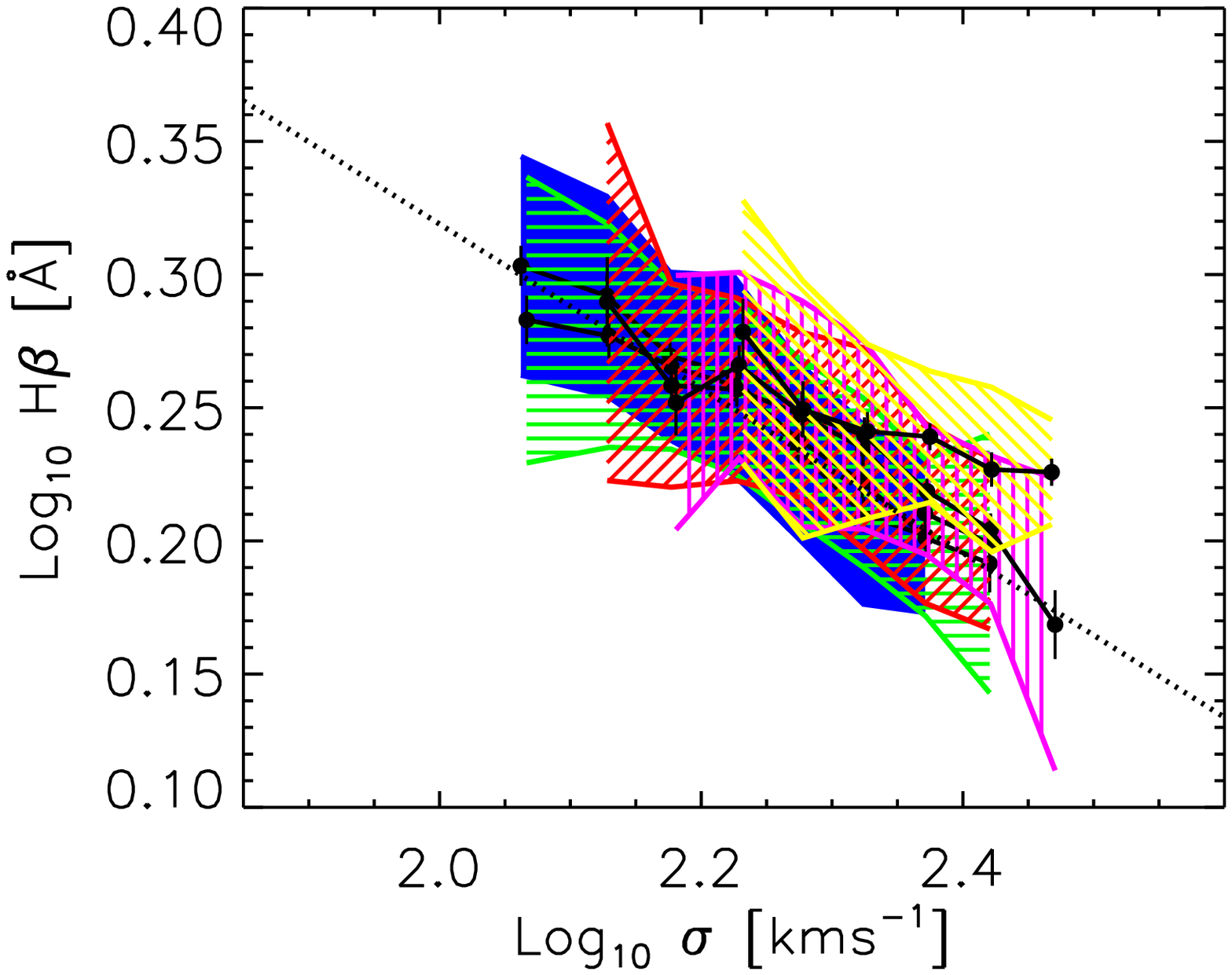}{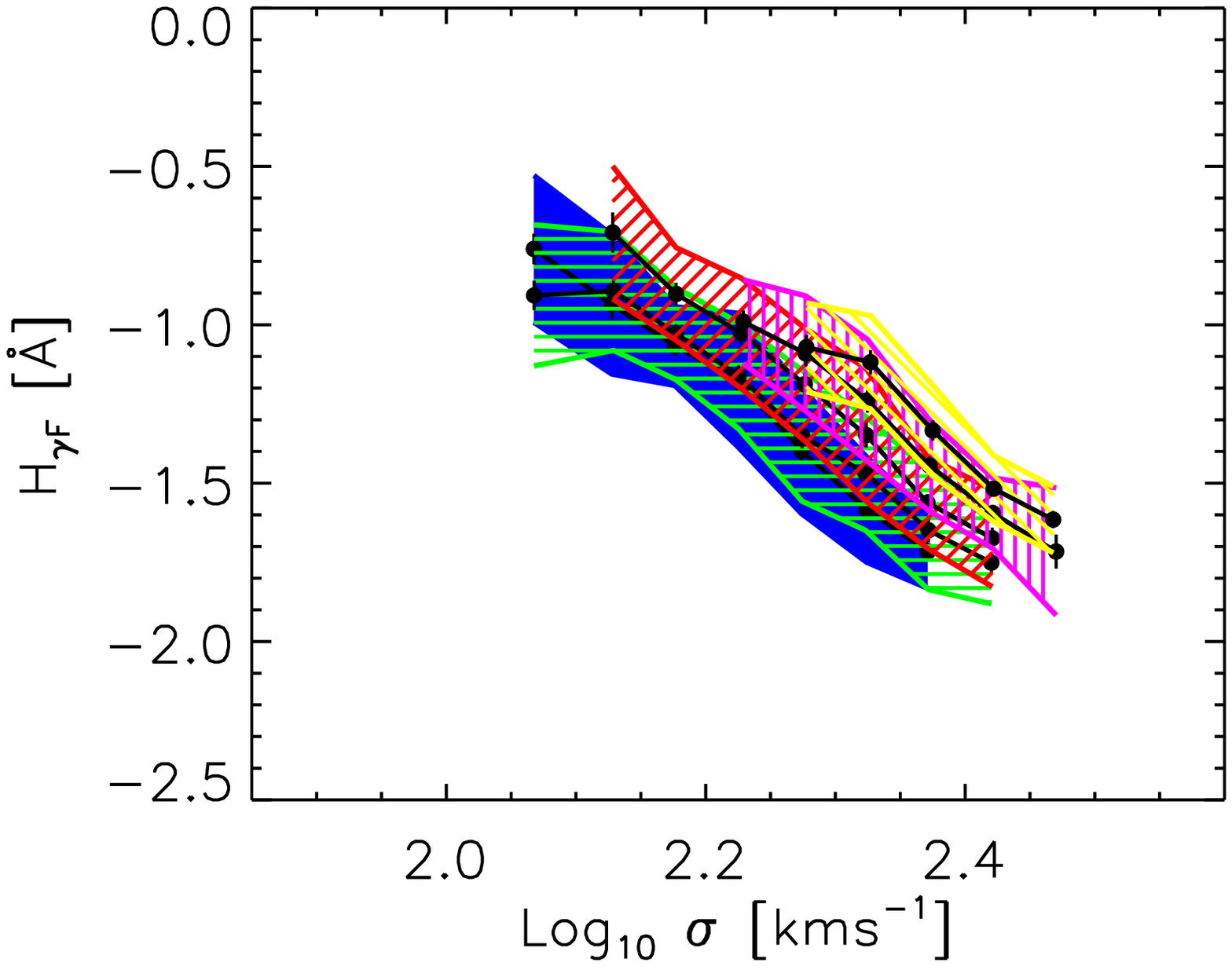}
 \plottwo{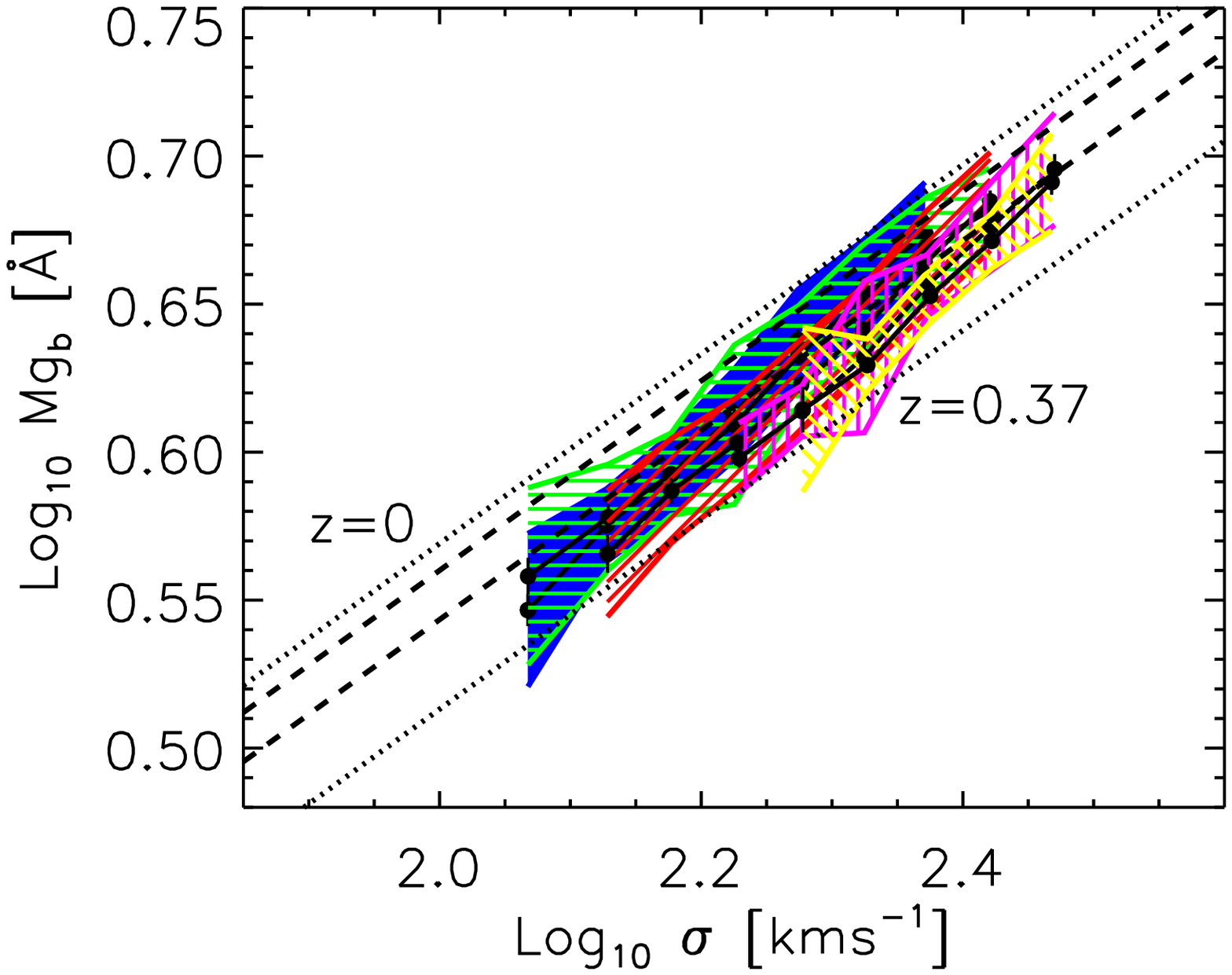}{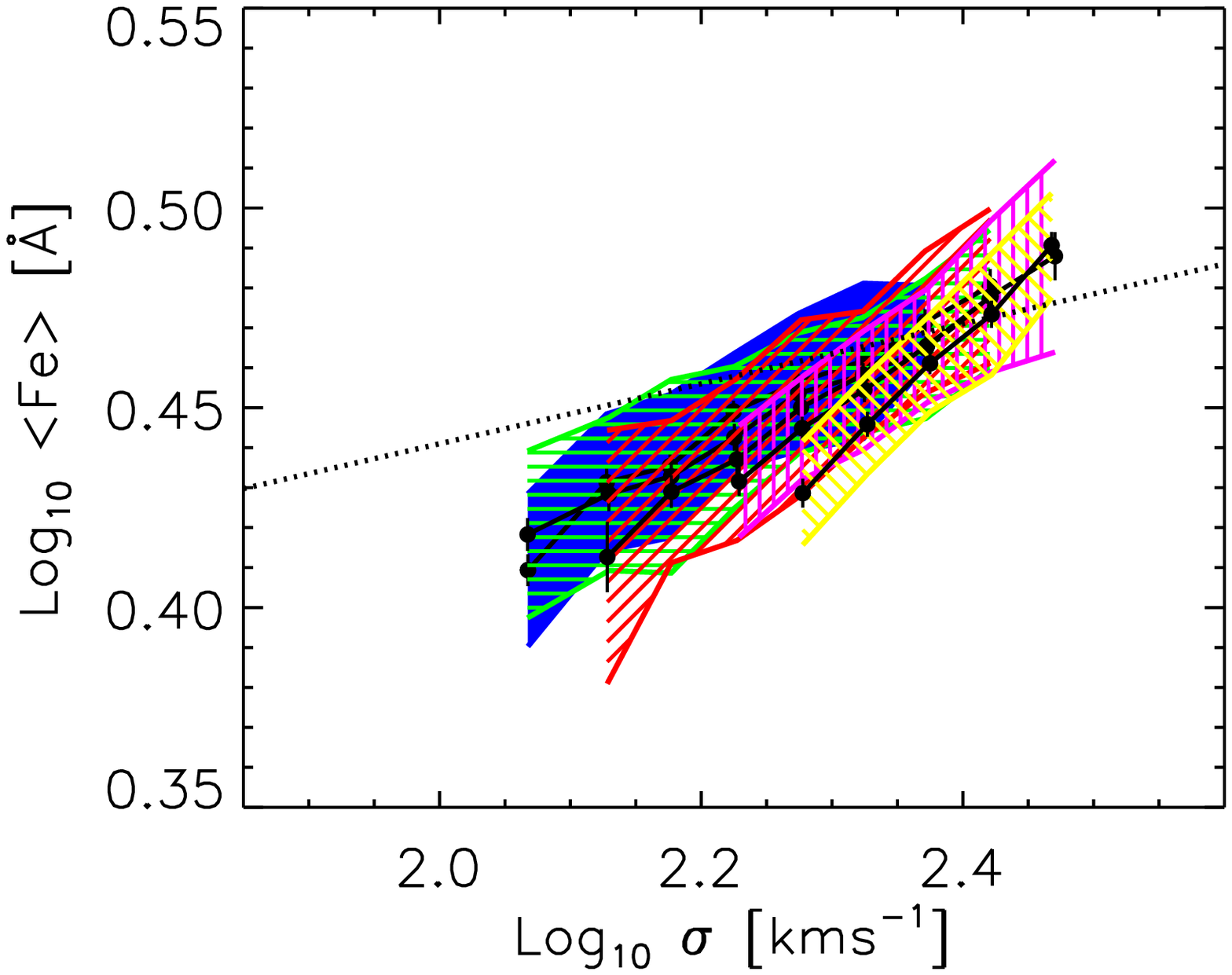}
 \caption{Correlation between the line-index strengths (smoothed 
          to Lick resolution) that are used in Section~\ref{sspmodels} 
          and velocity dispersion.  
          Top left and right panels show the anti-correlation 
          between H$\beta$ and H$\gamma_{\rm F}$ with $\sigma$.   
          Bottom left and right panels show 
          Mg$b$ and $\langle$Fe$\rangle$ are positively correlated 
          with $\sigma$.  
          Hashed colored regions show these relations in different 
          redshift bins.  
          Dotted lines for H$\beta$ and $\langle{\rm Fe}\rangle$ 
          show the correlations from J{\o}rgenson (1997).  
          For the Mg$b$-$\sigma$ relation, dotted lines show the 
          relation at $z=0$ and $z=0.37$ reported by Bender et al. (1996).  
          Dashed lines show their results interpolated to $z=0.06$ 
          and $z=0.17$ to simplify comparison with our results.
          The apparent rapid differential evolution of the 
          Fe-$\sigma$ relation is a selection effect.  }
 \label{HbHgMgFesigma}
\end{figure}

\begin{figure}
 \centering
 \plotone{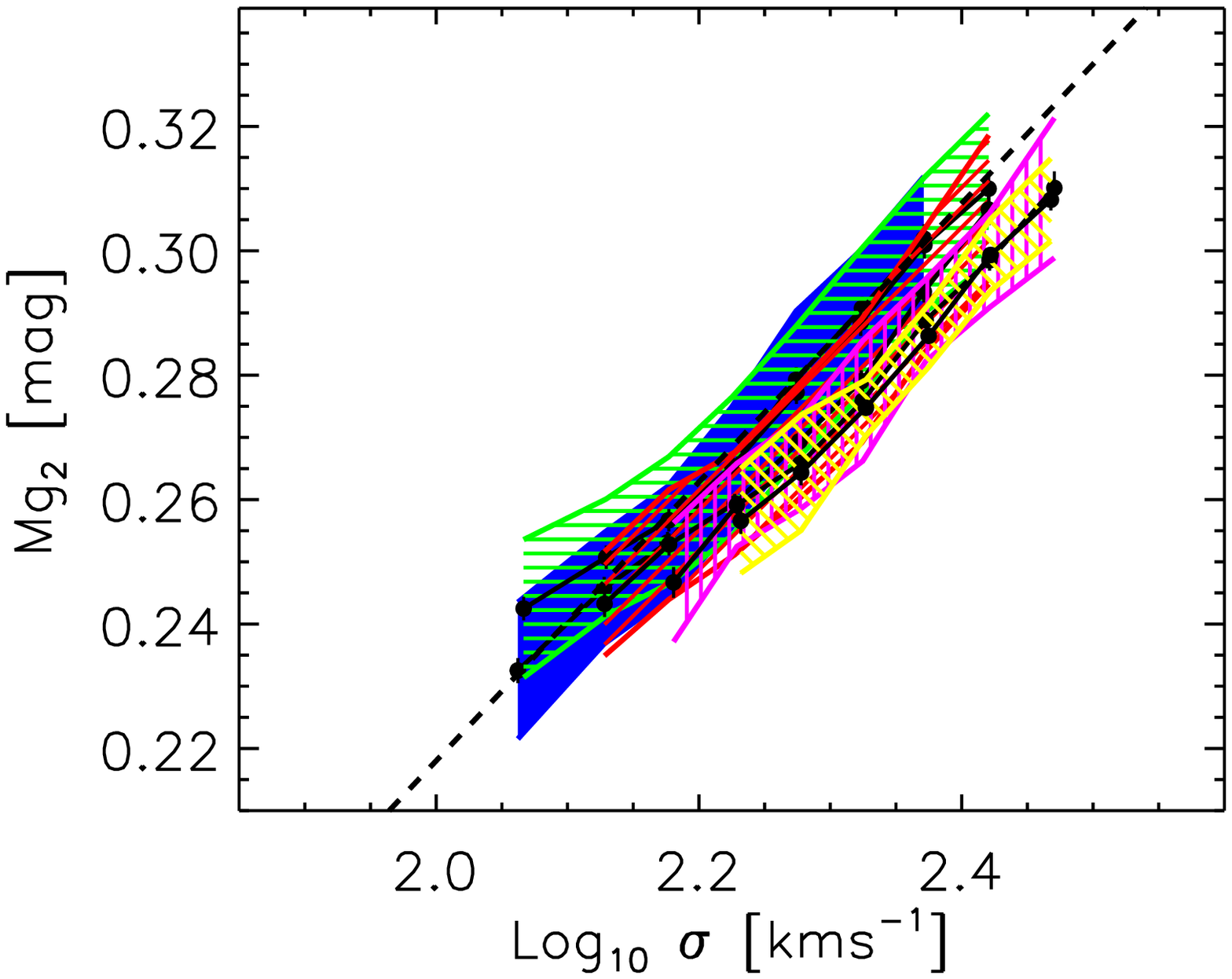}
 \caption{Comparison of correlation between Mg$_2$ and $\sigma$ 
          reported by Bernardi et al. (1998) (dashed line) and the 
          relation in the present data set (hashed colored regions 
          show the relation in different redshift bins).  }
 \label{Mg2sigma}
\end{figure}

\clearpage


\begin{table}
\centering
\caption[t]{Composite spectra.  
Mean values of redshift, evolution-corrected absolute magnitudes, 
velocity dispersion, effective radius, color,  fracDev$_r$, environment 
(0 and 1 indicate low and high density environments), as well as the 
number of galaxies and the mean S/N ratio for the objects which make up 
each composite are given.  This table is available in its entirety in 
the electronic edition of the Journal.\\}
 \begin{tabular}{ccccccccccccccc}
 \tableline
 \tableline
   ID & ID$_z$ & $\langle z\rangle$ & ID$_M$ & $\langle M_r\rangle$ & $\langle M_i\rangle$ & ID$_\sigma$ & $\langle {\rm log_{10}}\sigma\rangle$ & ID$_R$ & $\langle {\rm log_{10}}R \rangle$ & $\langle g-r\rangle$ & $\langle {\rm fDev} \rangle$ & Env & N$_{\rm g}$ & S/N \\
      & & & & [mag] & [mag] & & [km s$^{-1}$] & & [kpc $h^{-1}$] & [mag] & & & & \\
 \tableline
   1 & 1 & 0.0593 & 1 & $-19.873$ & $-20.240$ & 3 & 2.131 & 2 & 0.119 & 0.70 & 0.98 & 1 & 3 & 32 \\
   2 & 1 & 0.0615 & 2 & $-20.147$ & $-20.477$ & 3 & 2.120 & 1 & 0.058 & 0.72 & 0.98 & 1 & 3 & 37 \\
   3 & 1 & 0.0629 & 2 & $-20.204$ & $-20.513$ & 4 & 2.178 & 1 & 0.075 & 0.78 & 0.97 & 1 & 2 & 28 \\
   4 & 1 & 0.0634 & 2 & $-20.244$ & $-20.629$ & 5 & 2.228 & 1 & 0.047 & 0.79 & 0.97 & 1 & 5 & 53 \\
   5 & 1 & 0.0602 & 2 & $-20.349$ & $-20.720$ & 6 & 2.266 & 1 & 0.028 & 0.74 & 1.00 & 1 & 2 & 49 \\
 \tableline
 \end{tabular}
 \label{composites}
\end{table}

\begin{table}
\centering
\caption[t]{Line-strengths and uncertainties measured from the 
            composite spectra given in Table~\ref{composites}.
            This table is available in its entirety in the electronic 
            edition of the Journal.\\}
 \begin{tabular}{ccccccccccccc}
 \tableline
 \tableline
   ID & O{\small II} & err & H$\delta_A$ & err & H$\delta_F$ & err & H$\gamma_A$ & err & H$\gamma_F$ & err & D4000 & err \\
      & [\AA] & [\AA] & [\AA] & [\AA] & [\AA] & [\AA] & [\AA] & [\AA] & [\AA] & [\AA] &  & \\
 \tableline
   1 & $-1.036$ &  0.530 & $-1.579$ &  0.420 &  0.304 &  0.296 & $-4.967$ &  0.357 & $-0.723$ &  0.231 &  1.902 &  0.018\\
   2 & $-1.446$ &  0.484 & $-1.996$ &  0.362 &  0.362 &  0.247 & $-4.791$ &  0.291 & $-1.260$ &  0.191 &  1.865 &  0.015\\
   3 & $-0.210$ &  0.484 & $-1.919$ &  0.423 &  0.477 &  0.286 & $-6.164$ &  0.358 & $-1.873$ &  0.239 &  1.890 &  0.017\\
   4 & $-1.638$ &  0.384 & $-2.739$ &  0.265 &  0.699 &  0.190 & $-5.673$ &  0.222 & $-1.817$ &  0.142 &  1.963 &  0.013\\
   5 &  0.518 &  0.352 & $-2.508$ &  0.277 &  0.120 &  0.201 & $-6.861$ &  0.243 & $-2.527$ &  0.170 &  1.956 &  0.012\\
 \tableline
 \end{tabular}
 \label{linestrengths}
\end{table}

\begin{table}
\centering
\caption[t]{Lick index-strengths and uncertainties measured from the 
            composite spectra given in Table~\ref{composites}.  
            This table is available in its entirety in the electronic 
            edition of the Journal.\\}
 \begin{tabular}{ccccccccccccccccccccc}
 \tableline
 \tableline
   ID & CN$_1$ & err & CN$_2$ & err & Ca4227 & err & G4300 & err & Fe4383 & err & Ca4455 & err & Fe4531 & err & C$_2$4668 & err & H$\beta$ & err & Fe5015 & err \\
      & [mag] &  & [mag] &  & [\AA] &  & [\AA] &  & [\AA] &  & [\AA] &  & [\AA] &  & [\AA] &  & [\AA] &  & [\AA] &  \\
 \tableline
   1 &  0.092 &  0.011 &  0.141 &  0.011 &  1.758 &  0.185 &  5.974 &  0.289 &  6.733 &  0.306 &  2.572 &  0.199 &  4.561 &  0.241 &  6.346 &  0.309 &  1.983 &  0.164 &  5.959 &  0.271\\
   2 &  0.095 &  0.009 &  0.123 &  0.009 &  1.094 &  0.161 &  5.747 &  0.254 &  6.022 &  0.259 &  2.407 &  0.166 &  3.872 &  0.206 &  6.102 &  0.271 &  2.075 &  0.144 &  6.120 &  0.226\\
   3 &  0.146 &  0.012 &  0.201 &  0.012 &  1.347 &  0.203 &  6.044 &  0.305 &  5.452 &  0.340 &  1.756 &  0.193 &  4.144 &  0.277 &  6.264 &  0.353 &  1.326 &  0.192 &  6.071 &  0.298\\
   4 &  0.120 &  0.007 &  0.173 &  0.007 &  1.583 &  0.118 &  6.248 &  0.200 &  5.883 &  0.199 &  2.086 &  0.115 &  3.723 &  0.149 &  7.037 &  0.189 &  1.664 &  0.114 &  6.181 &  0.166\\
   5 &  0.126 &  0.008 &  0.171 &  0.008 &  1.384 &  0.117 &  6.693 &  0.185 &  6.066 &  0.203 &  2.188 &  0.118 &  4.735 &  0.158 &  6.668 &  0.205 &  1.654 &  0.109 &  4.713 &  0.190\\
 \tableline
 \end{tabular}
 \label{lick1}
\end{table}

\begin{table}
\centering
\caption[t]{Lick index-strengths and uncertainties measured from the 
            composite spectra given in Table~\ref{composites}.  
            This table is available in its entirety in the electronic 
            edition of the Journal.\\}
 \begin{tabular}{ccccccccccccccccccccccc}
 \tableline
 \tableline
  ID & Mg$_1$ & err & Mg$_2$ & err & Mg$b$ & err & Fe5270 & err & Fe5335 & err & Fe5406 & err & Fe5709 & err & Fe5782 & err & NaD & err & TiO$_1$ & err & TiO$_2$ & err\\
  & [mag] & & [mag] & & [\AA] & & [\AA] & & [\AA] & & [\AA] &  & [\AA] & & [\AA] & & [\AA] & & [mag] & & [mag] & \\
 \tableline
   1 &  0.137 &  0.004 &  0.256 &  0.006 &  3.702 &  0.186 &  3.613 &  0.361 &  2.543 &  0.193 &  2.071 &  0.156 &  1.365 &  0.118 &  1.367 &  0.102 &  3.104 &  0.140 &  0.052 &  0.005 &  0.095 &  0.003\\
   2 &  0.123 &  0.004 &  0.253 &  0.005 &  4.415 &  0.158 &  2.781 &  0.297 &  2.871 &  0.170 &  2.220 &  0.137 &  1.168 &  0.117 &  0.774 &  0.090 &  3.036 &  0.124 &  0.045 &  0.003 &  0.082 &  0.002\\
   3 &  0.149 &  0.005 &  0.284 &  0.007 &  4.711 &  0.209 &  3.308 &  0.307 &  2.340 &  0.209 &  1.874 &  0.186 &  1.132 &  0.129 &  1.182 &  0.117 &  3.678 &  0.175 &  0.038 &  0.004 &  0.086 &  0.003\\
   4 &  0.144 &  0.003 &  0.287 &  0.003 &  4.656 &  0.108 &  3.879 &  0.164 &  3.115 &  0.109 &  2.200 &  0.094 &  1.301 &  0.071 &  1.015 &  0.062 &  3.908 &  0.092 &  0.047 &  0.002 &  0.090 &  0.001\\
   5 &  0.154 &  0.003 &  0.291 &  0.004 &  4.429 &  0.128 &  3.440 &  0.172 &  3.512 &  0.119 &  2.055 &  0.106 &  0.885 &  0.077 &  1.024 &  0.068 &  3.983 &  0.093 &  0.050 &  0.003 &  0.092 &  0.002\\
 \tableline
 \end{tabular}
 \label{lick2}
\end{table}

\begin{table}
\centering
\caption[t]{Dependence of mean and rms residual from the Fundamental 
            Plane on environment, computed using the coefficients 
            of the orthogonal fit, in different bands.  \\}
 \begin{tabular}{ccccc}
 \tableline
 \tableline
   Band & $\langle\Delta\mu\rangle_{\rm high}$ 
        & $\langle\Delta\mu\rangle_{\rm low}$ 
        & rms$(\Delta\mu)_{\rm high}$    & rms$(\Delta\mu)_{\rm low}$ \\
        & [mag] & [mag] & [mag] & [mag] \\
 \tableline
   $g$  & $0.075\pm 0.006$ & $-0.006\pm 0.005$ & 0.345 & 0.355 \\
   $r$  & $0.055\pm 0.006$ & $-0.020\pm 0.005$ & 0.328 & 0.335 \\
   $i$  & $0.034\pm 0.006$ & $-0.035\pm 0.004$ & 0.324 & 0.322 \\
 \tableline
 \end{tabular}
 \label{fptable}
\end{table}

\begin{table}
\caption[]{Comparison of evolution and environment shown in 
  Figures~\ref{paramfbev}--\ref{lickbev}.   The first column shows 
  the name of the observable; if preceded by the symbol $^*$, 
  then $Y$ in subsequent columns is obtained from $\log_{10}$ of the 
  observable.
  $\langle Y\rangle$ is the (number weighted) mean of $Y$ over the 
  sample of composites; 
  rms$(Y)$ is the rms spread around this mean.  
  Both the mean and rms are obtained by summing over composite spectra 
  weighting each by the number of galaxies in it. 
  $\Delta Y$(Dens) = $Y$(High Dens)$-Y$(Low Dens).  
  Roman numerals I and II are for galaxies with $-22\le M_i\le -21$ in 
  redshift bins $0.05 < z < 0.07$ and $0.07 < z < 0.09$; 
  Roman numerals III and IV are for galaxies with $-23\le M_i\le -22$ in 
  redshift bins $0.09 < z < 0.12$ and $0.12 < z < 0.14$.  The final column 
  shows $\Delta Y$(Evol) $=Y(z\sim 0.06)-Y(z\sim 0.17)$ for galaxies with 
  $2.35\le \log_{10}\sigma\le 2.4$; if followed by the superscript `$^1$', 
  the reported value has been corrected for flux-calibration problems by 
  dividing the observed value by a factor of three 
  (c.f. Appendix~\ref{fluxcalib}).  \\}
\small
\begin{tabular}{lrrrrrrr}
 \tableline
 \tableline
  Parameter & $\langle Y\rangle$ & rms($Y$) & $\Delta Y$(Dens)-I & $\Delta Y$(Dens)-II & $\Delta Y$(Dens)-III & $\Delta Y$(Dens)-IV & $\Delta Y$(Evol)\\
  \tableline
$M_{r}$ [mag]                & -21.91& 0.71 & $ 0.054\pm 0.038 $ & $ 0.027\pm 0.033 $ & $ 0.087\pm 0.035 $ & $-0.021\pm 0.047 $ & $ 0.091\pm 0.028 $\\
$^*\sigma$ [kms$^{-1}$]      &  2.26 & 0.10 & $ 0.001\pm 0.011 $ & $-0.001\pm 0.010 $ & $ 0.002\pm 0.010 $ & $ 0.015\pm 0.010 $ & $ 0.007\pm 0.009 $\\
$^*R_{o,r}\sigma^2$        &  5.02 & 0.35 & $ 0.011\pm 0.027 $ & $ 0.071\pm 0.023 $ & $-0.003\pm 0.023 $ & $ 0.049\pm 0.026 $ & $-0.017\pm 0.021 $\\
$^*R_{o,r}$ [kpc $h^{-1}$] &  0.49 & 0.23 & $ 0.004\pm 0.021 $ & $ 0.024\pm 0.019 $ & $ 0.026\pm 0.020 $ & $ 0.006\pm 0.021 $ & $-0.011\pm 0.017 $\\
$g - r$ [mag]                &  0.73 & 0.04 & $ 0.016\pm 0.005 $ & $ 0.003\pm 0.004 $ & $ 0.024\pm 0.005 $ & $ 0.032\pm 0.006 $ & $ 0.031\pm 0.004 $\\
fracDev                      &  0.96 & 0.04 & $-0.003\pm 0.006 $ & $-0.004\pm 0.004 $ & $-0.002\pm 0.007 $ & $ 0.006\pm 0.007 $ & $ 0.006\pm 0.004 $\\
  \tableline

OII [\AA]              &  2.29 & 1.52 & $-0.757\pm 0.128 $ & $-0.759\pm 0.117 $ & $-0.793\pm 0.151 $ & $-0.469\pm 0.166 $ & $-0.702\pm 0.119 $ \\
H$\delta_{\rm A}$ [\AA]   & -1.72 & 0.93 & $-0.358\pm 0.111 $ & $-0.263\pm 0.080 $ & $-0.373\pm 0.103 $ & $-0.421\pm 0.135 $ & $-0.372\pm 0.073^1$ \\
H$\delta_{\rm F}$ [\AA]   &  0.66 & 0.46 & $-0.077\pm 0.062 $ & $-0.131\pm 0.038 $ & $-0.057\pm 0.049 $ & $-0.092\pm 0.066 $ & $-0.146\pm 0.035^{1} $ \\
H$\gamma_{\rm A}$ [\AA]   & -5.37 & 0.85 & $-0.287\pm 0.110 $ & $-0.280\pm 0.082 $ & $-0.432\pm 0.085 $ & $-0.399\pm 0.117 $ & $-0.235\pm 0.064^{1} $ \\
H$\gamma_{\rm F}$ [\AA]   & -1.41 & 0.54 & $-0.226\pm 0.071 $ & $-0.138\pm 0.051 $ & $-0.275\pm 0.051 $ & $-0.150\pm 0.070 $ & $-0.118\pm 0.040^{1} $ \\
D4000                  &  1.90 & 0.08 & $ 0.024\pm 0.009 $ & $ 0.026\pm 0.007 $ & $ 0.028\pm 0.008 $ & $ 0.036\pm 0.010 $ & $ 0.035\pm 0.007^{1} $ \\
$^*$[MgFe]$^{'}$       &  0.58 & 0.04 & $ 0.021\pm 0.005 $ & $ 0.008\pm 0.004 $ & $ 0.009\pm 0.004 $ & $ 0.017\pm 0.006 $ & $ 0.011\pm 0.003 $ \\
$^*$Mg$b$/$\langle$Fe$\rangle$    &  0.13 & 0.05 & $ 0.024\pm 0.005 $ & $ 0.003\pm 0.004 $ & $ 0.003\pm 0.005 $ & $ 0.012\pm 0.007 $ & $ 0.012\pm 0.003 $ \\
 \tableline

CN$_1$ [mag]        &  0.11 & 0.03 & $ 0.010\pm 0.004 $ & $ 0.002\pm 0.003 $ & $ 0.009\pm 0.004 $ & $ 0.010\pm 0.004 $ & $ 0.010\pm 0.004^{1} $\\
CN$_2$ [mag]        &  0.16 & 0.04 & $ 0.018\pm 0.005 $ & $-0.002\pm 0.004 $ & $ 0.018\pm 0.005 $ & $ 0.017\pm 0.006 $ & $ 0.010\pm 0.004^{1} $\\
$^*$Ca4227 [\AA]    &  0.19 & 0.06 & $ 0.026\pm 0.007 $ & $-0.001\pm 0.006 $ & $ 0.007\pm 0.007 $ & $ 0.019\pm 0.007 $ & $ 0.011\pm 0.004^{1} $\\
$^*$G4300 [\AA]     &  0.77 & 0.03 & $ 0.010\pm 0.003 $ & $ 0.003\pm 0.003 $ & $ 0.008\pm 0.003 $ & $ 0.013\pm 0.005 $ & $ 0.006\pm 0.002^{1} $\\
$^*$Fe4383 [\AA]    &  0.76 & 0.05 & $ 0.017\pm 0.005 $ & $ 0.009\pm 0.005 $ & $-0.014\pm 0.005 $ & $ 0.009\pm 0.007 $ & $ 0.010\pm 0.003 $\\
$^*$Ca4455 [\AA]    &  0.33 & 0.07 & $ 0.021\pm 0.007 $ & $ 0.020\pm 0.006 $ & $-0.001\pm 0.008 $ & $ 0.005\pm 0.010 $ & $ 0.005\pm 0.005 $\\
$^*$Fe4531 [\AA]    &  0.61 & 0.04 & $-0.002\pm 0.004 $ & $-0.001\pm 0.004 $ & $-0.003\pm 0.004 $ & $-0.006\pm 0.005 $ & $ 0.004\pm 0.002 $\\
$^*$C$_2$4668 [\AA] &  0.85 & 0.06 & $-0.003\pm 0.007 $ & $-0.002\pm 0.006 $ & $ 0.021\pm 0.007 $ & $ 0.006\pm 0.008 $ & $ 0.036\pm 0.005 $\\
$^*$H$\beta$ [\AA]  &  0.25 & 0.07 & $-0.005\pm 0.008 $ & $ 0.002\pm 0.007 $ & $-0.010\pm 0.008 $ & $-0.010\pm 0.009 $ & $-0.013\pm 0.005 $\\
$^*$Fe5015 [\AA]    &  0.77 & 0.05 & $ 0.007\pm 0.005 $ & $ 0.009\pm 0.004 $ & $ 0.020\pm 0.006 $ & $ 0.014\pm 0.007 $ & $-0.002\pm 0.003 $\\
Mg$_1$ [mag]        &  0.14 & 0.02 & $ 0.009\pm 0.003 $ & $ 0.004\pm 0.002 $ & $ 0.006\pm 0.002 $ & $ 0.011\pm 0.002 $ & $ 0.010\pm 0.002 $\\
Mg$_2$ [mag]        &  0.27 & 0.03 & $ 0.015\pm 0.004 $ & $ 0.007\pm 0.003 $ & $ 0.014\pm 0.004 $ & $ 0.011\pm 0.004 $ & $ 0.017\pm 0.003 $\\
$^*$Mg$b$ [\AA]     &  0.64 & 0.06 & $ 0.026\pm 0.007 $ & $ 0.020\pm 0.006 $ & $ 0.026\pm 0.006 $ & $ 0.016\pm 0.009 $ & $ 0.023\pm 0.005 $\\
$^*$Fe5270 [\AA]    &  0.52 & 0.05 & $ 0.003\pm 0.005 $ & $ 0.005\pm 0.004 $ & $-0.004\pm 0.005 $ & $ 0.009\pm 0.007 $ & $ 0.005\pm 0.003 $\\
$^*$Fe5335 [\AA]    &  0.50 & 0.06 & $ 0.005\pm 0.006 $ & $ 0.004\pm 0.006 $ & $ 0.007\pm 0.007 $ & $ 0.020\pm 0.008 $ & $ 0.004\pm 0.003 $\\
$^*$Fe5406 [\AA]    &  0.31 & 0.06 & $ 0.006\pm 0.006 $ & $ 0.009\pm 0.005 $ & $-0.004\pm 0.006 $ & $ 0.001\pm 0.009 $ & $ 0.005\pm 0.003 $\\
$^*$Fe5709 [\AA]    &  0.03 & 0.07 & $ 0.000\pm 0.006 $ & $ 0.001\pm 0.007 $ & $-0.011\pm 0.009 $ & $-0.018\pm 0.009 $ & $-0.015\pm 0.004 $\\
$^*$Fe5782 [\AA]    & -0.02 & 0.09 & $ 0.019\pm 0.009 $ & $ 0.008\pm 0.008 $ & $-0.023\pm 0.009 $ & $-0.011\pm 0.016 $ & $ 0.012\pm 0.005 $\\
$^*$NaD [\AA]       &  0.60 & 0.08 & $ 0.009\pm 0.009 $ & $ 0.002\pm 0.008 $ & $-0.005\pm 0.008 $ & $ 0.026\pm 0.009 $ & $ 0.049\pm 0.007 $\\
TiO$_1$ [mag]       &  0.04 & 0.01 & $ 0.003\pm 0.001 $ & $-0.003\pm 0.001 $ & $ 0.001\pm 0.001 $ & $ 0.003\pm 0.001 $ & $ 0.002\pm 0.001 $\\
TiO$_2$ [mag]       &  0.08 & 0.01 & $ 0.002\pm 0.001 $ & $ 0.004\pm 0.001 $ & $ 0.002\pm 0.001 $ & $ 0.002\pm 0.001 $ & $ 0.004\pm 0.001 $\\
 \tableline
\end{tabular}
\label{evenvtab} 
\end{table}
\normalsize

\begin{table}
\centering
\caption[t]{Lick index-strengths and uncertainties used to measure 
            age, [Z/H] and [$\alpha$/Fe] from SSP models.
            These were measured from the composite spectra given in 
            Table~\ref{composites} after convolving to Lick resolution.   
            This table is available in its entirety in the electronic 
            edition of the Journal.  \\}
 \begin{tabular}{ccccccccccc}
 \tableline
 \tableline
   ID & H$\gamma_F$ & err & H$\beta$ & err & Mg$b$ & err & Fe5270 & err & Fe5335 & err\\
      & [\AA] & [\AA] & [\AA] & [\AA] & [\AA] & [\AA] & [\AA] & [\AA] & [\AA] & [\AA]  \\
\tableline
   1 &  -0.650 &   0.232 & 1.887 &  0.164  & 3.524  & 0.187  & 3.086  & 0.363  & 2.018  & 0.195\\
   2 &  -0.983 &   0.192 & 1.935 &  0.144  & 4.138  & 0.160  & 2.511  & 0.298  & 2.226  & 0.172\\
   3 &  -1.664 &   0.239 & 1.346 &  0.191  & 4.436  & 0.211  & 2.843  & 0.309  & 1.883  & 0.211\\
   4 &  -1.524 &   0.143 & 1.654 &  0.114  & 4.457  & 0.108  & 3.474  & 0.165  & 2.531  & 0.110\\
   5 &  -2.248 &   0.170 & 1.643 &  0.109  & 4.370  & 0.129  & 2.966  & 0.173  & 2.839  & 0.121\\
 \tableline
 \end{tabular}
 \label{lickres}
\end{table}

\begin{table}
\centering
\caption[t]{Age, [Z/H] and [$\alpha$/Fe] obtained using the Lick 
            index-strengths reported in Table~\ref{lickres} and the 
            SSP models of TMB03-TMK04.
            Two sets of quantities are reported: the first set of 
            columns were obtained using H$\beta$, Mg$b$ and 
            $\langle{\rm Fe}\rangle$, while the final six 
            columns were obtained after substituting H$\gamma_{\rm F}$ 
            for H$\beta$.
            Objects with ages set to zero are those for which the 
            models did not return reliable estimates of age, [Z/H] and 
            [$\alpha$/Fe].  This can be either because they lie beyond 
            the model grids, or because the three ways of estimating 
            the ages gave very different answers.  
            This table is available in its entirety in the electronic 
            edition of the Journal.\\}
 \begin{tabular}{ccccccccccccc}
 \tableline
 \tableline
   ID & Age$_{{\rm H}\beta}$ & err & [Z/H]$_{{\rm H}\beta}$ & err & [$\alpha/$Fe]$_{{\rm H}\beta}$ & err &
        Age$_{{\rm H}\gamma_{\rm F}}$ & err & [Z/H]$_{{\rm H}\gamma_{\rm F}}$ & err & [$\alpha/$Fe]$_{{\rm H}\gamma_{\rm F}}$ & err \\
      & [Gyrs] & [Gyrs] &  & & & & [Gyrs] & [Gyrs] & & & & \\
\tableline
   1 & 6.9 & 2.7 & -0.017 &  0.089 &  0.102 &  0.065 & 4.8 & 2.4 &  0.059 &  0.082 &  0.123 &  0.044\\
   2 & 5.9 & 2.1 &  0.105 &  0.082 &  0.326 &  0.059 & 7.6 & 3.4 &  0.062 &  0.076 &  0.313 &  0.042\\
   3 & 0.0 & 0.0 &  0.000 &  0.000 &  0.000 &  0.000 & 0.0 & 0.0 &  0.000 &  0.000 &  0.000 &  0.000\\
   4 & 8.7 & 3.3 &  0.302 &  0.131 &  0.144 &  0.042 & 6.4 & 2.4 &  0.359 &  0.051 &  0.162 &  0.036\\
   5 & 9.6 & 3.4 &  0.225 &  0.107 &  0.150 &  0.069 & 0.0 & 0.0 &  0.000 &  0.000 &  0.000 &  0.000\\
 \tableline
 \end{tabular}
 \label{derivedTZA}
\end{table}

\begin{table}
\centering
\caption[t]{Correlations of derived parameters, age, [Z/H] and 
            [$\alpha$/Fe] with velocity dispersion.  The first 
            two rows show the fits (slope and zero-point) to the 
            high density sample at $z=0.06$.  
            The following three rows show the offsets with 
            respect to these fits for these relations at different 
            redshifts. The last four rows show the offsets between different
            environments at different redshifts; the slope of the fit was kept 
            fixed to the value given by the high density sample at $z=0.06$.\\}
 \begin{tabular}{lrrr}
 \tableline
 \tableline

 Using H$\beta$  & Log$_{10}$ Age - Log$_{10} \sigma$ & [Z/H] - Log$_{10} \sigma$ & [$\alpha$/Fe] - Log$_{10} \sigma$\\
slope &                         1.15                   &  0.38               &  0.32             \\
ZP &                            $-1.72\pm0.31$          &  $-0.64\pm0.013$    &  $-0.54\pm0.005$  \\
$\Delta$ZP$_{Evol(z\sim0.06 - z\sim0.08)}$ &           $0.27\pm0.17$ [Gyrs]   &  $0.008\pm0.009$   &  $0.001\pm0.004$  \\
$\Delta$ZP$_{Evol(z\sim0.06 - z\sim0.11)}$ &           $1.21\pm0.31$ [Gyrs]   &  $-0.011\pm0.013$   &  $-0.002\pm0.006$  \\ 
$\Delta$ZP$_{Evol(z\sim0.06 - z\sim0.13)}$ &           $0.81\pm0.47$ [Gyrs]   &  $0.008\pm0.019$   &  $0.005\pm0.010$  \\ 
$\Delta$ZP$_{Env(z\sim0.06)}$ &    $0.42\pm 0.38$ [Gyrs]  &  $0.018\pm 0.016$   &  $0.023\pm 0.006$ \\
$\Delta$ZP$_{Env(z\sim0.08)}$ &    $0.41\pm 0.28$ [Gyrs]  &  $0.016\pm 0.014$   &  $0.017\pm 0.005$ \\
$\Delta$ZP$_{Env(z\sim0.11)}$ &    $-0.27\pm0.37$ [Gyrs]  &  $0.044\pm0.016$    &  $0.025\pm0.007$  \\
$\Delta$ZP$_{Env(z\sim0.13)}$ &    $0.63\pm 0.55$ [Gyrs]  &  $0.025\pm 0.022$   &  $0.017\pm 0.011$ \\ 
\\
 \tableline
 Using H$\gamma_{\rm F}$  & Log$_{10}$ Age - Log$_{10} \sigma$ & [Z/H] - Log$_{10} \sigma$ & [$\alpha$/Fe] - Log$_{10} \sigma$\\
slope &                         0.81                   &  0.58               &  0.39             \\
ZP &                            $-0.98\pm0.23$          &  $-1.07\pm0.010$    &  $-0.68\pm0.005$  \\
$\Delta$ZP$_{Evol(z\sim0.06 - z\sim0.08)}$ &           $0.17\pm0.14$ [Gyrs]   &  $0.006\pm0.007$   &  $0.002\pm0.003$  \\
$\Delta$ZP$_{Evol(z\sim0.06 - z\sim0.11)}$ &           $0.95\pm0.20$ [Gyrs]   &  $-0.003\pm0.011$   &  $0.005\pm0.006$  \\ 
$\Delta$ZP$_{Evol(z\sim0.06 - z\sim0.13)}$ &           $1.17\pm0.41$ [Gyrs]   &  $0.001\pm0.017$   &  $0.006\pm0.010$  \\ 
$\Delta$ZP$_{Env(z\sim0.06)}$ &    $0.58\pm 0.28$ [Gyrs]  &  $0.008\pm 0.013$   &  $0.022\pm 0.006$ \\
$\Delta$ZP$_{Env(z\sim0.08)}$ &    $0.47\pm 0.21$ [Gyrs]  &  $0.016\pm 0.011$   &  $0.017\pm 0.005$ \\
$\Delta$ZP$_{Env(z\sim0.11)}$ &    $0.60\pm 0.24$ [Gyrs]  &  $0.015\pm0.013$    &  $0.014\pm0.007$  \\
$\Delta$ZP$_{Env(z\sim0.13)}$ &    $0.95\pm 0.46$ [Gyrs]  &  $0.010\pm 0.020$   &  $0.014\pm 0.011$ \\ 
 \tableline
 \end{tabular}
 \label{sigmafits}
\end{table}

\end{document}